\documentclass{emulateapj}
\usepackage{natbib}
\usepackage{graphicx}
\usepackage{graphics,epsfig}
\usepackage{tabularx}
\usepackage{adjustbox,lipsum}
\usepackage{subfigure}
\usepackage{amsmath,amsfonts,amsthm,bm} 
\usepackage{float}
\usepackage{courier}
\usepackage{color}
\usepackage[dvipsnames]{xcolor}
\usepackage{colortbl}
\usepackage{threeparttable}
\usepackage[colorlinks=true,linkcolor=red,citecolor=blue]{hyperref}
\usepackage{blindtext}
\usepackage{placeins}
\singlespace

\newcommand{\wmap}{{\sl WMAP}}

\newcommand{\planck}{{\it Planck}}
\newcommand{\ba}{\begin{eqnarray}}
\newcommand{\ea}{\end{eqnarray}}

\newcommand{\lsim}   {\mbox{$_<\atop^{\sim}$}}
\newcommand{\gsim}   {\mbox{$_>\atop^{\sim}$}}
\newcommand{\lt}     {\mbox{$<$}}
\newcommand{\gt}     {\mbox{$>$}}

\newcommand{\freqa}   {$98$}
\newcommand{\freqb}   {$150$}

\newcommand  \beq    {\begin{equation}}

\newcommand  \eeq    {\end{equation}}
\newcommand  \gtsim  {\lower.5ex\hbox{$\; \buildrel > \over \sim \;$}} 
\newcommand  \ltsim  {\lower.5ex\hbox{$\; \buildrel < \over \sim \;$}}

\newcommand{\LCDM} {$\Lambda$CDM}

\newcommand{\ukarcmin} {$\mu$K--arcmin}

\begin{document}

\title{The Atacama Cosmology Telescope: DR4 Maps and Cosmological Parameters}
%%%%%%%%%%%%%%%%%%%%%%%%%%%%%%%%%%%%%%%%%%%%%%%%%%%%%%%%%%%%%%%%%%%%%%%%%%%
% WARNING: This LaTeX block was generated automatically by authors.py
% Do not change by hand: your changes will be lost.

\author{
Simone~Aiola\altaffilmark{1,2},
Erminia~Calabrese\altaffilmark{3},
Lo\"ic~Maurin\altaffilmark{4,5},
Sigurd~Naess\altaffilmark{1},
Benjamin~L.~Schmitt\altaffilmark{6},
Maximilian~H.~Abitbol\altaffilmark{7},
Graeme~E.~Addison\altaffilmark{8},
Peter~A.~R.~Ade\altaffilmark{3},
David~Alonso\altaffilmark{7},
Mandana~Amiri\altaffilmark{9}
% ...
}
\altaffiltext{1}{Center for Computational Astrophysics, Flatiron Institute, 162 5th Avenue, New York, NY 10010 USA}
\altaffiltext{2}{Joseph Henry Laboratories of Physics, Jadwin Hall,
Princeton University, Princeton, NJ, USA 08544}
\altaffiltext{3}{School of Physics and Astronomy, Cardiff University, The Parade, 
Cardiff, Wales, UK CF24 3AA}
\altaffiltext{4}{Universit\'e Paris-Saclay, CNRS, Institut d'astrophysique spatiale, 91405, Orsay, France.}
\altaffiltext{5}{Instituto de Astrof\'isica and Centro de Astro-Ingenier\'ia, Facultad de F\`isica, Pontificia Universidad Cat\'olica de Chile, Av. Vicu\~na Mackenna 4860, 7820436 Macul, Santiago, Chile}
\altaffiltext{6}{Department of Physics and Astronomy, University of
Pennsylvania, 209 South 33rd Street, Philadelphia, PA, USA 19104}
\altaffiltext{7}{Department of Physics, University of Oxford, Keble Road, Oxford, UK OX1 3RH}
\altaffiltext{8}{Dept. of Physics and Astronomy, The Johns Hopkins University, 3400 N. Charles St., Baltimore, MD, USA 21218-2686}
\altaffiltext{9}{Department of Physics and Astronomy, University of
British Columbia, Vancouver, BC, Canada V6T 1Z4}
% ...

% End auto-generated block
%%%%%%%%%%%%%%%%%%%%%%%%%%%%%%%%%%%%%%%%%%%%%%%%%%%%%%%%%%%%%%%%%%%%%%%%%%%

\begin{abstract}
We present new arcminute-resolution maps of the Cosmic Microwave Background temperature and polarization anisotropy from the Atacama Cosmology Telescope, using data taken from 2013--2016 at \freqa\ and \freqb\,GHz. 
The maps cover more than 17,000\,deg$^2$, the deepest 600\,deg$^2$ with noise levels below $10$\ukarcmin. We use the power spectrum derived from almost 6,000\,deg$^2$ of these maps to constrain cosmology. 
The ACT data enable a measurement of the angular scale of features in both the divergence-like polarization and the temperature anisotropy, tracing both the velocity and density at last-scattering. From these one can derive the distance to the last-scattering surface and thus infer the local expansion rate, $H_0$. By combining ACT data with large-scale information from \wmap\ we measure $H_0=67.6\pm 1.1$\,km/s/Mpc, at 68\% confidence, in excellent agreement with the independently-measured \planck\ satellite estimate (from ACT alone we find $H_0=67.9\pm 1.5$\,km/s/Mpc).  
The \LCDM\ model provides a good fit to the ACT data, and we find no evidence for deviations: both the spatial curvature, and the departure from the standard lensing signal in the spectrum, are zero to within 1$\sigma$;
the number of relativistic species, the primordial Helium fraction, and the running of the spectral index are consistent with \LCDM\ predictions to within 1.5--2.2$\sigma$. 
We compare ACT, \wmap, and \planck\ at the parameter level and find good consistency; we investigate how the constraints on the correlated spectral index and baryon density parameters readjust when adding CMB large-scale information that ACT does not measure. The DR4 products presented here will be publicly released on the NASA Legacy Archive for Microwave Background Data Analysis.
\end{abstract}

\maketitle

\section{\label{sec:intro}Introduction}
\setcounter{footnote}{0}

The cosmic microwave background (CMB) provides the tightest constraints on our model of the universe of any current data source. Successive microwave background measurements, including \wmap, \planck, the South Pole Telescope (SPT), and the Atacama Cosmology Telescope (ACT), have produced increasingly precise and detailed maps of the microwave sky in both temperature and linear polarization \citep{bennett/etal:2013, planck2016-l01, henning/etal:2018, louis_atacama_2017}. The result has been a remarkably simple standard model of cosmology described by only six parameters quantifying the composition of the universe, the initial conditions of structure formation, and the reionization history, most of which are now determined at the percent level \citep{wmap_spergel_2003, planck2016-l06}.

Despite the simplicity of this model, known as \LCDM, and the consistency with many other cosmological probes \citep[e.g.,][]{2016ApJ...830..148C,alam/etal:2017,Abbott:2017wau,Hikage:2018qbn}, there are some tensions between its inferred parameter values and those estimated from low-redshift observations, most notably for the local expansion rate, $H_0$. While $H_0$ measured from Type Ia supernovae distances calibrated using tip-of-the-red-giant-branch (TRGB) stars is consistent with the \planck\ estimate \citep{freedman/etal:2019}, the rate measured from distances calibrated using Cepheid variable stars \citep{riess/etal:2019}, or by using time delays from the H0LiCOW strong gravitational lenses (\citealp{wong/etal:2019}, but see recent updates from \citealp{Birrer:2020tax} which give different results), is significantly higher than that estimated from \planck. \cite{riess/etal:2019} estimates that the deviation is at the 3--5$\sigma$ level. Determining whether or not a new component of the stardard model or new process is needed to explain all the current observations is of great interest, with new models being explored that could, for example, reduce the sound horizon in order to fit \planck\ CMB data with an increased local Hubble expansion rate \citep[e.g.,][]{cyr-racine/sigurdson:2014,poulin/etal:2019,Knox:2019rjx}.
One of ACT's goals is to make a second, experimentally-independent, measurement of the inferred Hubble constant with a precision comparable to \planck\ to help assess the robustness of the current estimate from the high-redshift CMB data.

The \planck\ satellite's full-sky maps in many frequency bands set the current standard for cosmological precision \citep{planck2016-l06}. The \planck\ temperature map does nearly as well as possible in measuring the angular power spectrum of temperature anisotropies: cosmic variance dominates measurement uncertainty at angular multipoles from $\ell=2$ to $\ell=1600$. In polarization \planck 's measurements are signal dominated up to $\ell=900$ in the $E$-mode polarization power spectrum, and up to $\ell=1150$ in the cross-correlation between temperature and $E$-mode polarization \citep{planck2014-a13}. Both SPT and ACT have been extending the reach of \planck\ by making new high resolution, high sensitivity temperature and E-mode polarization measurements \citep{ naess_atacama_2014, louis_atacama_2017,henning/etal:2018}. From its site in northern Chile, ACT has now mapped roughly half the sky. The gain in information beyond \planck\ is in new temperature measurements that constrain damping tail physics at $\ell>1500$ as well as secondary signals, and improved $E$-mode polarization measurements that directly map the velocity of the photon-baryon fluid at recombination, providing a second view of the physics at last scattering.

In this paper we present new maps made with ACT observations at \freqa\,GHz and \freqb\,GHz during 2013--2016. They incorporate data used to make the maps presented in \citet{naess_atacama_2014} (N14, hereafter) and \citet{louis_atacama_2017} (L17, hereafter), and additionally include data taken during nighttime observations in 2015 and 2016. Previous analyses used only 650\,deg$^2$ of sky; this analysis produces maps covering 17,000\,deg$^2$. We present the cosmological parameters derived from these new temperature and polarization sky maps from ACT, using the information contained in their angular power spectra. A companion paper, \citet{choi_atacama_2020} (C20, hereafter), details our methods for estimating and validating the angular power spectra and shows the power spectrum results, together with such technical considerations as noise characterization and a wide range of robustness tests both at the power spectrum and likelihood level. It also describes the procedure to extract foreground-marginalized CMB-only spectra starting from the frequency spectra and likelihood. The data products presented in this paper and in C20 comprise ACT's fourth data release, DR4, and will be available on the NASA Legacy Archive for Microwave Background Data Analysis.

This paper (elsewhere referred to as A20) is organized as follows. In \S\ref{sec:obs} and \ref{sec:data} we describe the observations and low-level data processing, including ancillary data products such as beams. \S\ref{sec:mapmake} presents our current methods for making maximum-likelihood maps. The resulting ACT DR4 maps are unveiled in \S\ref{sec:maps}. These maps are then employed in C20 to extract CMB power spectra. \S\ref{sec:like} of this paper describes our cosmological CMB-only likelihood function and the methods we use to apply it to the power spectra from C20. In \S\ref{sec:lcdm}, we show cosmological constraints from ACT data alone and from ACT combined with \wmap; these measurements of the CMB sky are independent from the \planck\ measurements. We explore extensions to the \LCDM\ model in \S\ref{sec:ext}. The concluding \S\ref{sec:concl} discusses the improvements we can expect from the substantially larger ACT data sets collected in 2017--2020.

\section{Overview of Nighttime Observations}
\label{sec:obs}
ACT is a 6-meter off-axis Gregorian telescope, located at an  elevation of 5190\,m on Cerro Toco in the Atacama Desert in Chile. In 2013 the ACTPol receiver was first deployed, upgrading ACT to a polarization sensitive instrument. Each year from 2013--2015 a new polarization array (PA) was deployed. The receiver holds up to three PAs, each contained in an optics tube and accompanied by a set of three mm-wave silicon optics, a set of corrugated feedhorns, and multiplexing readout for the detectors. The instrument is described in detail in \cite{thornton/2016}.

In this paper we analyse data taken by ACT with the ACTPol camera at nighttime, defined to be between 2300 and 1100 UTC. The observations in 2013 (s13, hereafter), initially presented in N14, consist of 648 hours, after data cuts, divided between three deep regions D1, D5, and D6 surveyed with one detector array, PA1, at \freqb\,GHz. In 2014 (s14, hereafter), observations conducted with two \freqb\,GHz detector arrays, PA1 and PA2, over a larger region called D56 amount to 391 hours after cuts, and were initially presented in L17. In 2015 (s15, hereafter) a third dichroic detector array, PA3, operating at \freqa\ and \freqb\,GHz was added. A total of 1271 post-cut hours were spent surveying the D56 region, a wider region called BN overlapping with the Baryon Oscillation Spectroscopic Survey (BOSS) \citep{boss_dr13:2017}, and a deep region in the southern equatorial sky called D8. Finally, observations in 2016 (s16, hereafter) consisted of 710 hours after cuts, spread over a wide region covering roughly $40\%$ of the sky, which we call AA (for AdvACT survey). As mentioned above, we restrict this analysis to only data taken with the ACTPol camera, meaning that for s16 observations we only consider data from PA2 and PA3. Data taken with the dichroic PA4 detector array of the AdvACT camera, at \freqb~and 220\,GHz, which replaced PA1 in 2016, will be part of a later analysis together with data taken from 2017 onwards in the AA region. We also neglect the small amount of test data taken with a spinning half-wave plate installed in front of the PA2 and PA3 arrays.

Table~\ref{table:obs} summarizes the ACT s13--s16 observations considered in this work, including the number of detectors used to make the sky maps, detector array sensitivities, and the number of hours of data used in the maps. A detailed list of region names and areas can be found in \S\ref{subsec:mapproducts}.
%Table~\ref{table:maps_stats}. 
More details regarding scan parameters and survey strategy are presented in \citet{debernardis/2016} and reviewed in C20 in the context of null tests.

\begin{figure}[t!]
    \centering
    \includegraphics[width=0.5\textwidth]{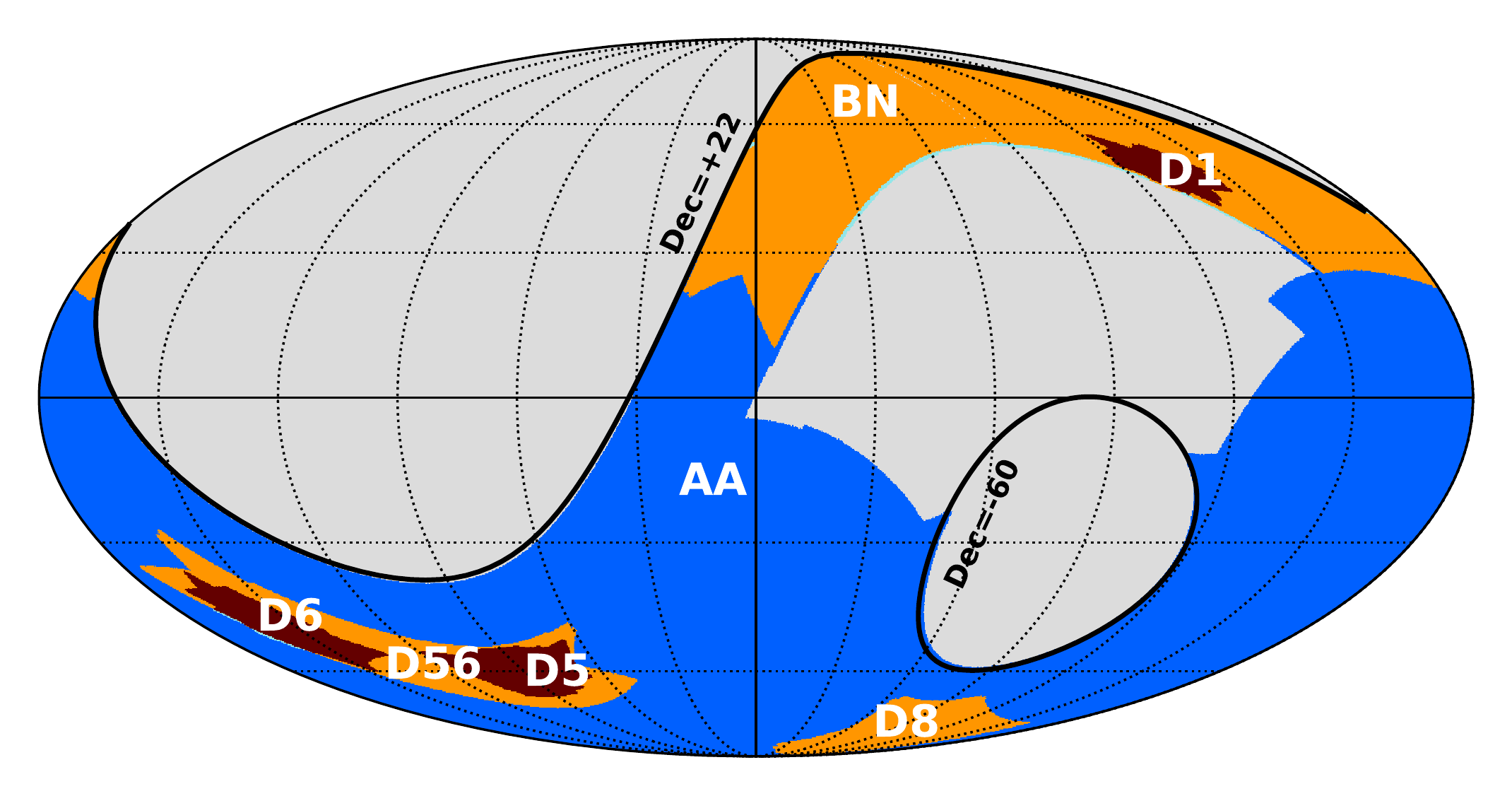}
    \caption{Mollweide projection of the ACT DR4 footprints in Galactic coordinates. The D1, D5, and D6 regions were observed in season s13, D56 in seasons s14 and s15, D8 and BN in s15. The AA region was observed in s16 and covers roughly 40\% of the sky. D1 overlaps with BN, D5 and D6 overlap with D56, and all regions overlap with AA. Two constant-declination lines are shown in black roughly indicating the observable sky from Chile. The grey region between ${\rm Dec}=[-60,+22]$ was not observed, although accessible, to maximize overlap with current optical surveys. The grid spacing in both Galactic longitude and latitude is $30^{\circ}$.
    \vspace*{0.15in}}
    \label{fig:mollview}
\end{figure}

Figure~\ref{fig:mollview} shows a Mollweide projection of the ACT DR4 region footprints in Galactic coordinates. The new ACT observation strategy allows us to independently map roughly 50\% of the clean sky used for the \wmap\ \citep{bennett/etal:2013} and \planck\ \citep{planck2016-l05} cosmological analyses at higher resolution and sensitivity, especially in polarization. In addition, we provide CMB data overlapping with on-going and future optical surveys, such as BOSS, DESI, DES, HSC, KiDS, and Rubin Observatory \citep{boss_dr13:2017, desi_2018, des_2018, hsc_2019, kids_2017, lsst_2019}.

\begin{table*}[t!]
\caption{Summary of s13--s16 observations, camera, and instrument performance considered in this work\label{table:obs}}
\vspace{-0.15in}
\begin{center}
\begin{tabular}{lccc|c|ccc|c}
\hline\hline
\\
Season&\multicolumn{3}{c}{\rm s13}&\multicolumn{1}{c}{\rm s14}&\multicolumn{3}{c}{\rm s15}&\multicolumn{1}{c}{\rm s16}\\
Regions & D1&D5&D6 & D56 & D56&BN&D8 & AA \\
\hline
& & & & & & & & \\
\emph{{\rm PA1} \freqb\,{\rm GHz}} & & & & & & & & \\
Median $\mathrm{N_{det}}$$^{\rm (a)}$ & 493&475&487 & 456 & 299&269&284 & \\
Array Sensitivity$^{\rm (b)}$ [$\mu$K$\sqrt{\rm s}$] & 16.7&20.2&17.6 & 20.2 & 24.7&29.1&27.9 & \\
$\mathrm{t_{obs}}$$^{\rm (c)}$ [hrs] & 114&256&278 & 380 & 521&498&121 & \\
\hline
& & & & & & & & \\
\emph{{\rm PA2} \freqb\,{\rm GHz}} & & & & & & & \\
Median $\mathrm{N_{det}}$  & & & & 753 & 617&689&717 & 671 \\
Array Sensitivity [$\mu$K$\sqrt{\rm s}$]  & & & & 13.3 & 16.5&17.6&17.1 & 16.5 \\
$\mathrm{t_{obs}}$ [hrs]  & & & & 402 & 527&621&172 & 912 \\
\hline
& & & & & & & & \\
\emph{{\rm PA3} \freqa\,{\rm GHz}} & & & & & & & & \\
Median $\mathrm{N_{det}}$ & & & & & 386&387&386 & 384 \\
Array Sensitivity [$\mu$K$\sqrt{\rm s}$] & & & & & 15.5&16.7&16.2 & 15.9 \\
$\mathrm{t_{obs}}$ [hrs] & & & & & 649&829&181 & 704 \\
\hline
& & & & & & & & \\
\emph{{\rm PA3} \freqb\,{\rm GHz}} & & & & & & & & \\
Median $\mathrm{N_{det}}$ & & & & & 369&397&396 & 399 \\
Array Sensitivity [$\mu$K$\sqrt{\rm s}$] & & & & & 21.1&22.5&21.6 & 20.6 \\
$\mathrm{t_{obs}}$ [hrs] & & & & & 491&715&152 & 515 \\
\hline\hline
\end{tabular}
\end{center}
\vspace{-0.05in}
Notes: $(a)$ Median number of detectors considered for map-making. We do not account for the fraction of samples masked within the single detector time-stream. In 2015 the addition of the third detector array, PA3, led to an increase of the bath temperature, which caused the number of working detectors in PA1 and PA2 to drop. In general, a small loss of detectors is expected from season to season due to cryogenic cycles deteriorating the quality of the connections between detectors and readout elements. $(b)$ The array sensitivity is estimated on a single 10-minute chunk of time-ordered data (TOD) from the map-maker noise model and inverse-variance averaged within the dataset. Variations across seasons for the same array and region are due to fluctuations in atmospheric loading and number of mapped detectors. Variations across regions within the same season and array are due to differences in the elevations of the scans (and thus atmospheric loading). These numbers can be compared with the ones estimated from planet observation presented in C20. $(c)$ Total amount of data mapped, measured in hours of exposure. This number is computed by summing the duration of each data chunk that passed TOD-level cuts. It does not consider the fraction of samples within the TOD that were flagged (thus not used), which would lead to only a small correction to the number provided.\\
\end{table*}

\section{Data Selection and Characterization}
\label{sec:data}

\subsection{\label{subsec:preproc}Data selection}

The data selection follows the same philosophy used for L17, and a detailed description can be found in \cite{dunner_atacama_2013}. Each 10-minute section of the time-ordered data from a single detector array (this data unit is called TOD) is analyzed in frequency space to characterize and assess its quality. In the low frequency band ($\sim[0.01-0.1]$ Hz), the signal is expected to be dominated by the atmosphere, which to good approximation is common for the whole array. The full correlation matrix is computed in the low-frequency band, and detectors that do not correlate well with the dominant mode are excluded. The response to changes in the focal plane temperature is the most important contaminant and must be handled carefully as it can reach comparable amplitude to the atmospheric signal for some TODs. Some detectors, called ``dark detectors,'' are deliberately not optically coupled and used to estimate and deproject thermal fluctuations. Additional cuts have been included with respect to L17. In the high frequency regime ($\sim[10-20]$ Hz), noise properties are computed and detectors are excluded based on their white noise level and if significant amounts of skewness and kurtosis are detected. 

We mask the brightest sources beforehand in order to avoid triggering the glitch-finder or biasing the high frequency noise analysis. Given the much larger survey area compared to L17, we mask 632 sources which are brighter than $2$\,mK (roughly 180\,mJy at \freqb\,GHz). This threshold is well below the noise level in the TODs.

Following \cite{dunner_atacama_2013}, eight quality criteria are computed and detectors are excluded if they do not meet the requirement for any of the criteria. A detector can be flagged for a short period of time or excluded for the entire TOD. If more than 40\% of the samples for a single detector are flagged, the detector is excluded. If more than 40\% of the detectors are excluded for a given TOD, the whole array is discarded.

The effective number of detectors being mapped varies from 60\% to 80\% depending on the array and season. Table~\ref{table:obs} shows the median number of detectors used in the map-making stage.

\subsection{Relative detector calibration}
The data are calibrated into physical units using the equation
\begin{equation}
d^\mathrm{pW}_{t,i} = d^\mathrm{DAQ}_{t,i} \times R_{t,i} \times f_i \times g^\mathrm{atm}_{t,i},
\end{equation}
where $i$ is a detector index within a single array and frequency band, $t$ is an index over 10-minute observations, $d^\mathrm{pW}_{t,i}$ and $d^\mathrm{DAQ}_{t,i}$ are the time-ordered data in physical (pico watts, $\mathrm{pW}$) and data acquisition units, respectively, $R_{t,i}$ is the responsivity of the detector measured from a bias step operation, $f_i$ is a stable optical flatfield factor of order unity, and $g^\mathrm{atm}_{t,i}$ is an atmospheric correction factor of order unity.

Bias step operations are performed at least every hour, and involve a brief modulation of the detector bias voltage while monitoring the detector current.  These calibration operations are analyzed to extract each detector's power--resistance relationship and time constant (see next section).  The factors $R_{t,i}$ have units of pW per data acquisition unit.  The calibration model always uses the most recent applicable bias step measurement \citep{grace_atacama_2014}.

The factors $f_i$ and $g^\mathrm{atm}_{t,i}$ are corrections computed based on each detector's response to the atmospheric common mode. Initially, the factors $f_i$ and $g^\mathrm{atm}_{t,i}$ are fixed to unity.  The common mode for each observation is computed as a simple mean over detectors, and each detector is then calibrated against the common mode, in the low-frequency regime, to obtain correction factors $g^\mathrm{atm}_{t,i}$.  The shape of $g^\mathrm{atm}_{t,i}$ at this stage is dominated by a time-invariant contribution due to the optical efficiency of the detectors.  This static component is factored out into the flatfield factors $f_i$, with mean $f_i$ forced to unity.  The process of computing a common mode and measuring $g^\mathrm{atm}_{t,i}$ is then iterated once, using the non-trivial flatfield from the first iteration.  This iteration is performed so that the factors $g^\mathrm{atm}_{t,i}$ will be (typically) more tightly distributed around a value of unity, which is useful for identifying calibration problems in individual detectors or observations.  In the \freqb\,GHz data, the $f_i$ and iterated $g^\mathrm{atm}_{t,i}$ factors are included in the calibration of data that goes into the map.  In the \freqa\,GHz data, the flatfield factors $f_i$ are included in the model, but we continue to fix $g^\mathrm{atm}_{t,i} = 1$.  This is appropriate because the inferred corrections $g^\mathrm{atm}_{t,i}$ are more tightly distributed around unity than is the case at \freqb\,GHz, except in cases of very good atmospheric conditions where the measurements of $g^\mathrm{atm}_{t,i}$ themselves become quite noisy.

We note that this process only corrects for relative differences in detector calibration.  The effects of atmospheric opacity are included in the absolute calibration to planets described in \S\ref{subsec:planetcal}.

In L17, detectors for which a responsivity could not be extracted from the most recent bias step measurement were excluded. We are now using the average responsivity of the array in such cases, with a correction from the atmospheric correlation. This allowed us to recover about 20\% of detectors for s14 with respect to L17, with a negligible impact on the quality of the relative calibration, evaluated using planet observations.

\subsection{Detector time constants}
Our method for characterizing the detector time constants is summarized in \citet{thornton_atacama_2016}. Time constants are measured using both planet observations and bias steps. Time constants computed from planet scans, operating using the same optical coupling as the CMB scans, are used as baseline values. From these measurements we find time constants in the range of [1.0--5.4] milliseconds (ms) with a median value of 1.8\,ms. While there are only $\mathcal{O}(100)$ planet scans throughout the season, bias steps, as an alternative, can be acquired more often, as frequently as per CMB scan. There is still a difference between bias step time constant and planet scan time constant, due to the fact that the planet measurements are a more direct probe of the relevant optical response speed compared with the bias step method. Thus a conversion factor is applied to the bias step results to account for this. Details are described in \citet{grace_atacama_2014}.

\newpage
\subsection{Polarization angles}
Following the same procedure used in N14 and L17, we compute the polarization angles of the detectors in PA1, PA2, and PA3 using a detailed optical model of the telescope mirrors and lenses in the optical design software CODE V\footnote{\url{https://www.synopsys.com/optical-solutions/codev.html}} \citep{koopman_spie_2016, thornton_atacama_2016}. The angles of individual detectors are lithographically defined during fabrication. 
We propagate the position of the detector array, as assembled in the telescope, to the sky using a ray trace and compare to observed detector positions derived from planet observations. This determines the physical angle
of the detectors within the telescope. The optical chain also introduces a polarization rotation of up to ${\sim} 1.7^{\circ}$ near the edge of the focal plane, furthest from the primary optical axis, which is
computed in CODE V using a polarization sensitive ray trace. This combined with the physical angle of the detectors in the telescope gives us the final
polarization angle calibration. We later test for any additional global angle offset using the EB power spectrum (see C20 for details regarding this test).

\subsection{Calibration with planets\label{subsec:planetcal}}
We obtain the absolute calibration of the instrument using measurements of Uranus. This calibration factor converts the data from physical units ($\mathrm{pW}$) to CMB brightness temperature units\footnote{Unless otherwise noted, the units $\mathrm{\mu K}$ refer to CMB differential temperature units, where the spectral radiance of a region with temperature $\Delta T$ is $S_\nu = (\partial B_\nu(T) / \partial T) |_{T_{\rm CMB}} \Delta T$, with $B_\nu(T)$ the spectral radiance of a blackbody and $T_{\rm CMB} = 2.725\,{\rm K}$ the mean temperature of the CMB.} ($\mathrm{\mu K}$) and is applied after the relative detector calibration. The factor  is allowed to vary between TODs in order to capture temporal fluctuations in atmospheric transmission but is the same for all detectors on an array in the same frequency band. The calibration procedure is essentially unchanged with respect to \cite{Hasselfield_atacama_2013} and \cite{dunner_atacama_2013}; here we provide a brief summary. 

The absolute calibration factor applied to each TOD in frequency band $\nu$ is modelled as
\begin{align}\label{eq:abscal}
g_{\nu}(w, \theta) = c_{\nu}  \exp \left(\frac{ \tau_{\nu} w }{ \sin \theta}\right) \, .
\end{align}
Here, $w$ and $\theta$ denote the average precipitable water vapor level (PWV) and elevation of the array center, respectively, during the observation. The values for $c_{\nu}$ and $\tau_{\nu}$ are obtained by fitting the model to measurements of the absolute calibration taken at different values of $w / \sin \theta$. The individual measurements of absolute calibration are given by: 
\begin{align}\label{eq:abscal_measured}
g_i = \frac{T_{\nu}}{A_{\nu, i}} \, ,
\end{align}
where  $T_{\nu}$  denotes the brightness temperature of Uranus in $\mathrm{\mu K}$ and $A_{\nu,i}$ is the observed amplitude of the planet in $\mathrm{pW}$. The index $i$ runs over Uranus observations. The temperature $T_{\nu}$ is obtained by modeling the planet as a disk with uniform brightness temperature that is scaled by the ratio of the solid angles of the planet, $\Omega_{\mathrm{U}}$, and the instrumental beam, $\Omega_{\mathrm{B}, \nu}$:
\begin{align}
T_{\nu} = T_{\mathrm{U}}(\nu)\frac{ \Omega_{\mathrm{U}}}{\Omega_{\mathrm{B}, \nu}} \, .
\end{align}
The assumed frequency dependence of Uranus' brightness temperature $T_{\mathrm{U}}(\nu)$ is described in \cite{Hasselfield_atacama_2013}. The amplitudes $A_{\nu,i}$ in Eq.~\eqref{eq:abscal_measured} are determined as the peak amplitude of Uranus in planet maps that are made using the maximum likelihood map-making approach described in that same paper.

The parameters $c_{\nu}$ and $\tau_{\nu}$ in Eq.~\eqref{eq:abscal} are fitted per observing season, detector array, and frequency band. Season s15 and s16 form an exception as they use a shared value for $\tau_{\nu}$ to improve the model fit. In addition, two separate values of $c_{\nu}$ are fitted for s15: one before and one after the telescope was refocused mid season. Finally, it should be noted that a correction to the absolute calibration is applied at a later stage in the analysis. This $\mathcal{O}(10 \%)$ correction is determined from the angular cross-correlation between the final ACT maps and the \emph{Planck} HFI maps and is determined per sky region, with overall uncertainty of $\mathcal{O}(1 \%)$; see C20 for further details on this correction.

\subsection{Pointing corrections\label{subsec:pointing}}
ACT has a blind pointing accuracy of about $1^\prime$ for the nighttime
observations considered here. This is comparable to our beam Full-Width at Half Maximum (FWHM) of $1.4^\prime$ at \freqb\,GHz, so if left uncorrected it would significantly broaden our effective beam. As in L17 we correct our array pointing\footnote{Detector positions with respect to the array center are well measured in the field with planet observations. This method would not improve those estimates as point sources are too weak to give a reasonable per-detector measurement.} by comparing the observed positions of bright point sources to their
known catalog positions for each TOD. However, due to our larger
data volume, for this analysis we now perform the fit in map-space instead of
in the time domain.
The fits are performed by building the measured map signal, $r=A^TN^{-1}d$, and an approximation for the inverse of its covariance, $C^{-1}$, for a 20$\times$20 arcminutes thumbnail around each bright source, where $d$ is the TOD data vector, $N$ is the noise matrix estimated from the data, and $A$ the pointing matrix (see \S\ref{sec:mapmake} for more information about this notation). We then minimize 
\begin{equation}
    \chi^2 = \sum_i (r_i-C_i^{-1}m(x,y,a_i))^TC_i(r_i-C_i^{-1}m(x,y,a_i)),
\end{equation}
where $i$ is the index of the source and $m(x,y,a_i)$ is a point source
model with a pointing offset of $(x,y)$ and a peak amplitude of $a_i$.
The resulting fit is 10--100 times faster than a TOD-level fit, at the cost of
slightly lower accuracy thus increasing the residual pointing jitter compared to L17 results (see \S\ref{subsec:beams}).

Our median single-TOD point source sensitivity is 30\,mJy, so only point
sources with a flux of $\gsim 150$\,mJy contribute usefully to the fit.
The likelihood of covering a bright point source in a single TOD is proportional to the sky area covered by the scan. The typical scan length\footnote{The ACT s13--s16 dataset is dominated by scans targeting regions with an extent of 10--20\,deg in the declination direction. The s16 AA region uses much longer scans, but they constitute only a small part of the data volume.} of 20\,deg means that about half of the s13--s16 TODs do not cover a bright source that can be used for pointing measurement. We interpolate the missing pointing measurements by fitting a Gaussian process model to the available ones.
After applying these pointing corrections we measure a residual per-map pointing
jitter and include it as a component in our final beam model (see \S\ref{subsec:beams} for the details on this procedure).

\subsection{Beams\label{subsec:beams}}
We use nighttime observations of Uranus to characterize the instrument's optical response function, or beam. For each observing season and detector array, we recover both the azimuthally-averaged radial beam profiles and the corresponding $\ell$-dependent window functions, the latter of which enter directly into the power spectrum analysis. While the pipeline by which beams are processed is similar to that employed in earlier ACT data releases such as N14 and L17, and is more generally based on methods described in \cite{Hasselfield_atacama_2013}, there are a number of important changes that warrant further discussion.

First, the method by which planet maps are made has been significantly improved by changing how large-scale correlated noise is treated in the time-domain prior to mapping. Instead of modeling and subtracting only a single detector common mode from the data (as was done for earlier datasets), the new map-making algorithm estimates and removes correlated detector modes within a 12$\arcmin$ radius of the planet's position in the same manner as the gap-filling technique described in \S\ref{sub-sec:gapfill}. This not only results in visibly cleaner maps dominated by mostly white noise, but it also considerably improves the effective signal-to-noise in the outer region (or wing) of the beam. The new mapping method does, however, induce a non-negligible transfer function in the form of a unique, but constant, offset in each map. This offset is estimated and removed prior to any subsequent beam analysis.

There has also been a change in the model used to fit the wing of the beam. For a diffraction-limited optical system, the wing is expected to decay radially as a power law with an exponent of $-3$, thus motivating a model of the form:
\begin{equation}\label{eq:beams_wing}
B_w(\theta) = \frac{a}{\sin^3\theta} + b,
\end{equation}
where the amplitude $a$ and mapping transfer-function offset $b$ are fit parameters. This model does not, however, account for the effects of scattering due to the surface roughness of the telescope's primary mirror. For an imperfect parabolic reflector with RMS surface deviations $\epsilon$, the antenna gain is given by Eq.~(9) in \cite{ruze_antenna_1966}; this may be recast in terms of a unit-normalized beam $B_n(\theta)$ and corresponding solid angle $\Omega$:
\begin{equation}\label{eq:beams_ruze}
B_n(\theta) = \frac{\Omega}{\Omega_0}B_{0}(\theta)e^{-\bar{\delta^2}} + S(\theta),
\end{equation}
\begin{equation}\label{eq:beams_scatter}
S(\theta) = \frac{\Omega}{4\pi}\left(\frac{2\pi c}{\lambda}\right)^2 e^{-\bar{\delta^2}} \sum_n \frac{\bar{\delta^2}^n}{n\cdot n!} e^{-(\pi c \sin\theta / \lambda)^2 / n},
\end{equation}
where $B_0(\theta)$ is the unit-normalized beam for an ideal reflector, $\hat{\delta^2} = (4\pi\epsilon/\lambda)^2$ is the phase variance, $\lambda$ is the effective wavelength of light, $c$ is the surface-error correlation length, and we have used the relation $G(\theta) = (4\pi/\Omega)B_n(\theta)$. While the first term in Eq.~\eqref{eq:beams_ruze} is expected to decay according to the same power law as a diffraction-limited system, the same does not hold true for $S(\theta)$. Using photogrammetry measurements of the telescope's primary mirror, we find that $\epsilon = 20$ $\mu$m and $c = 28$ cm, values for which the contribution of scattering to the wing cannot be neglected (especially at shorter wavelengths). Our new model is thus a modified version of Eq.~\eqref{eq:beams_wing} in which the scattering term $S(\theta)$ has been added.

Finally, there have been a number of modifications to how errors are estimated at various points in the beam analysis pipeline. When fitting the radial profiles, the covariance of the non-linear scaling parameter $\ell_{max}$ (as defined in \citealp{Hasselfield_atacama_2013}) with the core linear fit coefficients and wing amplitude is now incorporated in the overall model uncertainty. This is accomplished by using the solution from our original fitting method as the starting point for an MCMC sampling of the full posterior parameter distribution. The resulting chains are thinned (to reduce computational load) and transformed in the same manner as the radial fit parameters in order to recover the full $\ell$-space beam transform covariance. An independent `scattering' mode is subsequently added to this covariance to account for additional uncertainty in our new wing model due to possible geometric modeling errors\footnote{We only measure the positions of panel edges in the telescope's segmented primary mirror and must therefore model / interpolate the deformed surface in order to estimate the true RMS deviation.} and temporal fluctuations in the mirror surface RMS parameter $\epsilon$ in Eq.~\eqref{eq:beams_scatter}. 

Lastly, as in L17, the covariant errors are further adjusted for uncertainty in the pointing variance corrections. Unlike previous analyses, however, the contribution of position-dependent pointing variance to this uncertainty is now estimated by marginalizing over it as a parameter ($\sigma_p$) in the likelihood:
\begin{equation}\label{eq:beams_jitter}
\mathcal{L}(\mu_p,\sigma_p\,|\,\mathbf{d},\boldsymbol{\sigma}) = A(\boldsymbol{\sigma},\sigma_p)\,e^{-\sum_i^N (d_i - \mu_p)^2/2(\sigma_i^2+\sigma_p^2)},
\end{equation}
where $\mu_p$ is the overall pointing variance,  $A(\boldsymbol{\sigma},\sigma_p) = \Pi_i^N[2(\sigma_i^2+\sigma_p^2)]^{-1/2}$ is a normalization factor, and $\{\mathbf{d},\boldsymbol{\sigma}\}$ are pointing variance fits and corresponding errors for $N$ bright point sources in the maps. Note that by properly including the position dependence in the likelihood, we expect this method to yield improved estimates of both the total uncertainty and the pointing variance itself. Compared to L17, we find that overall pointing variance has increased (e.g., from 9.9 (8.4) to 12.1 (22.5) arcsec$^2$ in D56 PA1 (PA2)); this may be due to both the aforementioned changes as well as lower accuracy TOD-level pointing corrections (see \S\ref{subsec:pointing}).

\begin{figure*}[th!]
    \centering
    \includegraphics[width=0.9\textwidth]{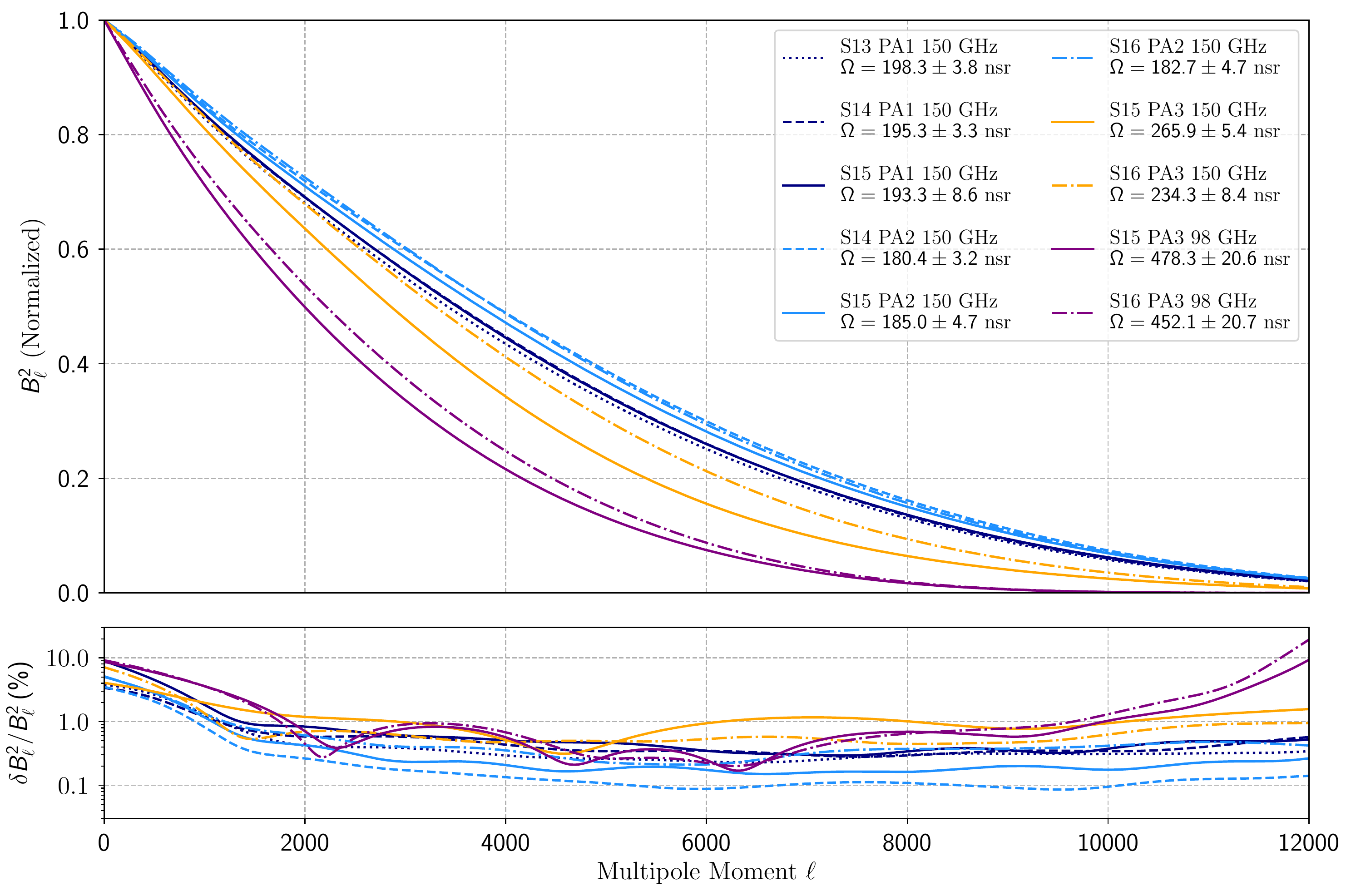}
    \caption{Window functions for the mean \emph{instantaneous} beam of each array and band in each season.  The window functions used for interpretation of the survey maps are slightly modified to account for residual pointing variance in the observations contributing to each map. The window function errors shown in the bottom panel are strongly correlated between multipoles.\\}
    \label{fig:beams}
\end{figure*}

The mean instantaneous $\ell$-space window functions for each season and array, as well as their corresponding errors, are shown in Fig.~\ref{fig:beams}. Additional corrections due to map-based pointing variance (see Eq.~\eqref{eq:beams_jitter}) and RJ-to-CMB spectral conversions are applied as in L17. A detailed treatment of how the broad passband modifies the effective beam depending on source spectral type may be found in \cite{Madhavacheril:2019nfz} in the context of component separation. We release both the mean instantaneous and map-effective (i.e., corrected) beam radial profiles and transforms as part of the current set of data products.

\begin{figure}
    \centering
    \includegraphics[width=\linewidth]{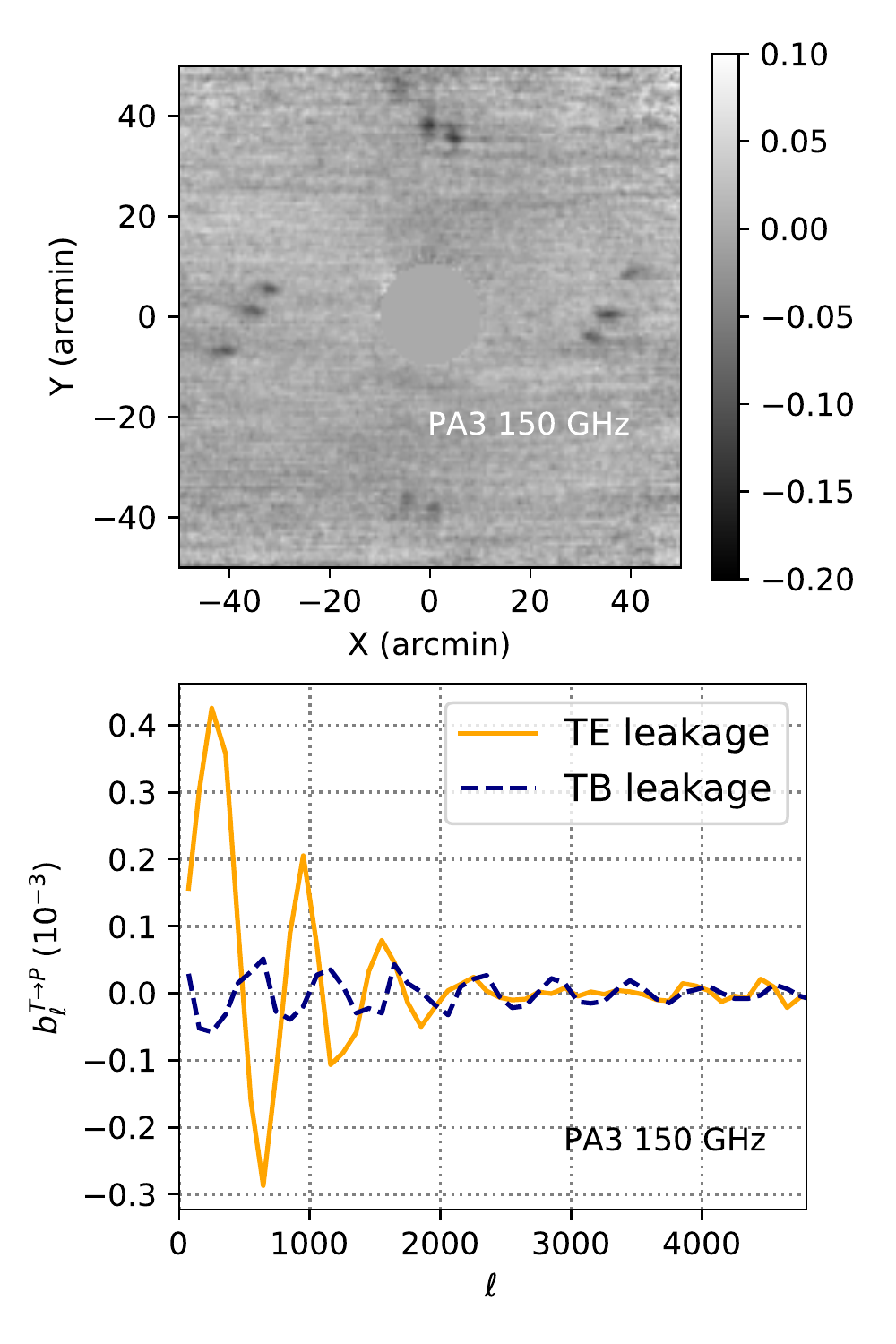}
    \caption{Weak polarized sidelobes in the PA3 \freqb\,GHz detectors.
    (Top panel) Map of the polarized sidelobes, obtained by stacking 20 observations of Saturn.  The main lobe of the beam is at the origin and is masked in this image. The grayscale is linear in units of 1/1000 of the main beam peak, with positive (negative) numbers indicating polarization parallel (perpendicular) to the radial direction. The sidelobes consist of four groups of compact features, at 30 to 40 arcminutes from the main beam. (Bottom panel) The effect of unmitigated sidelobes on CMB spectrum measurements, expressed as transfer beam functions $b_\ell^{T\to E}$ and $b_\ell^{T\to B}$. The sidelobes are strongly polarized in a direction perpendicular to the line connecting them to the main beam; this results in leakage that is primarily from temperature into E-mode polarization. Note that these sidelobes are projected out of the data prior to mapping the CMB, so the net leakage in the resulting spectra will be strongly suppressed relative to what is shown here.
    \vspace*{0.15in}}
    \label{fig:buddies}
\end{figure}

\subsection{Polarized Beam Leakage and Sidelobes}
We detect the presence of temperature-to-polarization (T-to-P) leakage in both the core of the main beam as well as its sidelobes. Although relatively small in magnitude, the leakage is significant enough that we must account for it at various steps in the analysis pipeline. For the main beam, we use the same set of Uranus observations from \S\ref{subsec:beams} to measure a polarization response in both $Q$ and $U$ Stokes parameters for each detector array and observing season. Since we do not expect Uranus to be polarized across our observing bands,\footnote{Measurements of Uranus at visible and near-infrared wavelengths have shown evidence of radial polarization patterns in the limbs, but a disk-integrated polarization of less than 0.05\%~\citep{schmid_2006}; while no equivalent data exists to our knowledge, the relevant scattering effects in the planetary atmosphere are expected to be much weaker at millimeter wavelengths, resulting in net polarization levels substantially lower than this bound.} we interpret these responses as leakages.

In order to extract the $\ell$-space $T$-to-$P$ leakage functions, we first convert each set of polarization maps to $\{Q_r,U_r\}$ (the third defined basis in \citealp{ludwig_1973}) via local map-space linear combinations of $Q$ and $U$. The new basis has the convenient property that its azimuthally averaged radial profiles have a direct correspondence to the $E$ and $B$ harmonic transforms. In the flat-sky approximation, this takes the form of a second-order Hankel transform:
\begin{equation}\label{eq:leakage_ell}
\{E(\ell),B(\ell)\} = -2\pi\!\int\!\{Q_r(\theta),U_r(\theta)\}J_2(\ell\theta)\,\theta\,d\theta.
\end{equation}
Since $Q$ and $U$ are formed by taking linear combinations of the individual detector beams within an array, we use the same framework to model radial profiles in $Q_r$ and $U_r$ as we do in temperature. This permits, with the exception of Eq.~\eqref{eq:leakage_ell}, the use of identical pipelines for processing the core beam response in both temperature and polarization. The resulting leakages and their associated covariances are then incorporated in the overall power spectrum analysis as described in C20.

In L17 we described polarized sidelobes of the main beam in the PA1 and PA2 arrays. The sidelobes, although weak in amplitude, were strong enough to cause noticeable T-to-P leakage (predominantly with E-mode structure). Although L17 did not include any data from PA3 detector array, we commented there that polarized sidelobes were not detected in PA3. However, with a more careful analysis using additional observations of Saturn, we confirm the presence of weak polarized sidelobes in the \freqb\,GHz detectors of PA3. These sidelobes are observed to be approximately one tenth as strong as the ones in PA1 and PA2. Maps of the sidelobes are shown in Fig.~\ref{fig:buddies}, along with the temperature to polarization leakage function.

For the present work, we apply the same sidelobe mitigation technique as was applied in L17, for all \freqb\,GHz detectors. This involves modeling the sidelobes as a sum of copies of the main beam lobe, and then using that model to deproject the sidelobe signal from the time-ordered data prior to map-making.

\newcommand{\uvec}[1]{\boldsymbol{\hat{\textbf{#1}}}}

The PA3 sidelobes have the same general features as the ones in PA1 and PA2; they consist of a small group of compact lobes, with approximate four-fold symmetry, strongly polarized perpendicular to the radius from the beam center. For any individual detector, a sidelobe will only map to the sky in cases where the main beam is within the array field of view. To explain this latter behavior more explicitly, imagine a fixed pointing of the boresight and consider the function $\uvec{n}(\vec{x})$ that gives the position on the sky of the camera's peak response, for a detector located at position $\vec{x}$ in the focal plane. The signal in detector $i$ originates mainly from the sky near $\uvec{n}(\vec{x}_i)$, but also contains contributions from each compact sidelobe $j$ at $\uvec{n}(\vec{x}_i +  \vec{s}_j)$.  However, sidelobe $j$ is observed for detector $i$ only in cases where $\vec{x}_i + \vec{s}_j$ is within the physical bounds of the focal plane.  This is consistent with an optical effect originating inside the camera rather than due to optical elements outside the cryostat.

The PA3 sidelobes are located at radii of approximately 30 to 40 arcminutes, leading to two significant differences relative to PA1 and PA2. First, the broader stride of the polarized sidelobes means that the angular scale of strongest T-to-P leakage is pushed to a lower multipole: $\ell \approx 300$ for PA3 as compared with 500 for PA1 and PA2. Second, because sidelobes only map to the sky when the main beam is also in the array field of view, the larger radii of the sidelobes leads to fewer detectors suffering from each sidelobe.

We do not detect sidelobes in PA3 \freqa\,GHz, even though we would have sufficient signal to detect them if they had similar amplitude and positions in this band. Preliminary lab and modeling work on similar systems suggest that the sidelobes arise from a diffraction effect induced by grids that make up the various optical filters. In such a case, one would expect sidelobes to occur at larger radius for lower frequencies. Sidelobes in \freqa\,GHz would appear starting at a distance of  approximately 47 arcminutes from the main beam. Since this is roughly the size of the focal plane, the sidelobes would map to the sky for only a small fraction of detectors, at the edges of the focal plane.

\section{Map-making}
\label{sec:mapmake}

As done for previous ACT data releases and described in \cite{dunner_atacama_2013}, N14, and L17, we adopt the maximum-likelihood map-making method to reduce several terabytes of TODs into high signal-to-noise sky maps. In summary, this is achieved by solving the mapping equation
\begin{equation}
    \label{eq:MLM}
    (A^T N^{-1} A) m = A^T N^{-1} d,
\end{equation}
where $A$ is the generalized pointing matrix projecting the data from map to time domain, $N$ is the detector-detector noise covariance matrix, which is defined in Fourier space and measured from the data, $d$ is the conditioned and slightly filtered time-ordered data, and $m$ is a set of $I$, $Q$, and $U$ Stokes parameter maps of the sky. In other words, by solving Eq.~\eqref{eq:MLM} we find the maximum-likelihood solution of ${\rm N_{pix}\cdot N_{Stokes}}$ degrees of freedom.\footnote{The concept of degrees of freedom is further used in \S\ref{subsec:pointsourcetreat}.} The slight filtering applied to the time-ordered data due to ground removal (see \S\ref{sub:groundsub}) is not handled in a maximum-likelihood framework. The details of the filtering and the resulting transfer function are discussed in the following sections.

The dataset described in \S\ref{sec:obs} is processed in separate bundles in order to produce four (or two for s16) independent split maps for each season, array, frequency, and region with uniform noise properties (see \S\ref{sec:maps} for a detailed description of the map-based products that are released with this work). The splitting of observations into independent bundles is the result of an optimization process to minimize the noise bias in the power spectrum. TODs are grouped into single-day blocks. The distribution of such blocks among the 2 or 4 splits is computed by optimizing for (i) even distribution of the data among the splits and (ii) uniformity of hit counts across the sky for each split. One important constraint imposed on this procedure is that observations from different arrays taken on the same day belong to the same split (e.g., s14-pa1-split0 connected to s14-pa2-split0) to confine cross-split correlations due to atmospheric noise to a single pair.

For this analysis, the \verb|Enki| map-maker,\footnote{GitHub repository: \url{https://github.com/amaurea/enlib}} which was developed and tested for N14 and L17 and easily scaled to handle large high-resolution maps, is used to produce the nominal maps. All previous ACT data releases employed the \verb|Ninkasi| map-maker.\footnote{GitHub repositories: \url{https://github.com/sievers/ninkasi_c} and \url{https://github.com/sievers/ninkasi_octave}}~These two software codes have been compared and shown to lead to consistent results at the power spectrum level when used to map data in the D56 region. In the following sections we describe improvements in the data processing  between this analysis and L17.

\subsection{Pixelization}
We continue to produce sky maps in equatorial cylindrical projection. We adopt the plate carr$\mathrm{\acute{e}}$e (CAR) pixelization, as opposed to the CEA pixelization which was used in previous releases. For CAR, pixels are equally spaced in latitude, and have the same number of pixels per latitude ring. Although the pixels are no longer equal area, CAR avoids the elongation of pixels at extreme declinations and thus requires fewer total pixels to achieve the necessary resolution across the full declination range of our maps. At the equator, the resolution continues to be 0.5 arcminutes for both \freqa\ and \freqb\ GHz. Exact weights for spherical harmonic transforms are provided in the \verb|libsharp| library\footnote{We use the pixell library (\url{https://github.com/simonsobs/pixell}) for manipulating CAR maps and performing curved sky Fourier operations.}\citep{reinecke/2013} for the CAR pixelization.

\subsection{Sidelobes and cuts}
As in L17, we estimate the contamination from the Moon being observed through the telescope sidelobes by mapping the data in Moon-centered coordinates. In this case, the data are bundled such that single-season, single-array, and single-frequency contamination maps are produced. Although the sky signal gradually averages away in such maps, we decide to not subtract the measured contamination, but rather to fully cut contaminated data. In L17, TODs that were partially contaminated were excluded from the map-making run. In this new analysis, individual samples that map onto a region of the corresponding moon-centered map with a polarization signal $> 200$\,$\mu $K are  cut. The recovered data were shown to not bias the signal power spectra when compared to previous results.

\subsection{Gap-filling and cuts\label{sub-sec:gapfill}}
Although the contaminated samples (see also \S\ref{subsec:preproc}) are not used to find the maximum-likelihood solution map, $m$, in Eq.~\eqref{eq:MLM}, estimating the noise model, $N$, requires continuity of the detector data. In this analysis, we adopt a new method to fill in the missing data samples for noise estimation purposes. We compute the detector-detector covariance matrix from the non-cut samples and we then use it to predict the value of the cut ones. This new method improved the estimation of the map-maker noise model and reduced the noise power spectra by roughly a factor of 1.5 in the $1000<\ell<2000$ region, while still leading to unbiased signal spectra. 

In addition, we use this filling method for glitch-like (short) cut regions. The original data from longer cut regions is kept to estimate the noise model, but still excluded when solving for the signal map.

\subsection{Ground subtraction\label{sub:groundsub}}
To characterize spurious emission from the ground and other scan-synchronous signals we adopt a similar analysis as in L17. Each of the original data splits (4 or 2 splits per season, array, frequency and sky region) are further split to separate observations at different elevations. For each detector, we then solve for a one-dimensional maximum-likelihood map with a one arcminute resolution along the azimuth coordinate. Any signal that changes while performing constant elevation scans, such as the CMB, will average away over many observations. 

These single-detector maps, called azimuth pickup maps, are used as a template (with scaling equal to unity) to subtract the contamination signal before mapping the data in the usual sky coordinates. In L17, the azimuth pickup maps were used to define data cuts (similarly to the treatment of sidelobes) for the D56 region, but the same procedure would cause large gradients in the hit counts for BN and AA, thus degrading the uniformity of the map's noise properties. Because we now use the full information of the azimuth pickup maps, rather than simply defining a threshold-based binary mask, we do not perform any scan-synchronous cleaning based on polynomial fitting as adopted in previous ACT data releases. In addition, we are able to capture and remove ground contamination on smaller scales. An aggressive polynominal fitting could achieve similar performance, but it would heavily bias the maps. 

As mentioned above, it is important to verify that the CMB signal is sufficiently suppressed in the pickup maps. Any residual CMB signal will be removed from the TODs and bias the sky maps. To characterize the transfer function induced by this new ground subtraction, as well as testing for possible convergence issues and noise bias, we inject a simulated $I,Q,U$ map into the TODs and perform an end-to-end mapping run (including estimating the pickup maps). For C20, the power spectra of the input map and the processed map are compared to quantify the bias as function of the angular scale, showing a small $0.3\%$ bias for $\ell>350$ (see C20 for more details).

\begin{table*}[!ht]
\caption{Summary of region sizes and map inverse-variance averaged white-noise levels in \ukarcmin \label{table:maps_stats}}
\vspace{-0.15in}
\begin{center}
\begin{tabular}{lc|cc|cccc|ccc}
\hline\hline
\\
Season & \multicolumn{1}{c}{s13} & \multicolumn{2}{c}{s14} & \multicolumn{4}{c}{s15} & \multicolumn{3}{c}{s16}\\
Array and Frequency & PA1$_{150}$ & PA1$_{150}$&PA2$_{150}$ & PA1$_{150}$&PA2$_{150}$&PA3$_{98}$&PA3$_{150}$ & PA2&PA3$_{98}$&PA3$_{150}$ \\
\hline
\textbf{D1} (131 deg$^2$) & 18.6 & &&&&&& \\
\textbf{D5} (157 deg$^2$) & 16.4 & &&&&&& \\
\textbf{D6} (135 deg$^2$) & 12.6 & &&&&&& \\
\textbf{D56} (834 deg$^2$) & & 32.3&20.5 & 33.3&21.9&18.4&28.6 & && \\
\textbf{D8} (248 deg$^2$) & & & & 42.0&21.5&20.0&29.0 & & \\
\textbf{BN} (3157 deg$^2$) & & & & 76.8&41.3&33.9&49.1 & & \\
\textbf{AA} (17044 deg$^2$) & & & & &&& &72.9&78.7&118.5 \\
\hline\hline
\end{tabular}
\end{center}
\vspace{-0.05in}
Notes: The region areas include pixels that (i) have been observed in all seasons and by all arrays of interest and (ii) have an inverse-variance weight $\sigma^{-2}_{\rm pix} \gsim 0.1\overline{\sigma}^{-2}_{\rm pix}$ in each dataset (single-array, single-season, single-frequency) separately. These areas are somewhat bigger (depending on the region) than those reported in C20 and use to compute power spectra. The inverse-variance weighted map noise levels are computed within the areas defined above and reported in units of \ukarcmin. Combined map noise levels for each region are provided in the text in \S\ref{sec:maps}. Sub-regions of these areas are selected for the power spectrum analysis in C20.\\
\end{table*}

\subsection{Point source treatment\label{subsec:pointsourcetreat}}
Small errors in our data model, such as residual pointing jitter, gain errors, intrinsic variability of point sources or the implicit assumption that the sky is pixelized, can introduce bias in the maximum-likelihood sky maps. For ACT, these biases result in $\mathcal{O}(10^{-4})$ of the signal leaking by a noise correlation length ($\approx 1 \deg$) away from high-contrast regions, typically manifesting as a faint X-like pattern around strong point sources. We previously handled this by subtracting a point source model from the time-ordered data before making the maps, and then optionally adding them back again in pixel space. However, this procedure only eliminates one of the sources of model error: the assumption that the sky is pixelized. Due to having more bright point sources, with the addition of the BN and AA regions, a new detector array with a less symmetric beam (PA3), and a faster but slightly less accurate pointing model, we chose a more thorough approach this time.

For point sources with a peak amplitude brighter than $10\,\textrm{mK}$ (roughly 800\,mJy at \freqb\,GHz), we add to the maximum-likelihood solution in Eq.~\eqref{eq:MLM} an extra degree of freedom for each TOD sample where the contribution from the source is brighter than $100\,\mu\rm{K}$, which typically is true for samples mapping 5 arcminutes around the source. By having a degree of freedom for each data point, no explicit model is fit to these point sources, and thus the uncertainties associated to the beam model, the amplitude of the source as function of time, and the pointing model do not introduce any error into our treatment.
This procedure is applied to 64 (86) sources at \freqa\,(\freqb)\,GHz, corresponding to roughly 0.01\% of the sky. In theory, the extra degrees of freedom come at the cost of worse handling of correlated noise in such regions. However, since the noise is approximately white on their length scale there is in practice no impact on our map noise properties. See \cite{naess_mm_2019} for a more thorough discussion of model errors and biases in maximum-likelihood map-making, and the section on source sub-sampling in particular for the method used here.

For dimmer point sources, but still detected with a signal to noise ratio SNR$\gt 5$ in each per-season, per-array map, we perform the point-source model subtraction at the TOD level, as mentioned above and described in L17 and \cite{dunner_atacama_2013}. This method only removes the season-averaged point sources signal. Although this approximation is sufficient to guarantee an unbiased map-making procedure, further map-space refinements are adopted as described in \S\ref{subsec:mapproducts}, \cite{naess_atacama_2020}, and \cite{Madhavacheril:2019nfz}.

\subsection{Noise iterations}
Because of the wide sky area covered over four different seasons, especially s16, and the increased large-scale sensitivity due to integration time, we require stricter control of the biases induced by the map-making procedure. The estimation of the noise covariance matrix, $N$, using the data $d$ introduces a small bias at intermediate and large scales. This bias is iteratively removed, as described in \cite{dunner_atacama_2013}, by multi-pass map-making where corrections to the map estimate $m_i$ from pass $i$ are computed by running noise estimation and mapping on modified time-ordered data $d - Am_i$.
In this new analysis as a precaution measure, we perform three iterations to correct for such a bias as opposed to two as performed in L17. 

\section{DR4 Map products}
\label{sec:maps}

ACT DR4 maps provide the first signal-dominated arcminute-resolution $E$-mode fluctuation maps over thousands of deg$^2$ and additional small-scale temperature anisotropies over 40\% of the sky. 
The products are described in this section and will be publicly released on the NASA Legacy Archive Microwave Background Data Analysis (LAMBDA) website.\footnote{\url{https://lambda.gsfc.nasa.gov/product/act/}} iPython notebooks developed to produce tables and figures in this section can be found on the ACT GitHub.\footnote{\url{https://github.com/ACTCollaboration/act_dr4_sm}}   
\subsection{Products and definitions\label{subsec:mapproducts}}
The map-making procedure, described in \S\ref{sec:mapmake}, requires  roughly 8.3 core-years on a modern high-performance computing cluster to reduce 11.3~TB of time-ordered data into roughly 300~GB of science-quality sky maps. The resulting maps have observed-pixel counts ranging roughly from 1.9 million (for D1) to 278 million\footnote{Each map has an extra 30-40\% overhead of unobserved (zero) pixels due to the scan-strategy and due to RA-Dec region center offsets between different detector arrays.} (for AA).

The released maps can be grouped into 20 different sets, one for each region, season, array, and frequency combination (see Table~\ref{table:maps_stats} for details). Each set contains 4 (2 for AA) separately-processed independent splits and one map-based co-add for a total of 94 data units. For each unit we release 4 FITS files that fully describe the data and its properties: 

\begin{figure*}
    \centering
    \includegraphics[width=\textwidth,trim=0 5.5mm 0 0,clip]{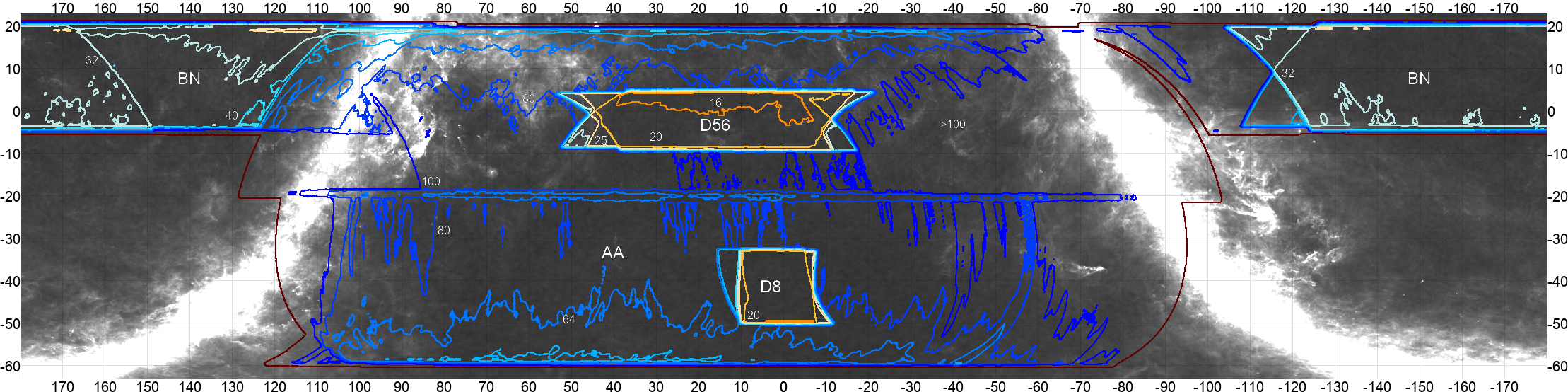}
    \includegraphics[width=\textwidth,trim=0 0 0 6.5mm  ,clip]{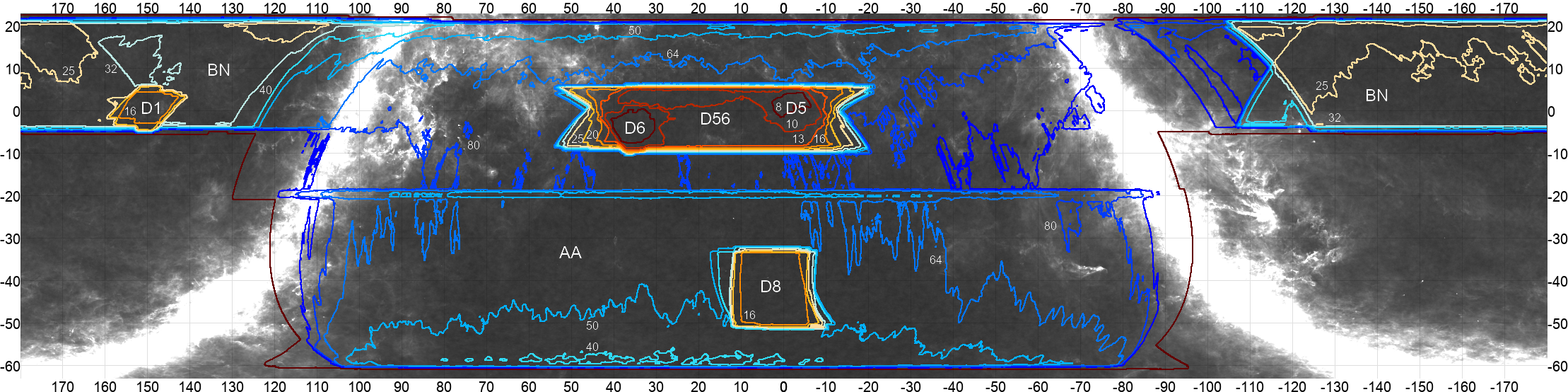}
    \caption{ACT DR4 survey depth and sky coverage maps for \freqa\,GHz (top panel) and \freqb\,GHz (bottom panel) in Equatorial coordinates. Lines at constant map-noise are plotted ranging from 7.5\,\ukarcmin (red) to 101\,\ukarcmin (blue). The \planck\ 353 GHz intensity map is shown in the background. The x-axis (y-axis) shows the RA (Dec) coordinates in degree.}
    \vspace*{0.2in}
    \label{fig:ivar_f090150}
\end{figure*}

\begin{itemize}
    \item {\it Source-free maps}: $I$ (or $T$), $Q$, and $U$ Stokes components of the sky in units of microkelvin. The polarization components are defined by the IAU convention\footnote{This is explicitly specified in the FITS header by the keyword \texttt{POLCCONV=IAU}. The ACT DR3 maps were released following the \texttt{COSMO} convention~\citep{healpix_2005}.} \citep{hamaker/bregman:1996}. Although the signal from point sources is removed at the TOD level to avoid biases in the maximum-likelihood maps (see \S\ref{subsec:pointsourcetreat}), a second more accurate map-space point source subtraction is adopted to reduce artifacts that could be induced by map-level Fourier filtering often used in cosmological analyses ~\citep[e.g., C20;][]{darwish_atacama_2020, Madhavacheril:2019nfz, han_atacama_2020}. Optimally co-added single-frequency maps \citep{naess_atacama_2020}, made from nighttime ACT data from 2013 to 2017, are used to measure the positions of point sources. For each source detected with a SNR$\gt 5$, a symmetric beam-shaped profile at the measured location is used as map-space template. The amplitude of the template is fit for in each split map both in temperature and polarization. This procedure better captures small variations in the sources amplitudes in each split (see \citealp{naess_atacama_2020} for more details).  
    
    \item {\it Source maps}: $I$ (or $T$), $Q$, and $U$ Stokes components of the point-source signal subtracted following the procedure described above in units of microkelvin. The simple sum of the source-free and source maps represent the observed sky. 
    
    \item {\it Inverse-variance map}: temperature--temperature (or $II$) component of the $3\times 3$ per-pixel inverse covariance matrix in units of $\mu$K$^{-2}$. The $3\times 3$ matrix is computed by projecting the diagonal component of the map-maker noise model onto the sky (thus dominated by the high-frequency part of the noise spectrum). Although the inverse-variance map is roughly proportional to a hit count map, it further takes into account the variability in detector performances and is thus better suited for map-based weighting and map co-addition procedures. Because of our scan-strategy and large number of detectors, the off-diagonal elements ($IQ$, $IU$, and $QU$) are negligible. In addition, the $QQ$ and $UU$ elements are each well approximated as having half the inverse variance of $II$. For these reasons we only release the $II$ component of the full 3x3 covariance matrix. This product is used in C20 to exclude noisy regions and to weight the spectra.
    
    \item {\it Cross-linking maps}: $T$-, $Q$-, $U$-like components used to describe the cross-linking properties of a dataset. These components are computed by projecting the time-domain vector, $d={\rm const}$, onto the sky using a projection matrix constructed by artificially setting all detector angles to zero (i.e parallel to the azimuthal motion of the telescope). The $T$-like component is the usual inverse-variance map (see above). For a small-throw perfectly azimuthal scan over a setting (rising) region of the sky around ${\rm Dec}=0$, the resulting footprint in equatorial coordinates is a nearly-straight scan tilted by roughly $+45$ $(-45)$ degrees with respect to the equator. In such configuration the observed pixels in the $Q, U$-like maps will have $Q\sim0$ and $U\sim+I$ for the setting scan and $Q\sim0$ and $U\sim-I$ for the rising scan. In other words, the $Q$ component of the cross-linking map mainly encodes the local angle between the scanning direction and north, while the $U$ component mainly encodes the balance between rising and setting scans. These maps are used in C20 to select regions with uniform cross-linking, which translates into uniform noise properties.
\end{itemize}

Although the time-ordered data are calibrated before entering the map-making step, an additional calibration factor is estimated for each map via a spectrum-level comparison with the \planck\ temperature maps (see \S\ref{subsec:planetcal}), as described in C20. The released maps have been recalibrated, but no correction for polarization efficiency has been applied. Similarly, the power spectra of these maps entering the likelihood analysis have been corrected for the calibration factors and the covariance matrix includes calibration uncertainties. Polarization efficiency is then applied correcting the model within the likelihood code (see C20).

\subsection{Sky maps}
Figure~\ref{fig:ivar_f090150} shows the s13--s16 survey area and the temperature map noise levels for both the \freqa\ and \freqb\,GHz channels. Although observations at \freqa\,GHz only started in 2015, the spatial distribution of the noise is similar between the two frequencies, with the exception of the D1, D5, and D6 regions observed only in s13.\footnote{D5 was also observed in s16 and mapped together with the AA wide scans for simplicity.} The \planck\ 353 GHz intensity map is also shown, highlighting that with the AA survey we have observed a considerable fraction of the Galactic plane. The Galactic center is part of the survey but sparsely sampled in time.
 
\begin{figure}[t]
    \centering
    \includegraphics[width=\columnwidth]{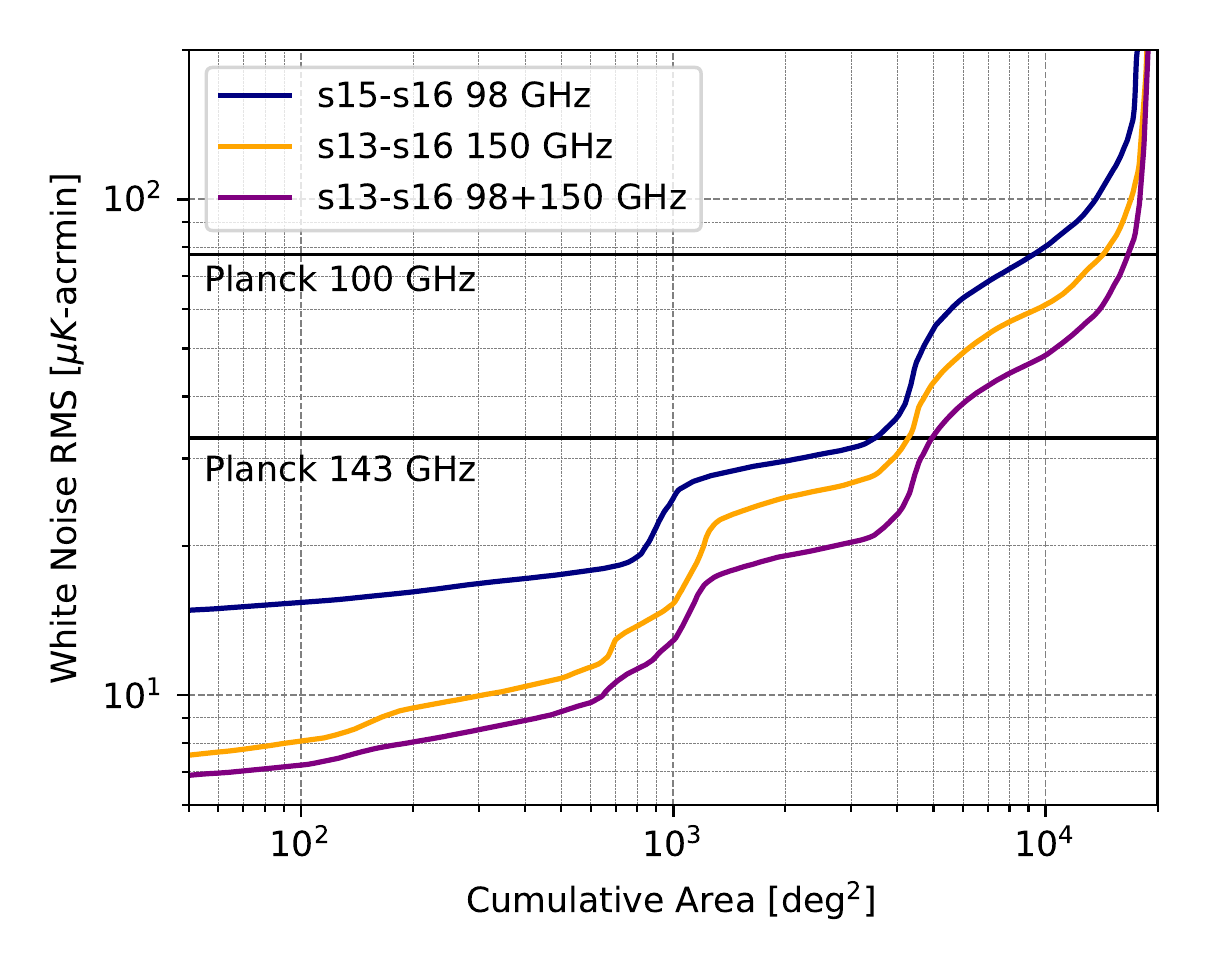}
    \caption{Map noise cumulative distribution function for \freqa~and~\freqb\,GHz and their combination. Two horizontal lines show the \planck\ full-sky averaged white-noise level for the 100\,GHz and 143\,GHz channels.\\}
    \label{fig:depth_curves}
\end{figure}

More quantitatively, Fig.~\ref{fig:depth_curves}\, shows the cumulative distribution function of the map white noise. The information about the spatial distribution of the different regions is lost in this representation; however it is a useful tool for forecasting studies. When combining \freqa\ and \freqb\,GHz data, roughly 600\,deg$^{2}$ of the sky have been observed at a noise level $\le 10$\,\ukarcmin. Approximately 4000\,deg$^{2}$ of the s13--s16 \freqb\,GHz survey have a noise level that is lower than the full-mission whole-sky average 143\,GHz \planck\ channel. The corresponding number for ACT \freqa\,GHz compared to \planck\ 100\,GHz is 9000\,deg$^{2}$. It is worth noting that (i) some of the deepest regions in the \planck\ survey are observed by ACT, (ii) at large angular separations the ACT temperature noise strongly departs from a white spectrum due to atmospheric correlations; much less so for the polarization channels, and (iii) the comparison does not account for differences in the beam size, which significantly increase the \planck\ uncertainties at small scales compared to ACT. A comparison of the ACT and \planck\ uncertainties at the power spectrum level is presented in C20.

\begin{figure*}[!htp]
    \centering
    \includegraphics[width=\textwidth,trim=0 3.8mm 0 0,clip]{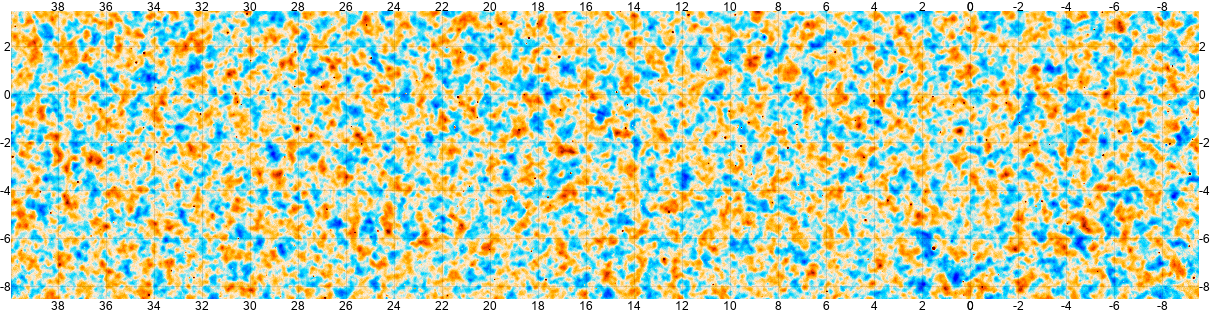}
    \includegraphics[width=\textwidth,trim=0 3.8mm 0 3.8mm,clip]{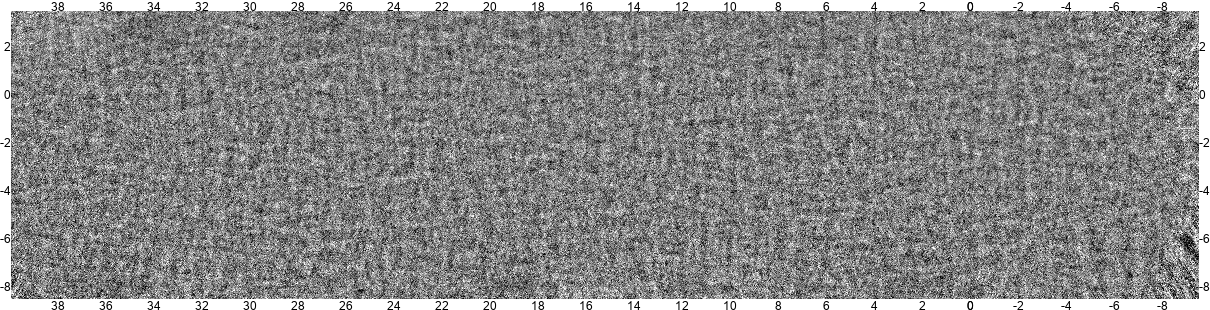}
    \includegraphics[width=\textwidth,trim=0 3.8mm 0 3.8mm,clip]{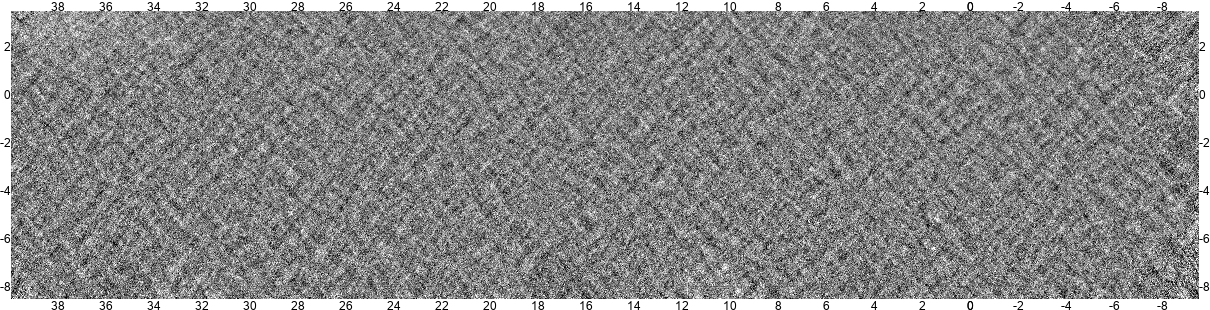}
    \includegraphics[width=\textwidth,trim=0 3.8mm 0 3.8mm,clip]{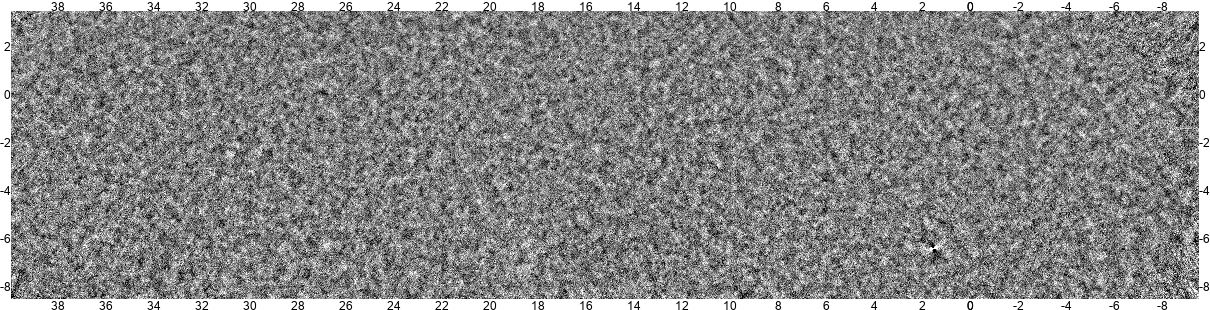}
    \includegraphics[width=\textwidth,trim=0 0 0 3.8mm,clip]{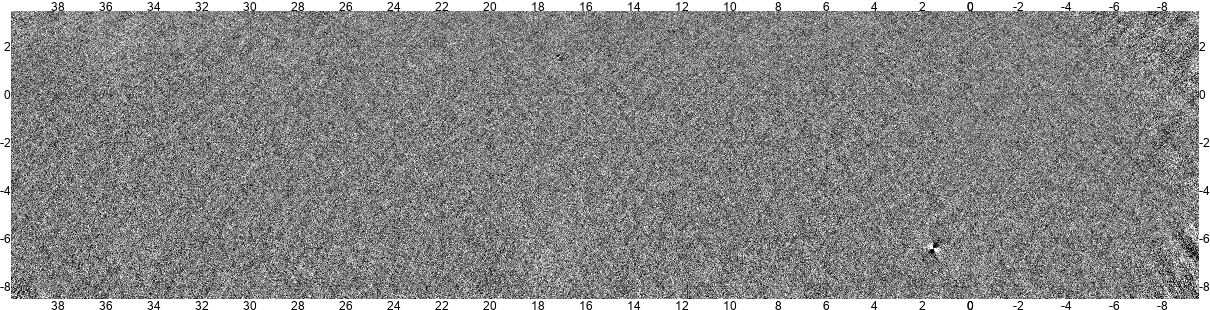}
    \caption{Rectangular cutout of the s13--s16 \freqa\,GHz maps in the D56 region. The selected area covers 630\,deg$^2$ of the sky. To visually highlight the wide range of measured angular scales, modes with $|\ell|\lsim 150$ and $|\ell_x|\lsim5$ have been filtered out. The top panel shows temperature fluctuations in a range of $\pm 250$\,$\mu$K. The remaining four black-and-white panels show (top to bottom) $Q$, $U$, $E$-mode, and $B$-mode polarization measurements in a range of $\pm 30$\,$\mu$K. The x-axis (y-axis) shows the RA (Dec) coordinates in degree.}
    \label{fig:d56_f090}
\end{figure*}
\begin{figure*}[!htp]
    \centering
    \includegraphics[width=\textwidth,trim=0 3.8mm 0 0,clip]{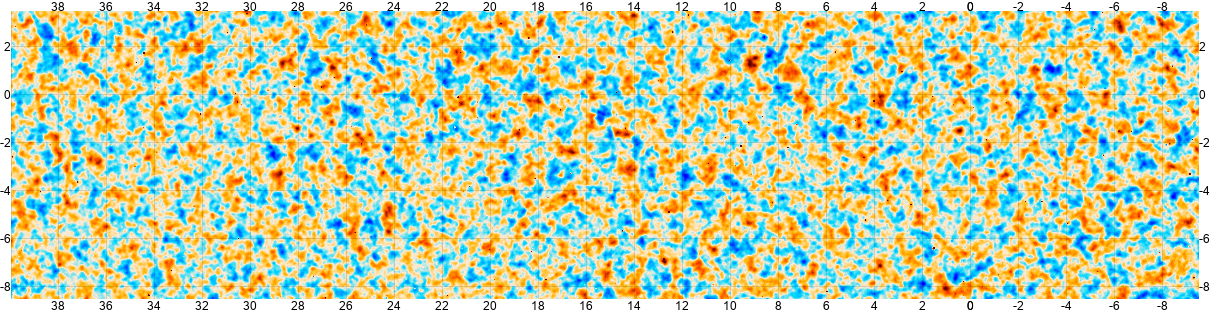}
    \includegraphics[width=\textwidth,trim=0 3.8mm 0 3.8mm,clip]{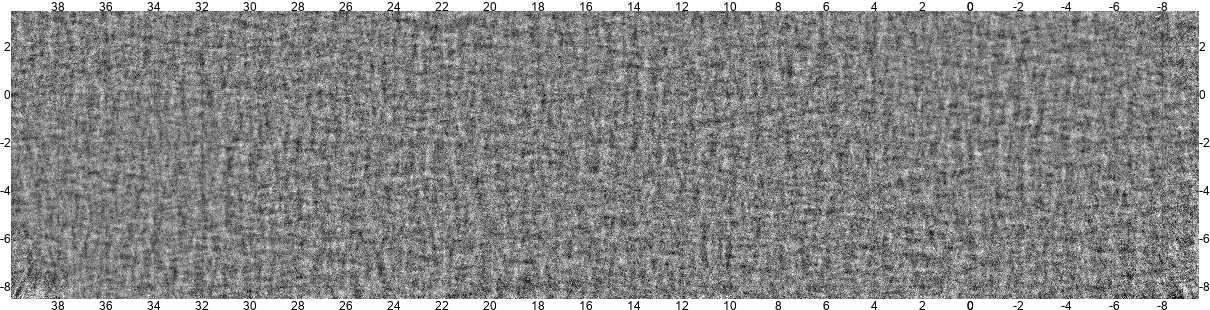}
    \includegraphics[width=\textwidth,trim=0 3.8mm 0 3.8mm,clip]{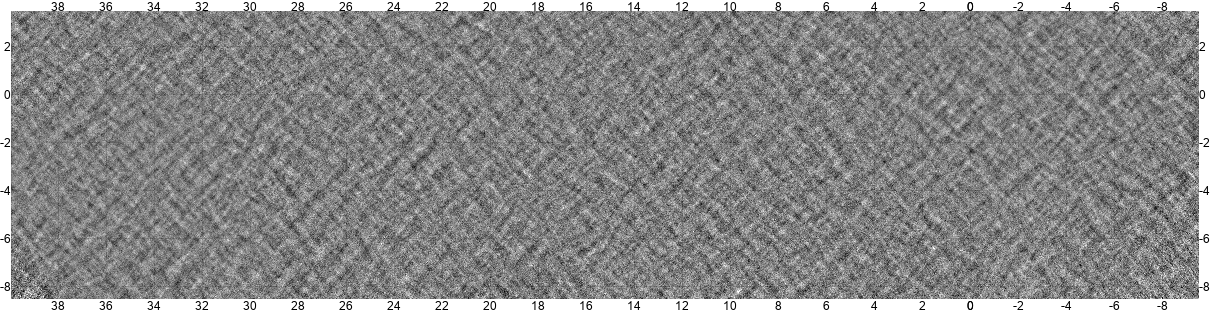}
    \includegraphics[width=\textwidth,trim=0 3.8mm 0 3.8mm,clip]{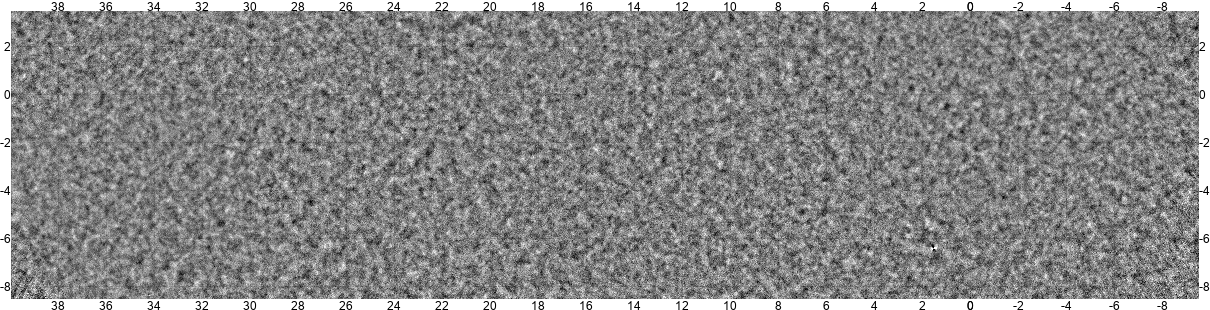}
    \includegraphics[width=\textwidth,trim=0 0 0 3.8mm,clip]{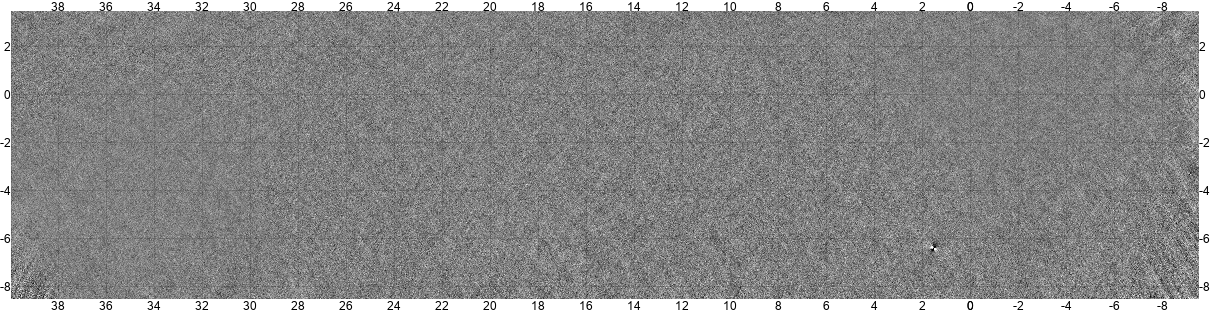}
    \caption{Rectangular cutout of the s13--s16 \freqb\,GHz maps in the D56 region. The selected area covers 630\,deg$^2$ of the sky. To visually highlight the wide range of measured angular scales, modes with $|\ell|\lsim 150$ and $|\ell_x|\lsim5$ have been filtered out. The top panel shows temperature fluctuations in a range of $\pm 250$\,$\mu$K. The remaining four black-and-white panels show (top to bottom) $Q$, $U$, $E$-mode, and $B$-mode polarization measurements in a range of $\pm 30$\,$\mu$K. The x-axis (y-axis) shows the RA (Dec) coordinates in degree.}
    \label{fig:d56_f150}
\end{figure*}
\begin{figure*}[!htp]
    \centering
    \includegraphics[width=\textwidth,trim=0 3.8mm 0 0,clip]{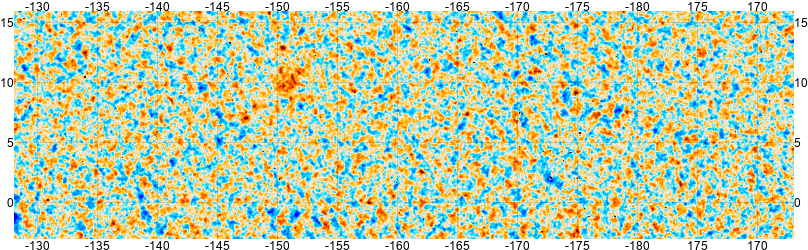}
    \includegraphics[width=\textwidth,trim=0 3.8mm 0 3.8mm,clip]{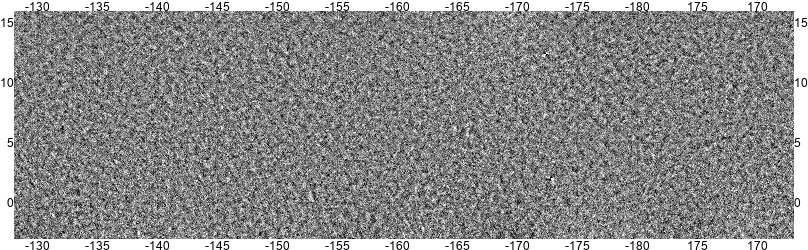}
    \includegraphics[width=\textwidth,trim=0 0 0 3.8mm,clip]{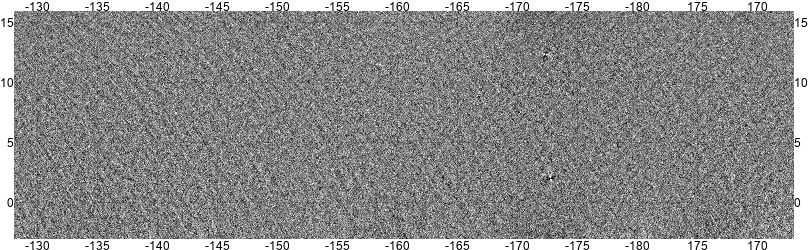}
    \caption{Rectangular cutout of the s13--s16 \freqa\,GHz maps in the BN region. The selected area covers 1230\,deg$^2$ of the sky. To visually highlight the wide range of measured angular scales, modes with $|\ell|\lsim 150$ and $|\ell_x|\lsim5$ have been filtered out. The top panel shows temperature fluctuations in a range of $\pm 250$\,$\mu$K. The remaining two black-and-white panels show (top to bottom) $E$-mode and $B$-mode polarization measurements in a range of $\pm 30$\,$\mu$K. The x-axis (y-axis) shows the RA (Dec) coordinates in degree.}
    \vspace*{0.2in}
    \label{fig:bn_f090}
\end{figure*}
\begin{figure*}[!htp]
    \centering
    \includegraphics[width=\textwidth,trim=0 3.8mm 0 0,clip]{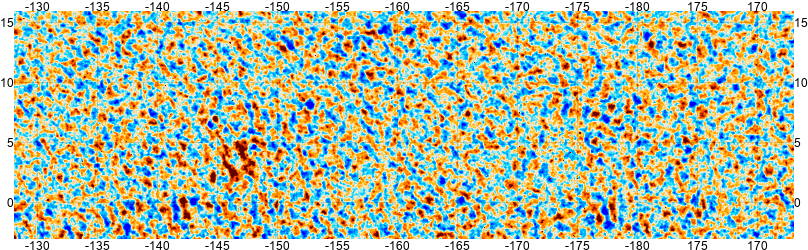}
    \includegraphics[width=\textwidth,trim=0 3.8mm 0 3.8mm,clip]{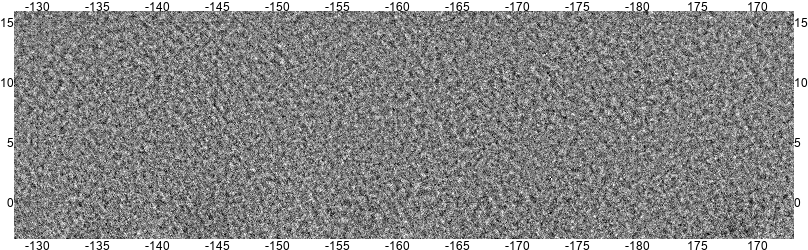}
    \includegraphics[width=\textwidth,trim=0 0 0 3.8mm,clip]{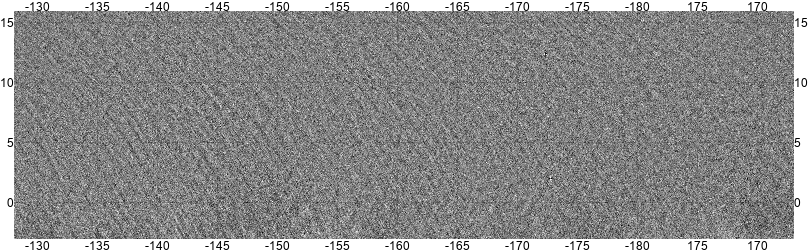}
    \caption{Rectangular cutout of the s13--s16 \freqb\,GHz maps in the BN region. The selected area covers 1230\,deg$^2$ of the sky. To visually highlight the wide range of measured angular scales, modes with $|\ell|\lsim 150$ and $|\ell_x|\lsim5$ have been filtered out. The top panel shows temperature fluctuations in a range of $\pm 250$\,$\mu$K. The remaining two black-and-white panels show (top to bottom) $E$-mode and $B$-mode polarization measurements in a range of $\pm 30$\,$\mu$K. The x-axis (y-axis) shows the RA (Dec) coordinates in degree.}
    \vspace*{0.2in}
    \label{fig:bn_f150}
\end{figure*}

\begin{figure*}[th!]
    \centering
    \includegraphics[width=\textwidth]{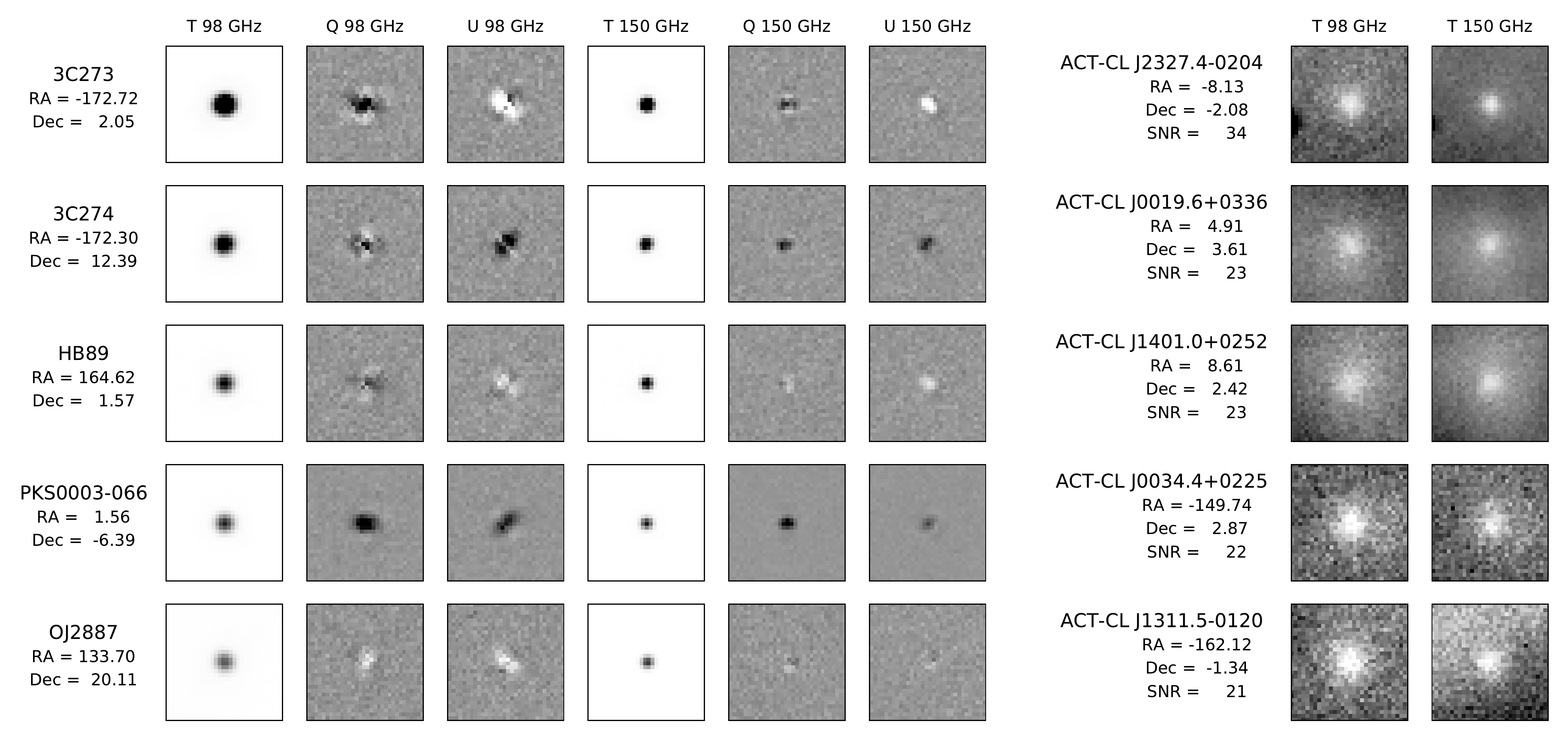}
    \caption{Sample of compact arcminute-scale objects detected in the ACT DR4 maps. All thumbnails are extracted by reprojecting a subset of the s13--s16 single-frequency map onto the plane tangent to the source coordinates to  remove declination-dependent distortions. Each thumbnail has an extent of $15\times15$ arcminutes and a resolution of 0.1\,arcmin. (Left panel) Top-five highest signal-to-noise point sources detected in temperature at \freqb\,GHz, ordered (top to bottom) by decreasing peak temperature. The colors (white to black) cover a range of $[0, 5]\times10^4 \mu$K in temperature and $\pm 1.5 \times 10^3\mu$K in polarization. No correction for the T-to-P leakage from the main-beam was applied. (Right panel) Top-five highest signal-to-noise SZ-selected galaxy clusters detected in a combination of the \freqa\ and \freqb\,GHz temperature maps, ordered (top to bottom) by decreasing signal-to-noise ratio. The colors (white to black) cover a range of $[-500,250]$\,$\mu$K. An offset computed for each thumbnail (excluding the central source) is removed.\\}
    \label{fig:sources}
\end{figure*}

The list of effective region sizes, computed by excluding noisy parts at the edge of each region, and inverse-variance averaged white noise levels are presented in Table~\ref{table:maps_stats}. D56 is the best cross-linked region with a map-noise level\footnote{We ignore the contribution of the D5 and D6 regions.} of 11.5 (18.4)\,\ukarcmin\ at \freqb\ (\freqa)\,GHz over 834\,deg$^2$.
Figures~\ref{fig:d56_f090} and~\ref{fig:d56_f150} show a slightly-filtered cutout of the D56 region observed at \freqa\ and \freqb\,GHz, respectively. The top three panels show the measurement of the three Stokes parameters $T$, $Q$, and $U$ (from top to bottom) highlighting scales between 1 degree to a few arcminutes. The temperature fields look identical, as expected from a high signal-to-noise measurement of a low-foreground part of the sky. The $Q$ and $U$ measurements are also clearly signal dominated and highly correlated between frequencies with with plus-like and cross-like patterns for $Q$ and $U$ that are characteristic of E-mode polarization. The two bottom panels show a combination of the $Q$ and $U$ fields into $E$ and $B$ fields. The $E$ field is clearly signal dominated, at a signal-to-noise much higher than previously reported in L17 and with the addition of the \freqa\,GHz channel. The $B$ field map is clearly noise dominated, thus providing a proxy to understand the noise properties of the dataset. The D5 (in the top-right corner) and the D6 (in the bottom-left corner) regions are now barely visible in the co-added map compared to L17. This shows that the s15 data have greatly improved the overall depth of the D56 region. In addition, a localized 45$^{\circ}$-tilted quadrupole is clearly visible at ${\rm RA}=1.56$ and ${\rm Dec}=-6.39$, with a corresponding 0$^{\circ}$-tilted quadrupole in the $E$ map, showing the polarized emission of the bright active galactic nucleus (AGN) PKS0003-066.

The BN region has an extent of 3157\,deg$^2$ with a noise level of 29.2 (33.9)\,\ukarcmin at \freqb\ (\freqa)\,GHz, respectively. The degree of cross-linking of this region is bi-modal: roughly 1800\,deg$^2$, between right ascensions of 170 and 235 (or $-125$)\,deg, have been observed both during rising and setting, i.e. with the sky at two different orientations with respect to the constant-elevation scans (see Fig.~\ref{fig:ivar_f090150}). This subset of BN is used in C20,~\cite{Madhavacheril:2019nfz},~\cite{Namikawa:2020ffr} and~\cite{darwish_atacama_2020}. The remaining 40\% of the region was observed only during rising scans, and is currently used for point-source and cluster searches. Figures~\ref{fig:bn_f090}~and~\ref{fig:bn_f150} show a slightly-filtered cutout of the $T$, $E$, and $B$ fields measured in the BN region at \freqa\ and \freqb\,GHz, respectively. 
Both frequencies show a signal-dominated measurement of the $E$ field. 

The AA region covers roughly 17,000\,deg$^2$ of the sky reaching a noise level of 62.1 (78.7)\,\ukarcmin\ at \freqb\,(\freqa)\,GHz. The degree of cross-linking varies substantially across the observed area. Constant-elevation scans from the ACT site project to arcs on the equatorial sky, and near declination $-35^{\circ}$, these arcs tend to run nearly parallel to lines of constant RA, making it difficult to achieve strong cross-linking.  Some variety in trajectory is achieved by observing at different elevations, but the net result is that mid-latitudes have somewhat different noise properties than the higher and lower latitudes where the cross-linking angle is closer to 90 degrees. In general, it is possible to define stripes at constant declination in the AA maps that have uniform noise properties (modulo differences due to exposure time). This feature has been used in the power spectrum analysis presented in C20.  

The D8 region has comparable map noise levels to the D56 region but over a smaller region of 248\,deg$^2$ of the sky. Given its location on the sky, the degree of cross-linking is very low.

Figure~\ref{fig:sources} illustrates some of the high-resolution information in the s13--s16 ACT temperature and polarization measurements. The left-hand panel shows the emission of the five highest signal-to-noise point sources detected at \freqb\,GHz ordered (top to bottom) by decreasing unpolarized flux. The size difference, especially visible in the temperature thumbnails, is due to the different resolution at \freqa\ and \freqb\,GHz (see \S\ref{subsec:beams} for details). The polarization signal, in $Q$ and $U$, is also clearly correlated between frequencies. The polarized flux from bright AGNs was previously measured using the ACT DR3 \freqb\,GHz data and showed to be on average roughly $3\%$ of the unpolarized flux \citep{actpol_datta_2018}. The right-hand panel shows the five highest signal-to-noise SZ-detected clusters which, as expected from the tSZ effect, produce a temperature decrement both at \freqa\ and \freqb\,GHz. A detailed analysis of SZ-selected clusters will be presented in \cite{hilton_atacama_2020}.

In summary, the combination of the ACT DR4 map products provides the first signal-dominated arcminute-resolution temperature and $E$-mode fluctuation maps over 4000\,deg$^2$ of the sky. They also add substantial small-scale temperature information over 40\% of the sky, complementing existing \planck\ measurements.

\newpage
\subsection{Point source masks}
Although interesting from the astrophysical point of view, the presence of bright point sources in the maps is a source of bias for many cosmological analyses. For this reason we release a set of point source masks as part of the DR4 data products. Each of the 16 per-region, single-season, single-array maps at \freqb\,GHz is matched filtered using a radially-symmetric beam as signal template\footnote{Each map is paired with the corresponding jitter-corrected beam (see \S\ref{subsec:beams}).} and a 2D correlated noise model measured from the same map. We then look for positive temperature fluctuations with a detection threshold of SNR $\ge 2$ $(2.8$ for AA). At \freqb\,GHz, each region of the full s13--s16 survey is observed by at least 2 independent datasets (maps). We use this feature to clean the single-map catalogs from random noise fluctuations by cross-matching all the 16 catalogs and requiring a source to be observed at least twice. A final catalog of source positions, SNR, and fluxes is then formed by inverse-variance weighting the individual matches.\footnote{The weighting is done by assuming independent errors. This is a good approximation at arcminute scales where the noise is greater than the residual CMB signal.}   

The catalog is effectively complete for sources detected at a combined SNR$ \ge 5$, but because of the varying map depth, the corresponding flux limit is a function of sky position. To take full advantage of the deep regions and to keep the number of masks at a minimum, we produce two sets of masks: (i) one set with all sources with flux $\ge 100$\,mJy over the whole AA footprint including BN, and (ii) another set with all sources with a flux $\ge 15$\,mJy only covering the D1, D56, and D8 regions. Both sets are produced for 5, 8, 10, 35 arcmin hole radii. The first set contains 1215 holes over roughly 17,000\,deg$^2$, the second one 1012 holes over roughly 1,200\,deg$^2$.  

A visual inspection of the DR4 maps shows the presence of a few extended astrophysical objects\footnote{A description of such objects can be found in~\cite{naess_atacama_2020}.}. In some cases the center of the objects is detected by the point source finder, but the extended structures may not be fully masked. In addition, some objects are too faint to be robustly detected. We therefore additionally include an extended-source mask. Using the HyperLeda catalog \citep{hyperleda_2014}, we select 414 objects with a mean surface brightness $\lt 25$ mag in the optical $B$-band and with a linear extent $d \gt 4^{\prime}$.\footnote{We pick as linear extent the maximum between the RA and Dec angular extent provided in the catalog.} Extended objects identified by this procedure include members of the Virgo cluster. We then mask a radius $d/2+2$ arcmin at the source location. Both versions of the masks are used as input in the C20 power spectrum analysis.

\section{Likelihood and parameter estimation methods}
\label{sec:like}
\subsection{ACT DR4 power spectra and likelihood}

C20 describes the estimation of the power spectra of the DR4 maps and their covariance matrix. Table~\ref{table:summary_obs} summarizes the regions, maps and frequencies that are used, and the number of individual cross-spectra that enter each region's power spectrum. As described in C20, the spectra for regions which have sources masked to a 15~mJy level are co-added into a `deep' product, with the `wide' product composed of spectra from regions masked to a 100~mJy level. The foreground-marginalized CMB bandpowers (in TT, TE, EE) are then estimated from the multi-frequency spectra, including data from ACT--MBAC DR2~\citep{Das:2013zf}, for these two subsets of the data. The resultant DR4 coadded spectra are shown in Fig.~24 of C20 and in \S\ref{sec:lcdm} of this paper.

\begin{table*}[htp!]
\begin{center}
\caption{Regions, areas, maps and frequencies that are used in the DR4 power spectra (described in C20), and the number of individual cross-spectra that enter each region's power spectrum\label{table:summary_obs}.}
\begin{tabular}{llrc|rl|rlrlrl}
\hline
\hline
\rule{0pt}{3ex}&	{\bf Region} & {\bf Area}& {\bf Maps\tablenotemark{a}} &\multicolumn{2}{c|}{\boldmath $n_{\rm maps}$} & \multicolumn{2}{c}{\boldmath $98\times98$} &  \multicolumn{2}{c}{\boldmath $98\times150$}&  \multicolumn{2}{c}{\boldmath $150\times150$} \\
&	&   ($\rm{deg}^2$)  &  & 98\,GHz & 150\,GHz & $N_{TT}$\tablenotemark{b} & $N_{TE}$ & $N_{TT}$ &$N_{TE}$ &  $N_{TT}$ &$N_{TE}$ \\
	\hline
{\bf Deep\tablenotemark{c}}  &	D1  & 23 & s13 PA1 & 0 & 1  & 0 & 0 & 0 & 0 & 1 & 1 \\
&	D5  & 20 &s13 PA1& 0&1 & 0&0 & 0&0& 1&1\\
&	D6   &  20  &s13 PA1&  0&1  & 0 &0 &   0&0  &    1&1\\
&	D56 & 340 &s14 PA12, s15 PA123 &  1&5  & 1 &1 &   5&10&15&25\\
&&&&&&&	\\
{\bf Wide}&	BN  &  1400& s15 PA123& 1&3 & 1&1 &3&6&6&9\\
&	AA-w0  &  570  & s16 PA23 & 1&2 & 1&1 &2&4&3&4\\
&	AA-w1  &  1170  &  s16 PA23&1&2 & 1&1 &2&4&3&4\\
&	AA-w3  &  880  &  s16 PA23& 1&2 & 1&1 &2&4&3&4\\
&	AA-w4  &  1070  &  s16 PA23& 1&2 & 1&1&2&4&3&4\\
&	AA-w5  &  210  &  s16 PA23 & 1&2  & 1&1 &2&4&3&4\\
\hline
\hline
\end{tabular}
\begin{tablenotes}
\vskip -0.5in
\item \textsuperscript{a} The maps are labeled by season of observation (s13, s14, s15 or s16) and by detector array (PA1, PA2, PA3). When multiple arrays are used, PA123 indicates PA1+PA2+PA3, for example. Only PA3 has detectors at 98~GHz. The number of maps is $n_{\rm maps}$.
\item \textsuperscript{b} The number of TT cross-spectra, $N_{TT}$, is equal to $N_{EE}$.
  \item \textsuperscript{c} The `deep' regions have a point source mask with a 15~mJy threshold; the `wide' regions have a 100~mJy threshold mask.
\end{tablenotes}
\end{center}
\end{table*}

The likelihood of the derived CMB-only ACT bandpowers can be well described by a simple Gaussian distribution, summing the contribution from the deep and wide bandpowers as
\begin{eqnarray}
  -2\rm{ln}\mathcal{L}_{\rm ACT}=-2\rm{ln}\mathcal{L}_{\rm ACT, d}-2\rm{ln}\mathcal{L}_{\rm ACT, w} \,, 
\end{eqnarray}
with 
\begin{eqnarray}
-2\rm{ln}\mathcal{L_{\rm ACT, d/w}} = (C_b^{\rm th}-C_b^{\rm d/w})^T {\bf \Sigma}_{\rm d/w}^{-1}
(C_b^{\rm th}-C_b^{\rm d/w}) \,,\nonumber\\
\end{eqnarray} 
where $C_b^{\rm th}$ is the binned CMB theory, $C_b^{\rm d/w}$ are the deep and wide CMB bandpowers and {$\bf \Sigma_{\rm d/w}$} are their covariance matrices. Although the deep and wide spectra are both estimates of the same underlying cleaned CMB power spectrum, their bandpower window functions that bin the theory are different (see C20), and so are kept separate in the likelihood.   

The covariance of the CMB bandpowers already accounts for foreground uncertainty, and includes calibration and beam uncertainty. Only one nuisance parameter, $y_p$, is included in the likelihood to marginalize over an overall polarization efficiency. It scales $C^{\rm TE}=C^{\rm th, TE}\times y_p$ and $C^{\rm EE}=C^{\rm th, EE}\times y_p^2$ and we allow it to vary in a flat symmetric range centred at $y_p=1$.\footnote{We choose to work with symmetric ranges as this acts as an overall polarization calibration for the CMB bandpowers; we tested with the full multi-frequency likelihood that restricting polarization efficiencies to be less than unity does not impact cosmological results.} This parameter is different from the frequency-dependent polarization efficiencies of C20, the frequency-dependent $y_p$s have been combined in the CMB-only extraction step to retain the single overall polarization efficiency used here. The difference in efficiency amplitude across frequencies has also been accounted for in that step and therefore we here expect $y_p=1$. 

We provide both Fortran--90 and Python versions of the likelihood software, \texttt{actpollite\_dr4}, with the DR4 data release on LAMBDA. The lite likelihood product has been tested and validated against the full multi-frequency analysis for \LCDM\ and all extended models considered here; small differences in cosmological constraints are due to the additional information from MBAC incorporated in the \texttt{lite} likelihood.

\subsection{Additional CMB data}
\subsubsection{Large-scale polarization}
While our primary goal is to make a measurement of cosmological parameters that is independent of the \planck\ data, ACT by itself cannot constrain the optical depth to reionization, $\tau$. There is also evidence that the \wmap\ estimate of the large-scale polarization has dust contamination~\citep{planck2013-p08}. We 
choose to include an estimate of $\tau$ obtained in recent analyses combining \planck\ and \wmap\ large-scale polarization measurements. We use as a baseline a Gaussian prior of $\tau=0.065 \pm 0.015$, a conservative choice based on {\it Planck}-HFI, {\it Planck}-LFI \citep{planck2016-l05, 2019arXiv190809856P} and {\it Planck}-LFI combined with \wmap\,  estimates~\citep{Natale:2020owc}.\footnote{For reference, the optical depth estimated from the full \LCDM\ cosmological analysis of \planck-HFI in the legacy release results is $\tau = 0.054\pm 0.007$ ~\citep{planck2016-l06}; using only low multipole data $\tau = 0.0566^{+0.0053}_{-0.0062}$  from improved HFI maps~\citep{2019arXiv190809856P} and $\tau = 0.051\pm 0.006$ from joint processing of LFI and HFI maps~\citep{Plancknpipe}. The combination of WMAP and \planck-LFI incling also TE data yields $\tau = 0.069\pm 0.012$ in~\cite{Natale:2020owc}. The change in $\tau$ due to the prior we use here causes a 1.6$\sigma$ shift in mean and 60\% widening of error in the correlated spectrum amplitude parameter and has very little impact on other \LCDM\ parameters.}

\subsubsection{\wmap\ data}
ACT has a minimum multipole of 600 in TT, and 350 in TE and EE, and so lacks data around the first two acoustic peaks in TT and first full peak in TE/EE. Therefore in our nominal results (see \S\ref{sec:lcdm}) we include the full power spectrum information from \wmap\, using the public 9-year observations  \citep{bennett/etal:2013} on large and intermediate scales ($2<\ell<1200$) in temperature and at intermediate scales in TE ($24<\ell<800$) using the public {\it WMAP} 9-year likelihood software. We do not use the $2<\ell<23$ polarization likelihood\footnote{Setting to false the {\texttt use\_WMAP\_lowl\_pol} flag.} which is replaced by the $\tau$ prior.

ACT and \wmap\, only overlap on angular scales that are noise dominated in at least one of the two experiments and therefore we ignore correlations between the two datasets. We tested and confirmed this hypothesis with the same calculations described for \planck\ in the next subsection. The likelihood can then be well approximated as $-2\rm{ln}\mathcal{L}_{\rm ACT}-2\rm{ln}\mathcal{L}_{\rm WMAP}$. 

\subsubsection{Use of Planck data}
A goal of this paper is to give an estimate of the cosmological parameters using measurements that are independent of \planck. However, we also study the consistency of the ACT and \planck\ data, and the impact of combining the two datasets.
For \planck\, we use the latest 2018 TT/TE/EE high-$\ell$ {\texttt {plik\_lite}}  likelihood~\citep{planck2016-l05} ($30<\ell<2508$ in TT; $30<\ell<1996$ in TE/EE) in combination with the {\texttt {commander}} low-multipole likelihood in temperature ($2<\ell<29$). In all cases we impose the same $\tau$ prior used for ACT.\footnote{We note that because of this prior all \planck-derived results reported here will be slightly different from the \planck\ legacy results in \cite{planck2016-l06}.} 

To combine ACT with \planck\, we neglect the covariance between the two datasets and simply multiply the likelihoods. For this to be valid we truncate the multipole range used for ACT, to keep the effect of double counting information below a chosen threshold. To choose these cuts we compute the Fisher-forecasted $\Lambda$CDM parameter errors derived by truncating ACT spectra below some minimum multipole range. We compute the ratio of errors when ignoring the ACT-\planck\ covariance, to those where the full off-diagonal covariance is included.\footnote{The off-diagonal covariance was estimated assuming that the CMB information contained in the ACT regions is a subset of that in \planck, ignoring differences due to data filtering or source masking.}
We then select the minimum multipole for which this ratio is below 1.05, i.e., where the impact of neglecting the ACT-\planck\ covariance is less than a 5\% underestimation of errors. We find that multipoles below $1800$ in TT should not be included for the ACT data in this case, but no cut in TE/EE is required. This $\ell$ range cut was found to also be valid for single-parameter \LCDM\ extensions. To account for this we introduce a flag in the ACT likelihood to select the TT spectra at $\ell>1800$ when combining with \planck, and then set the joint likelihood to be $-2\rm{ln}\mathcal{L}_{\rm ACT}-2\rm{ln}\mathcal{L}_{\rm Planck}$. In future analyses we expect to include the cross-correlation between ACT and \planck\ in a joint likelihood, to exploit the full angular range of both datasets.

\subsection{Parameter estimation methods}

To sample cosmological parameters we use the publicly available \texttt{CosmoMC} software \citep{Lewis:2002ah}\footnote{The results were also reproduced with different codes for theory predictions and sampling, i.e., using \texttt{CLASS+Cobaya}, \texttt{CAMB+Cobaya} and \texttt{CAMB+CosmoSIS}~\citep{CLASS, CAMB,cosmosis2015,Cobaya}.} and estimate the basic six $\Lambda$CDM cosmological parameters: the baryon and cold dark matter densities, $\Omega_b h^2$ and  $\Omega_c h^2$, an approximation to the angular scale of the acoustic horizon at decoupling, $\theta_{\rm MC}$, the reionization optical depth, $\tau$, the amplitude and the scalar spectral index of primordial adiabatic density perturbations, $A_s$ and $n_s$, both defined at a pivot scale $k_0=0.05$\,Mpc$^{-1}$. From these we derive and present constraints on the Hubble constant, $H_0$ in km/s/Mpc, and on the amplitude of matter fluctuations on scales of 8\,$h^{-1}$Mpc, $\sigma_8$. We assume a single family of massive neutrinos with a total mass of 0.06\,eV and impose Big Bang Nucleosynthesis consistency relations to calculate the primordial Helium abundance. 

We also explore a set of extended models beyond $\Lambda$CDM that have a single parameter extension. A set of these are sensitive to the primary CMB temperature and polarization anisotropies: the effective number of relativistic species, $N_{\rm eff}$, the primordial Helium abundance $Y_{\rm HE}$, and the running of the spectral index $d n_s/d ln k$. We also explore models that affect the degree of lensing in the power spectra: the total sum of neutrino masses $\sum m_\nu$, and the spatial curvature $\Omega_k$.

MCMC chains are run with theory predictions computed using $\ell_{\rm max}=6000$ and stopped once the \texttt{CosmoMC} Gelman-Rubin convergence parameter, $R-1$, reaches values smaller than 0.01.

\section{$\Lambda$CDM and the Hubble constant}
\label{sec:lcdm}
\begin{table*}[ht!]
\centering
\newcolumntype{C}{ @{}>{${}}c<{{}$}@{} }
\caption{$\Lambda$CDM parameters with marginalized mean and 68\% confidence level from ACT, ACT+WMAP, ACT+Planck\tablenotemark{a} and Planck\tablenotemark{b}. The best-fit parameters from our baseline result are shown in Column 4.}  \label{table:lcdm}
\centering
\begin{tabular}{l *2{rCl} c *2{rCl}}
\hline\hline
\rule{0pt}{3ex}Parameter & \multicolumn{3}{c}{ACT} & \multicolumn{3}{c}{\bf ACT+WMAP} & ACT+WMAP & \multicolumn{3}{c}{ACT+Planck} & \multicolumn{3}{c}{Planck} \\
& \multicolumn{3}{c}{} & \multicolumn{3}{c}{} & best-fit & \multicolumn{3}{c}{} & \multicolumn{3}{c}{} \\
\hline
\rule{0pt}{2ex}Basic: & &&&&&&& \\
$100\Omega_b h^2$ & $2.153$&\pm&$0.030$ & ${\bf 2.239}$&\pm&${\bf 0.021}$ & $2.223$ & $2.237$&\pm&$0.013$ & $2.241$&\pm&$0.015$\\
$100\Omega_c h^2$ & $11.78$&\pm&$0.38$ & ${\bf 12.00}$&\pm&${\bf 0.26}$ & $12.12$ & $11.97$&\pm&$0.13$ & $11.97$&\pm&$0.14$\\
$10^4 \theta_{\rm MC}$ & $104.225$&\pm&$0.071$ & ${\bf 104.170}$&\pm&${\bf 0.067}$ & $104.180$ & $104.110$&\pm&$0.029$ & $104.094$&\pm&$0.031$\\
$\tau$ & $0.065$&\pm&$0.014$ & ${\bf 0.061}$&\pm&${\bf 0.012}$ & $0.061$ & $0.072$&\pm&$0.012$ & $0.076$&\pm&$0.013$\\
$n_s$  & $1.008$&\pm&$0.015$ & ${\bf 0.9729}$&\pm&${\bf 0.0061}$ & $0.9714$ & $0.9691 $&\pm&$0.0041$ & $0.9668$&\pm&$0.0044$\\
ln$(10^{10} A_s)$ & $3.050$&\pm&$0.030$ & ${\bf 3.064}$&\pm&${\bf 0.024}$ & $3.068$ & $3.086$&\pm&$0.024$ & $3.087$&\pm&$0.026$\\
\hline
\rule{0pt}{2ex}Nuisance: & &&&&&&& \\
$y_p$ & $1.0008$&\pm&$0.0047$ & ${\bf 1.0033}$&\pm&${\bf 0.0042}$ & $1.0028$ & $1.0019$&\pm&$0.0046$ & & -- &\\
\hline
\rule{0pt}{2ex}Derived: & &&&&&&& \\
$H_0$~[km/s/Mpc] & $67.9$&\pm&$1.5$ & ${\bf 67.6}$&\pm&${\bf 1.1}$ & $67.1$ & $67.53$&\pm&$0.56$ & $67.51$&\pm&$0.61$\\
$\sigma_8$ & $0.824$&\pm&$0.016$ & ${\bf 0.822}$&\pm&${\bf 0.012}$ & $0.828$ & $0.8287 $&\pm&$0.0099$ & $0.8279$&\pm&$0.011$\\
\hline
\rule{0pt}{2ex}Additional derived\tablenotemark{c}: & &&&&&&& \\
$\Omega_\Lambda$ &$0.696$&\pm&$0.022$ &${\bf 0.687}$&\pm&${\bf 0.016}$ & $0.680$ & $0.6871$&\pm&$0.0078$& $0.6867$&\pm&$0.0084$\\
$t_0 $ & $13.832$&\pm&$0.047$ & ${\bf 13.772}$&\pm&${\bf 0.039}$ & $13.786$ & $13.791$&\pm&$0.021$ & $13.791$&\pm&$0.025$\\
$S_8 $ & $0.830$&\pm&$0.043$ & ${\bf 0.840}$&\pm&${\bf 0.030}$ & $0.855$ & $0.846$&\pm&$0.016$ & $0.846$&\pm&$0.017$\\
$10^4\theta^*$ & $104.252$&\pm&$0.071$ & ${\bf 104.189}$&\pm&${\bf 0.067}$ & $104.199$ & $104.128$&\pm&$0.029$& $104.112$&\pm&$0.031$ \\
$D_{\rm A}^*$ & $13.972$&\pm&$0.091$ & ${\bf 13.860}$&\pm&${\bf 0.058}$ & $13.839$ & $13.879$&\pm&$0.027$& $13.878$&\pm&$0.028$ \\
$r_{\rm drag}$ & $148.6$&\pm&$1.0$ & ${\bf 147.06}$&\pm&${\bf 0.63}$ & $146.89$ & $147.18$&\pm&$0.29$& $147.14$&\pm&$0.30$ \\
\hline\hline
\end{tabular}
\begin{tablenotes}
\item \textsuperscript{a}
The ACT+Planck combination discards ACT temperature data at $\ell<1800$ to avoid double counting of information.
\item \textsuperscript{b} 
Parameters estimated from \planck\ data alone (using TTTEEE and the same prior on $\tau$ used for ACT) are reported for comparison; the official \planck\ results (e.g., Table 2 of~\citealp{planck2016-l06} for TT,TE,EE+lowE) have a stronger constraint on $\tau$ and correlated parameters, including $A_s$ and $\sigma_8$.
\item \textsuperscript{c} 
Dark Energy density, $\Omega_\Lambda$, the age of the Universe in Gyr, $t_0$, a combination of $\sigma_8$ and the matter density, $S_8=\sigma_8\sqrt{\Omega_m/0.3}$, the acoustic scale at the CMB last-scattering, $\theta^*$, the angular diameter distance to last-scattering in Gpc, $D_{\rm A}^*$, and the sound horizon at the end of the baryonic-drag epoch in Mpc, $r_{\rm drag}$.
\vspace*{0.2in}
\end{tablenotes}
\end{table*}

\subsection{$\Lambda$CDM from ACT alone}\label{subsec:basic_lcdm} 

\begin{figure}[ht!]
    \centering
    \includegraphics[width=\columnwidth]{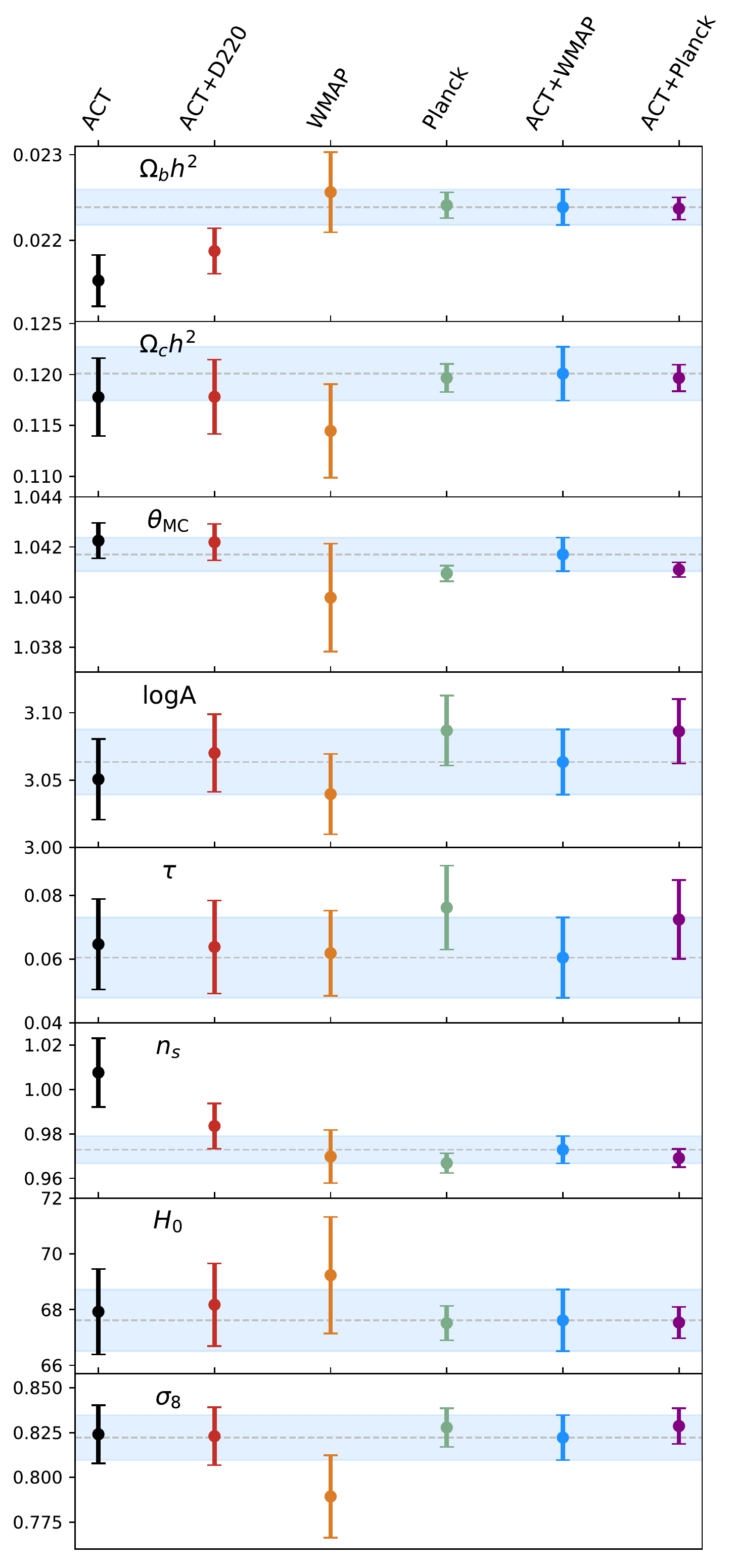}
    \caption{Comparison of $\Lambda$CDM parameters (mean and 1$\sigma$ error) estimated from different datasets (different colors): ACT alone (aslo shown in C20), ACT with the D220 prior, WMAP, Planck, ACT+WMAP and the combination of a subset of ACT data with Planck, ACT+Planck. The light blue horizontal band highlights the 1$\sigma$ measurement in our nominal case for ACT+WMAP. Before including information from the first acoustic peak, ACT prefers a best-fit value with higher $n_s$ and lower $\Omega_bh^2$ than WMAP and \planck, at 2.3--2.7$\sigma$ significance. We see excellent consistency between ACT+WMAP and Planck.} 
    \label{fig:lcdmcomp}
\end{figure}

\begin{figure*}[tp!]
    \centering
    \includegraphics[width=0.95\textwidth]{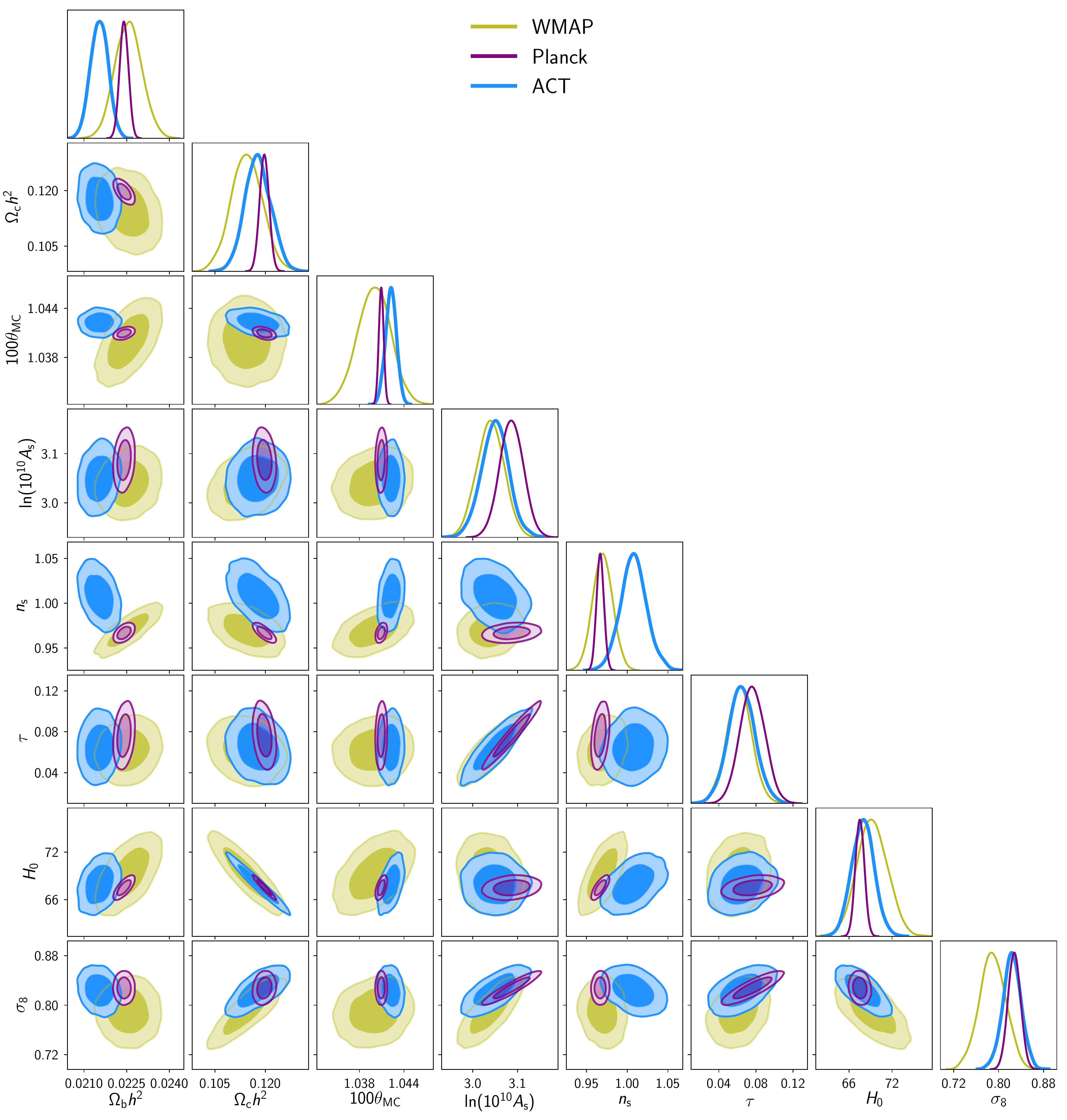}
    \caption{Constraints on the six basic and two derived $\Lambda$CDM parameters from ACT, \wmap\, and \planck\ using a common $\tau$ prior. The diagonal shows the 1-dimensional posterior distributions and the contours show the 68\% and 95\% confidence regions. Parameters from ACT are consistent with \wmap\ (\planck) to within 2.3 (2.7)$\sigma$; the largest difference is in the $\Omega_b h^2$-$n_s$ plane. The Hubble constant agrees to within 1$\sigma$.}
    \vspace*{0.15in}
    \label{fig:lcdmtri}
\end{figure*}

\begin{figure*}[htp!]
    \centering
    \includegraphics[width=0.9\textwidth]{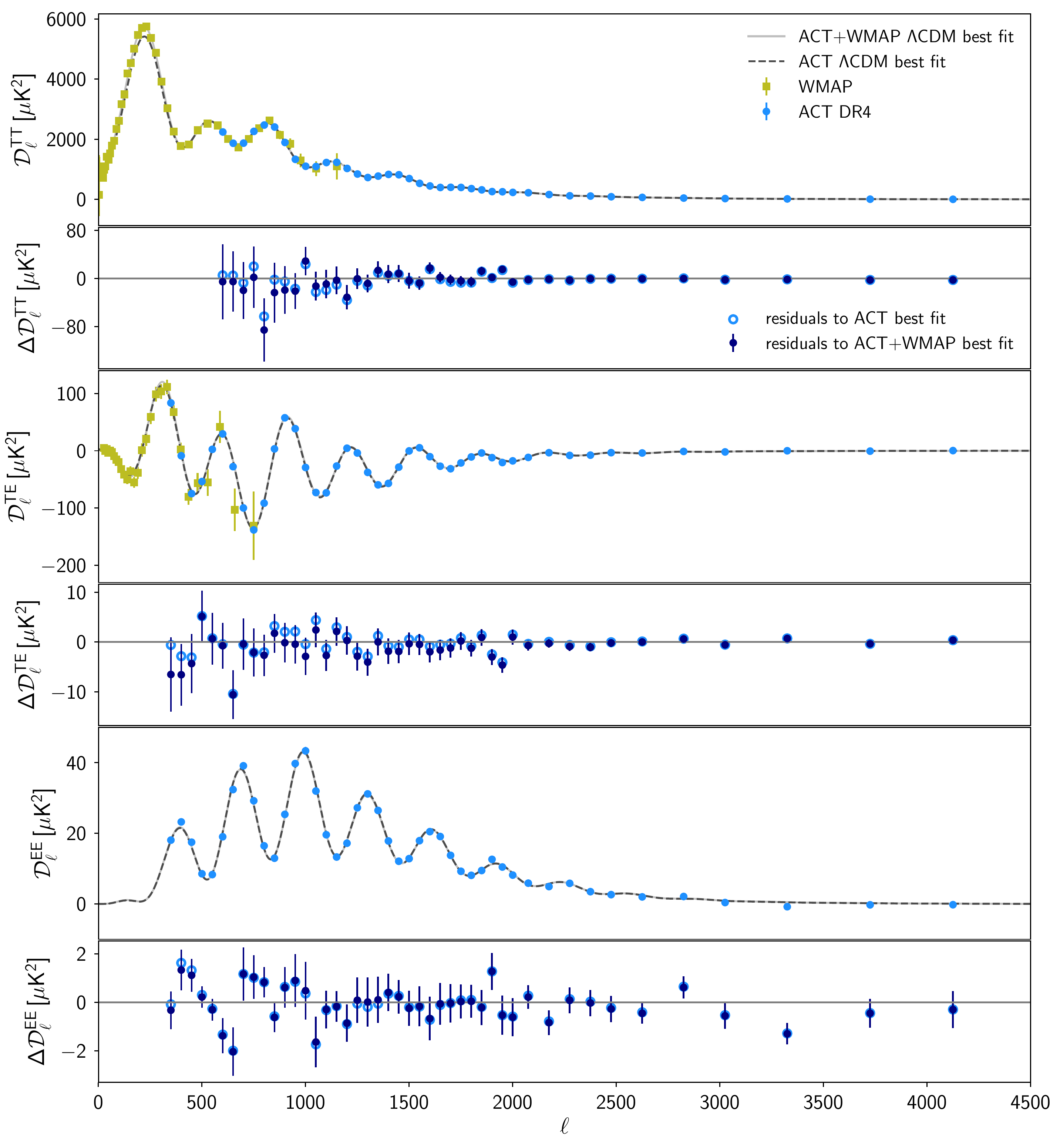}
    \caption{Panel 1, 3, 5: ACT CMB TT, TE and EE power spectra for co-added deep and wide spectra complemented by \wmap\, large scale TT and TE measurements, together with the best-fitting $\Lambda$CDM theoretical model fit to either ACT or ACT+WMAP data. Panel 2, 4, 6: residuals with respect to the best-fitting models, with $\chi^2=279\ (288)$ for ACT (ACT+WMAP). 
    The addition of WMAP data pushes the best-fit model towards a lower (higher) value of $n_s$ ($\Omega_bh^2$) than for ACT-alone, which produces overall negative TE residuals for ACT in the range $1200 < \ell < 1700$, raising the TE $\chi^2$ by 9. This feature may be a statistical fluctuation.\\}
    \label{fig:lcdmres}
\end{figure*}

\begin{figure}[htp!]
    \centering
    \includegraphics[width=0.9\columnwidth]{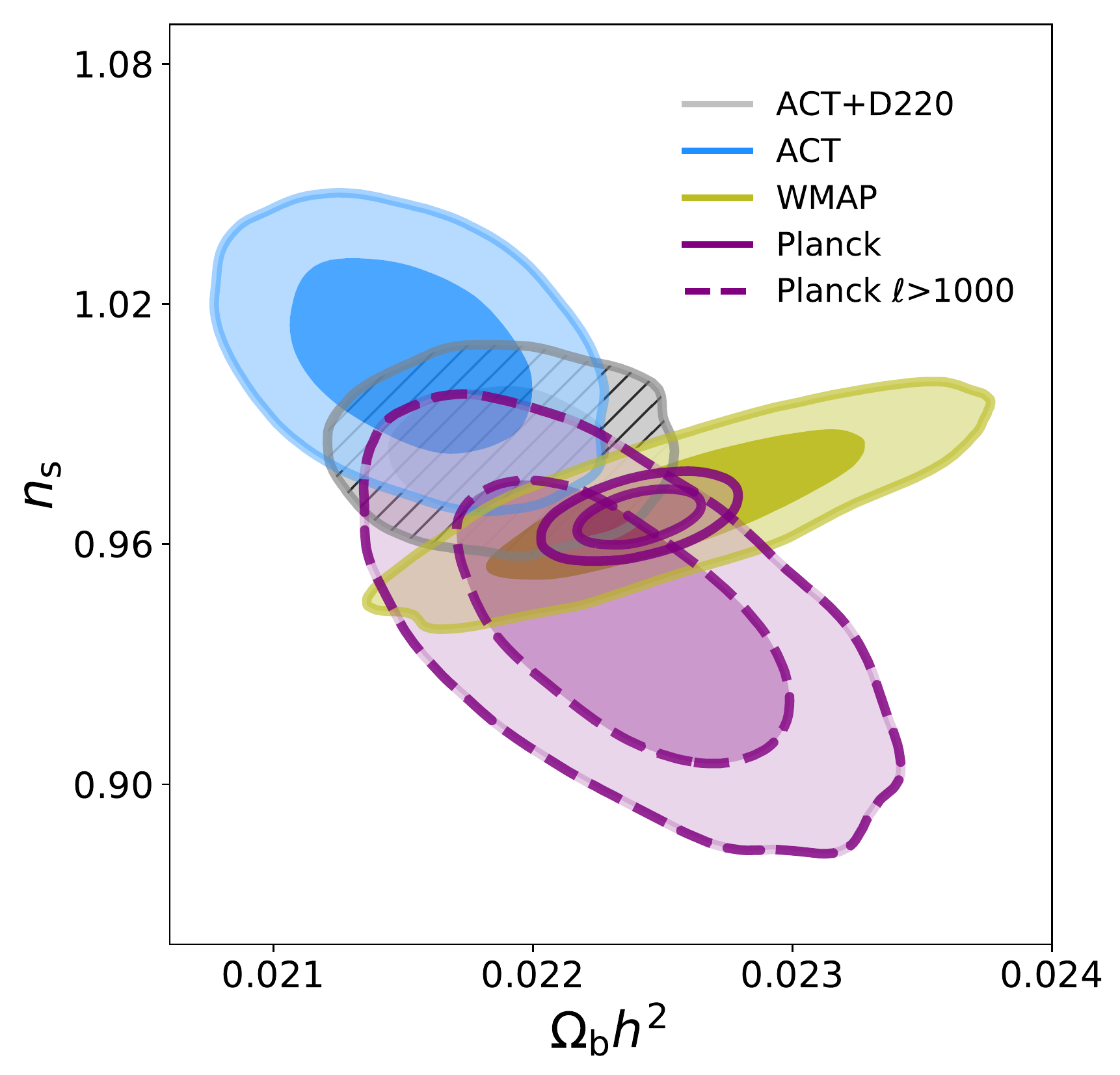}
    \includegraphics[width=0.9\columnwidth]{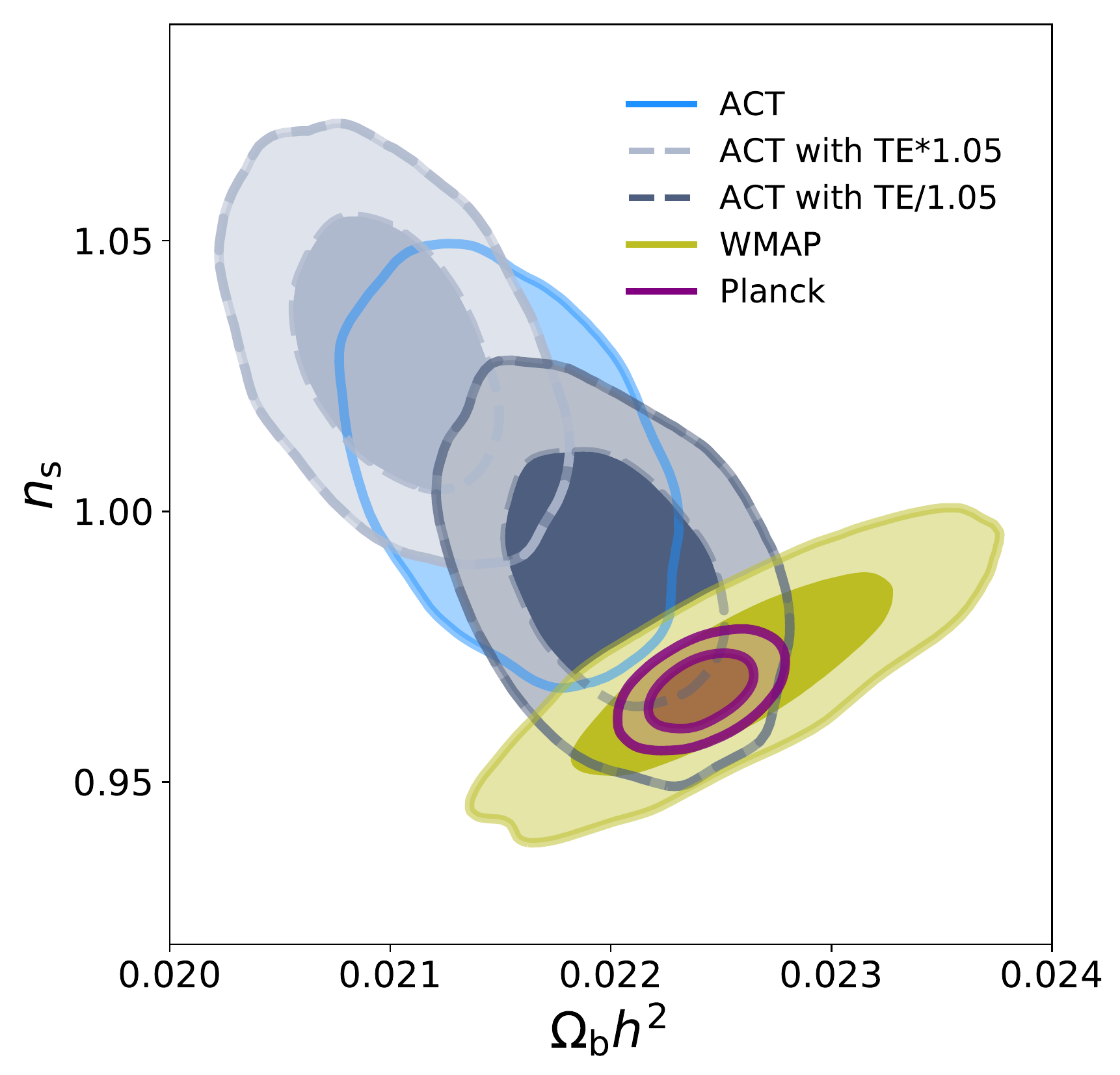}
    \caption{Top panel: The direction of the correlation between $\Omega_b h^2$ and $n_s$ changes when TT large-scale information is included (68\% and 95\% confidence levels shown): for \wmap\ and \planck\ a larger $n_s$ increases the 2nd to 1st peak height ratio, which can be compensated by increasing the baryon density. For ACT, and \planck\ limited to $\ell>1000$, the increase in $n_s$ boosts the small-scale power, now compensated by decreasing the baryon density. Including a prior on the first peak height (D220, grey contours with hatches) brings the ACT, \wmap\ and \planck\ best-fit models into close agreement. Bottom panel: The ACT constraints move along the $\Omega_b h^2-n_s$ degeneracy line if an artificial 5\% calibration factor is applied just to the TE data.\\}
    \label{fig:nsob}
\end{figure}

We find that the ACT data are well fit by the $\Lambda$CDM model (as shown in C20), and our constraints on the six basic and two derived  $\Lambda$CDM parameters are reported in Table~\ref{table:lcdm} and shown in  Fig.~\ref{fig:lcdmcomp} and Fig.~\ref{fig:lcdmtri}, compared to constraints from \wmap\ and from \planck. The polarization efficiency is tightly constrained around unity.

The residuals of the ACT data with respect to the best-fit $\Lambda$CDM model are shown in Fig.~\ref{fig:lcdmres} for the TT, TE and EE spectra, together with the \wmap\ data in TT and TE at larger scales not measured by ACT. This best-fitting model has a $\chi^2$ for ACT of 279 for 254 (260 bandpowers minus 6 freely-varying parameters) degrees of freedom corresponding to a PTE=0.13. The model provides a good fit to the TT, TE and EE spectra.

The estimated $\Lambda$CDM parameters from ACT are broadly consistent with both the \wmap\ and \planck\ constraints, and the ACT data by themselves now put tighter constraints on all the $\Lambda$CDM parameters than the \wmap\ data do, significantly so for the peak angle, $\theta_{\rm MC}$, from which parameters including the Hubble constant are derived. We find excellent consistency in the estimated Hubble constant between ACT and \planck, to be discussed further in \S\ref{subsec:hubble}. ACT measures many more acoustic peaks than \wmap\ and its current power spectrum signal-to-noise ratio is about 1/3 of \planck's (see \S\ref{sec:ACTmodes}). The ACT DR4 significant improvement in constraining power compared to earlier ACT analyses  \citep{sievers/etal:2013,louis_atacama_2017} comes from measuring the CMB peaks to higher precision in both temperature and polarization due to the increased sky area and higher sensitivity (see \S\ref{sec:ACTmodes}). As shown in C20, TT and TE now carry similar weight in constraining $\Lambda$CDM parameters. 

We find that the dominant difference between parameters estimated from ACT compared to \wmap\ and \planck\ lies in the spectral index, $n_s$, and the baryon density, $\Omega_b h^2$. In the absence of data constraining the Sachs-Wolfe plateau and the first acoustic peak, there is a strong anti-correlation between these two parameters, as shown in Fig.~\ref{fig:nsob}. This was also noted in ~\citet{planck2016-l05} and ~\citet{planck2016-l06} when constraining cosmology from \planck\ data using only the smaller angular scales. A lower value of the baryon density, which damps the small-scale power spectrum, can be partially compensated by a higher value of $n_s$, which tilts the spectrum to restore the small-scale power. Within that degeneracy direction, the ACT data prefer a model that has a lower value of the baryon density, and higher spectral index, than \wmap\ or \planck.  

\begin{figure}[htp!]
    \centering
    \includegraphics[width=\columnwidth]{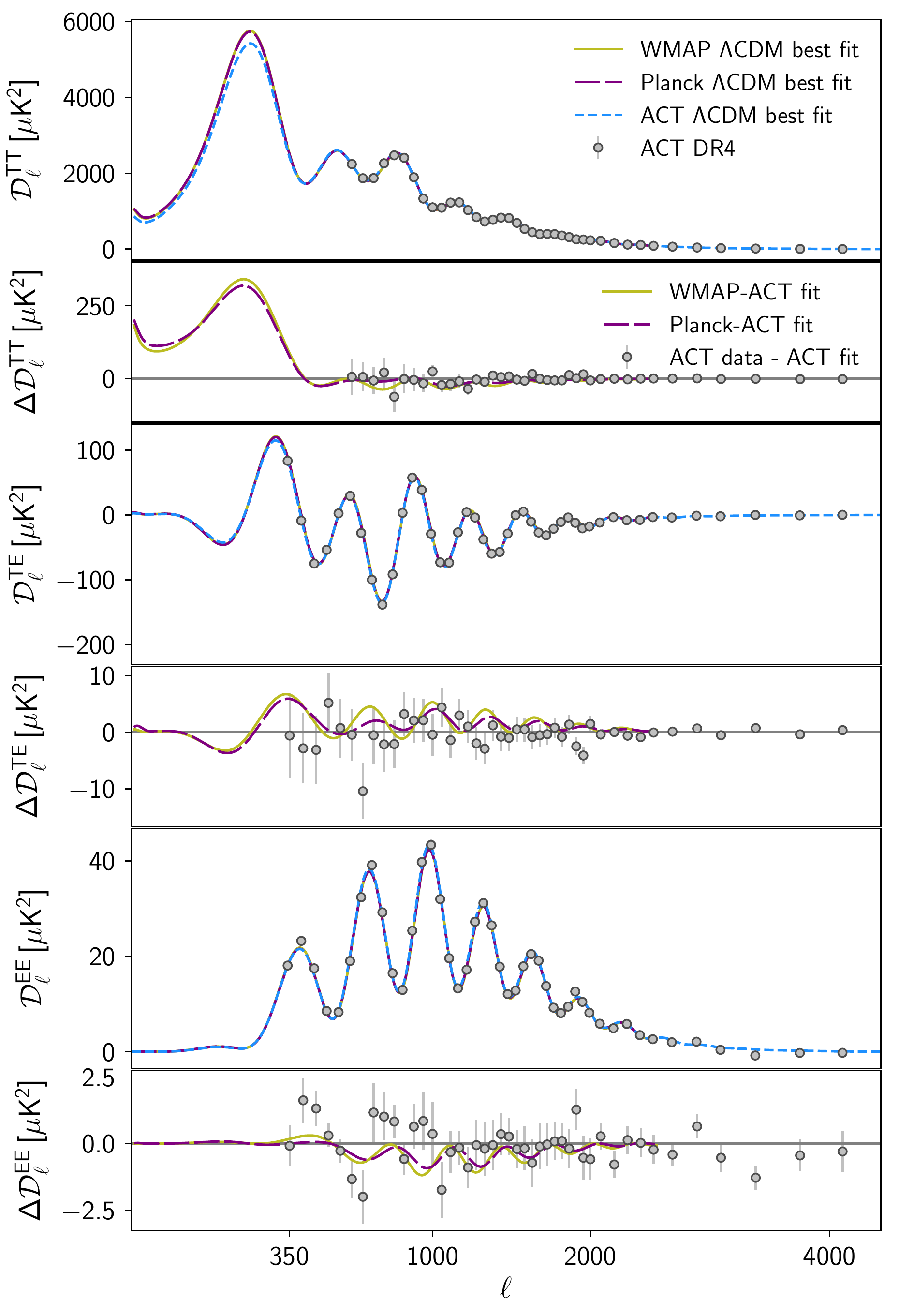}
    \caption{Comparison of the ACT, \wmap\ and \planck\ best-fitting cosmological model spectra and residuals, plotted against the ACT data. The models have consistent peak positions. The ACT-alone model under-estimates the first TT peak amplitude by $2.4\sigma$, and has on average a lower (higher) TE (EE) best-fit spectrum than \wmap\ and \planck, at less than $3\sigma$ level; also seen in Fig.~\ref{fig:lcdmres}.\\}
    \label{fig:lcdm_mod_res}
\end{figure}

 Figure~\ref{fig:lcdm_mod_res} shows the best-fitting spectra for each dataset: the $\Lambda$CDM solution obtained from ACT by itself has a large enough tilt that it underestimates the amplitude of the first peak in TT compared to either \wmap\ or \planck. The amplitude of temperature fluctuations at $\ell=220$ (D220, hereafter) inferred from the ACT best-fit model is $2.4\sigma$ lower than the \wmap\ measurement. 

Since these large scales have been well measured, and consistently, by both \wmap\ and \planck, it is natural to include this information as a prior. If we simply impose a weak prior on the amplitude of the first peak, choosing a conservative Gaussian prior that doubles the \wmap\ uncertainty on the amplitude D220, with D220$=5750.6\pm 71.1$\,$\mu$K$^2$,\footnote{This is the value from the re-estimate of \wmap\, results obtained with the $\tau$ prior used here. We note that D220 from \planck\, yields the same result.} we find that this truncates the $\Omega_b h^2$--$n_s$ degeneracy and brings the parameters estimated from ACT in closer agreement with \wmap\ and \planck, as shown in the top panel of Fig.~\ref{fig:nsob} with the ACT+D220 contours.

A stronger prior is to include the full \wmap\ information. Before doing so we quantify the consistency of $\Lambda$CDM parameters estimated individually from ACT and \wmap\ data.
As in e.g.,~\citet{Addison:2015wyg} we use a quadratic estimator to test whether the parameter differences are consistent with zero, computing
\begin{equation}
\chi^2 = 
({\rm mean}_1-{\rm mean}_2)^T
({\rm Cov}_1+{\rm Cov}_2)^{-1}
({\rm mean}_1-{\rm mean}_2),
\end{equation}
where mean$_i$ and Cov$_i$ are the mean and covariance of parameters estimated from each experiment, $i$. To account for the common information coming from the $\tau$ prior we compute the statistic in a 5-dimensional parameter space using $\Omega_b h^2$, $\Omega_c h^2$, $\theta_{\rm MC}$, $n_s$ and $A_s\exp(-2\tau)$, and we determine significance assuming that the parameter posteriors are Gaussian. We find that the ACT and \wmap\ $\Lambda$CDM parameters agree at the $2.3\sigma$ significance level. Repeating the same test for ACT and \planck\ cosmologies, we find a slightly larger difference, at the $2.4\sigma$ level, or at the $2.7\sigma$ level if we remove common modes between the two datasets (i.e., by excluding ACT TT at $\ell<1800$).

The \wmap\ and \planck\ best-fit models prefer a somewhat higher (lower) amplitude of the spectrum in TE (EE) compared to ACT, in addition to the higher first peak in TT as shown in Fig.~\ref{fig:lcdm_mod_res}. This difference cannot be explained by calibration or by our current model for instrumental beam leakage. The bottom panel of Fig.~\ref{fig:nsob} shows that an artificial 5\% change in the TE calibration, relative to the TT calibration, could shift the ACT constraints towards the \wmap\ and \planck\ measurements. However, at present we have no reason to introduce such a correction, and such a calibration of the polarization would be inconsistent with the EE data. We also note that a similar shift is not generated by any of the \LCDM\ model extensions considered here.

The mild tension between the datasets may simply be a statistical fluctuation, or could be an unaccounted-for systematic effect in the ACT data, or a hint of a preferred model beyond \LCDM. We expect to be able to draw stronger conclusions from additional data from ACT that has already been collected. 

\begin{figure}[htp!]
    \centering
    \includegraphics[width=\columnwidth]{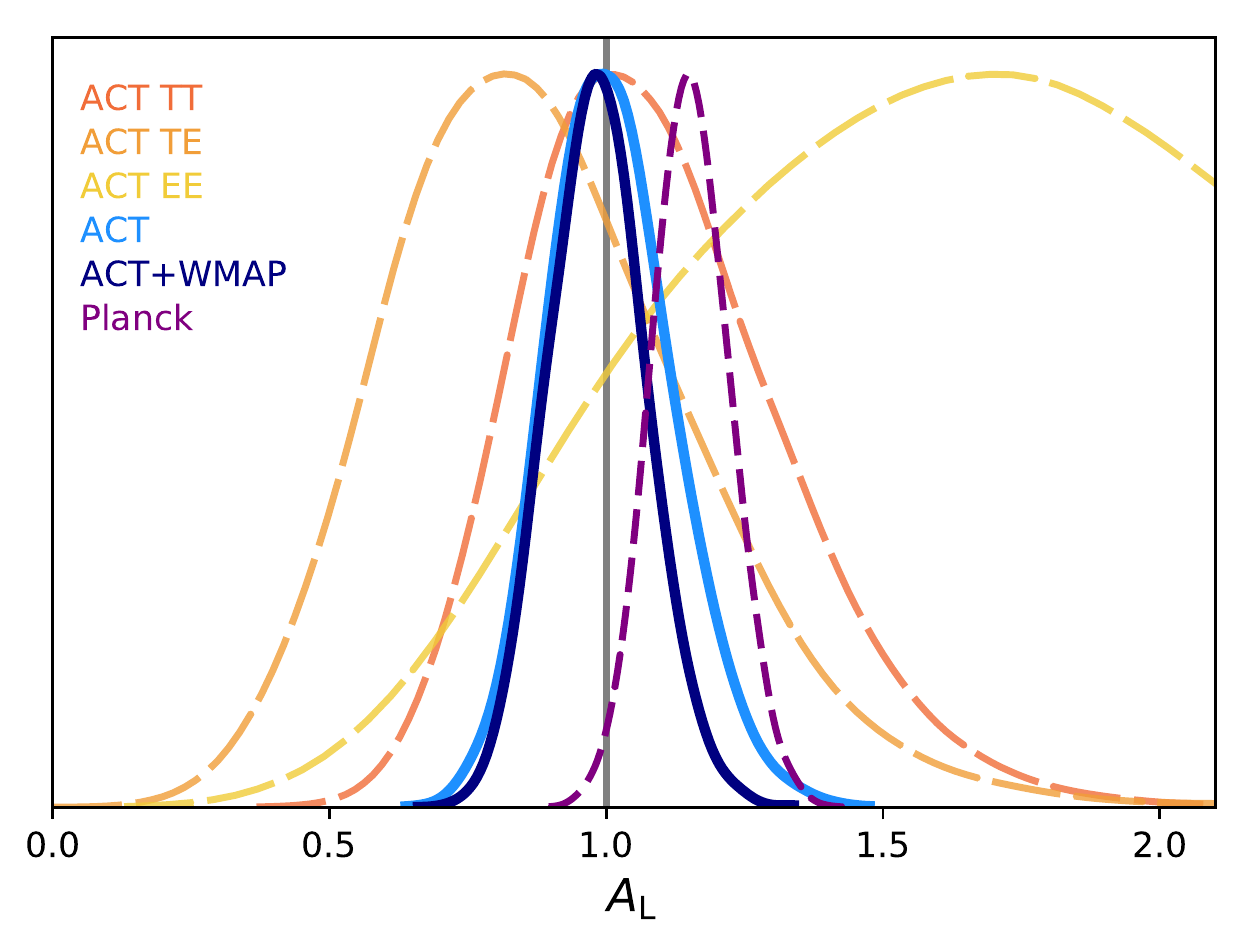}
    \caption{Constraints on the peak smearing amplitude, $A_L$, as measured by ACT, by the individual TT, TE and EE spectra from ACT, and ACT combined with \wmap. The ACT data are consistent with standard gravitational lensing: the parameter $A_L$ artificially scales the lensing potential relative to the standard $A_L=1$ model prediction (shown as a vertical line). The \planck\ measurement is shown for comparison.\\}
    \label{fig:alens}
\end{figure}

\subsubsection{The smoothing of the acoustic peaks}
In the analysis of the \planck\ data, an unexpected feature was a preference for an excess amount of gravitational lensing in the power spectrum compared to the amount expected for the given cosmological model, quantified by the parameter $A_L$~\citep{Calabrese:2008}. Here $A_L=1$ if the degree of lensing matches the model prediction; $A_L=0$ would correspond to no lensing. The effect of lensing is to smooth the features of the acoustic peaks, and add small-scale power. This lensing parameter was found by \planck\ to be higher than what would be expected  at the $2.8\sigma$ level ($A_L=1.180\pm0.065$)~\citep{planck2016-l06}, and higher than what was found using their reconstructed lensing signal. The significance of this excess was reduced with the addition of lensing potential data or, to a lesser extent, with different likelihood versions~\citep{Efstathiou:2019mdh}.  

With the new ACT power spectra data, we find no deviation from the standard lensing effect predicted for a $\Lambda$CDM model, with
\begin{equation}
A_L=1.01\pm 0.11 \, \quad {\rm ACT},
\end{equation}
consistent with unity to within $1\sigma$. The marginalized distribution for $A_L$ is shown in Fig.~\ref{fig:alens}, together with the constraints from the TT, TE and EE spectra alone. These each give consistent results, with the main constraining power from both TT and TE. 

This result is consistent with a companion paper \citep{han_atacama_2020} that studies the degree of lensing in the power spectrum for a subset of the ACT data, in the D56 region; by `delensing' the spectrum to remove some of the impact of lensing directly from the maps, they find no deviation from the expected lensing signal.

\subsection{\label{subsec:awp} Baseline results: $\Lambda$CDM from ACT and \wmap}

\begin{figure*}[htp!]
    \centering
    \includegraphics[width=0.95\textwidth]{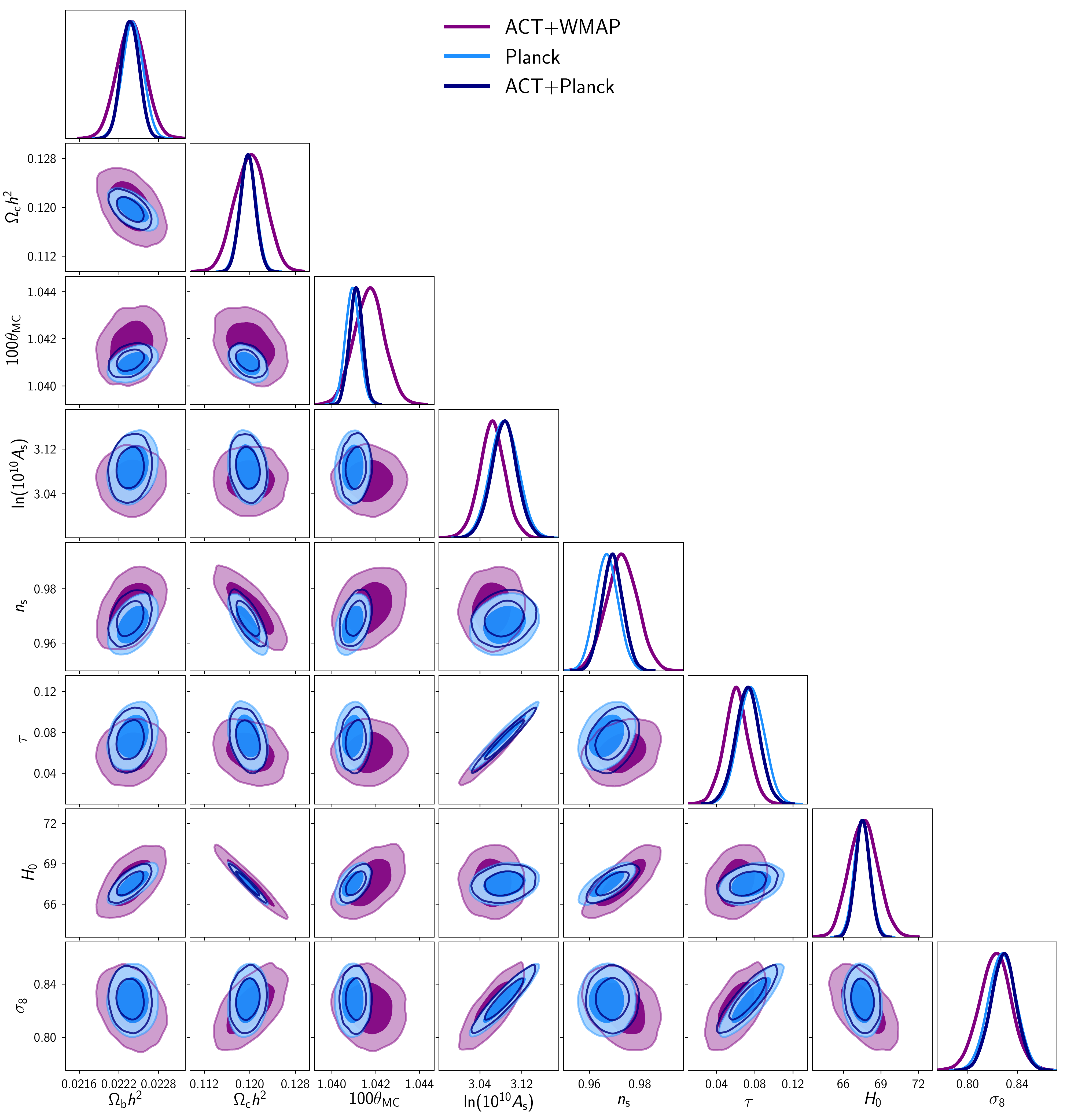}
    \caption{Parameter constraints as in Fig.~\ref{fig:lcdmtri} for ACT+WMAP versus Planck $\Lambda$CDM and derived parameters. There is close agreement between these two independently-measured CMB datasets, including the inferred Hubble constant. ACT+Planck constraints are also shown for comparison.\\}
    \label{fig:lcdmwaptri}
\end{figure*}

With high confidence in the large scale measurements where \wmap\ and \planck\ are signal-dominated and agree well, we take the approach of supplementing the ACT data with the \wmap\ data for our baseline results. As an independent dataset to \planck\, we can assess how the preferred cosmological model compares. 

We report constraints from ACT+WMAP in Table~\ref{table:lcdm} and Figs.~\ref{fig:lcdmcomp} and \ref{fig:lcdmwaptri}. We find that ACT+WMAP together provide stringent new constraints on the cosmological parameters and are in excellent agreement with the \planck\, legacy results~\citep{planck2016-l06}. Each individual parameter differs by no more than $1.1\sigma$, and similar degeneracy directions are seen between pairs of parameters. The scalar spectral index is less than unity at more than 4$\sigma$ significance, consistent with the \planck\ detection of a departure from scale invariance of the primordial fluctuations. As shown in the appendix \S\ref{sec:ACTmodes}, these two independently-measured datasets do not have the same balance of statistical power among different parts of the data: \planck\ has comparatively more weight from the TT spectrum, while ACT has most weight from the TE spectrum, and comparatively more constraining power from the EE spectrum.

The best-fitting ACT+WMAP $\Lambda$CDM model has $\chi^2_{\rm ACT}$=$288$,\footnote{If we allocate to ACT the same degrees of freedom as for the ACT-alone fit, i.e., assuming ACT and \wmap\ both contribute to constraining all the varying parameters, this corresponds to a PTE=0.07. If we instead assume that half of the \LCDM\ parameters are constrained by \wmap\ the ACT PTE is 0.09.} with residuals shown in Fig.~\ref{fig:lcdmres}. The worsening of the ACT $\chi^2$ relative to ACT-alone ($\Delta \chi^2=9$:  +2 in TT, +9 in TE and -2 in EE) comes, as described above, from the TT at large scales and the TE at intermediate multipoles ($800<\ell<2000$, see Fig.~\ref{fig:lcdmres}). 
As mentioned above, the significance of this effect is not yet large enough to draw conclusions about whether it is a statistical fluctuation, or a hint of a systematic effect, or potentially a preference for a different model.

\subsection{The Hubble constant}\label{subsec:hubble}
\begin{figure}[htp!]
    \centering
    \includegraphics[width=\columnwidth]{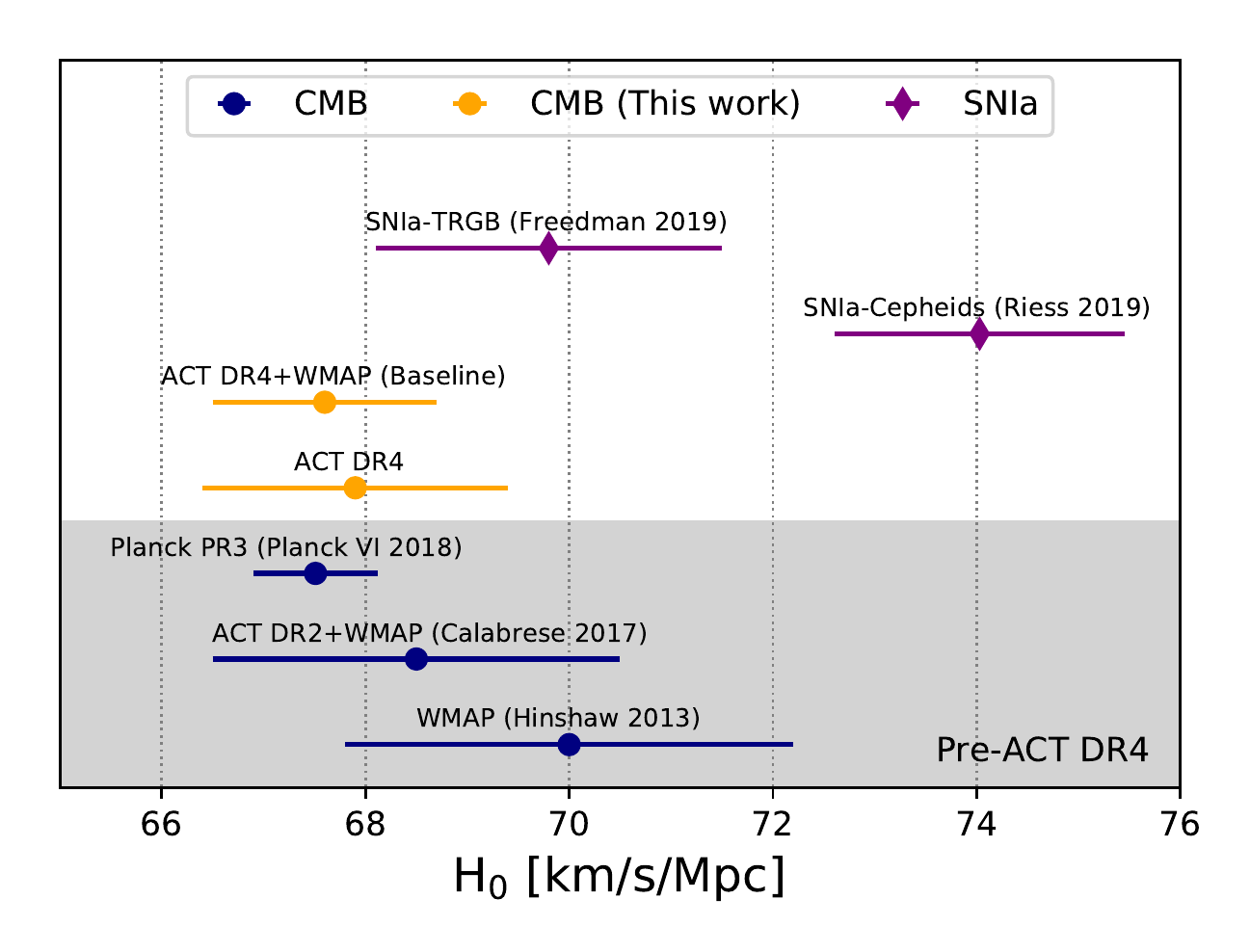}
    \caption{The Hubble constant estimated from ACT and ACT+WMAP (this work) is in excellent agreement with the measurement from \planck~\citep{planck2016-l06}. The constraints are compared to the \wmap-alone estimate~\citep{Hinshaw_2013} and combination of \wmap\ with previous ACT temperature data~\citep{Calabrese:2017ypx} to show the additional information coming from ACT DR4. Within the $\Lambda$CDM model these measurements agree with local measures using distances calibrated with TRGB stars~\citep{freedman/etal:2019}, but disagree with those calibrated using Cepheid variable stars~\citep{riess/etal:2019}.}
    \label{fig:H0comp}
\end{figure}
By combining ACT and \wmap\ we obtain a CMB-derived estimate of the Hubble constant, within the $\Lambda$CDM model, that is derived from the same sky seen by \planck\ but is independently measured.\footnote{The only information retained from \planck\, is the conservative estimate of $\tau$.} We find 
\begin{equation}
    H_0=67.6\pm 1.1 \, \ {\rm km/s/Mpc} \quad {\rm ACT+WMAP}. 
\end{equation}
This 1.6\% inference of the local expansion rate of the universe is remarkably consistent with the \planck\ measurement:
\begin{equation}
    H_0=67.5\pm 0.6 \, \ {\rm km/s/Mpc} \quad {\planck}\,.\footnote{This is our estimate using the same $\tau$ prior as ACT; \citealp{planck2016-l06} finds $67.3\pm0.6$ using their `lowE’ likelilhood to constrain $\tau$.}
\end{equation}
We also find a 2.2\% precision measurement of $H_0$ from ACT alone:
\begin{equation}
H_0=67.9\pm 1.5 \, \ {\rm km/s/Mpc} \quad {\rm ACT} \,,
\end{equation}
or $H_0=68.2\pm1.5$ km/s/Mpc with the D220 prior, and note that the ACT-derived estimate is stable to a number of analysis choices (see C20). Given that TE provides significant constraining power for $H_0$ (see \S\ref{sec:ACTmodes}) we also note that our estimate is stable to tests run on TE; for example, applying an artificial 5\% calibration factor to TE moves the $H_0$ mean by only $0.15\sigma$.
 
Without the ACT data, the \wmap-derived estimate has an uncertainty twice as large ($H_0=70.0\pm2.2$~km/s/Mpc, \citealp{bennett/etal:2013,Hinshaw_2013}). 

This agreement between two independently-measured CMB datasets at a more comparable precision level, adds weight to the robustness of the CMB-inferred measurement.  This new estimate of the Hubble constant, like \planck, is inconsistent with the $z<1$ SH$_0$ES Cepheids/supernovae based measurement \citep{riess/etal:2019} at $>4\sigma$ within the $\Lambda$CDM model (see Fig.~\ref{fig:H0comp}). It also disagrees at $3\sigma$ with the H0liCOW estimate from strong lenses \citep{wong/etal:2019} but agrees with the more recent TDCOSMO/H0LiCOW revised estimate including more flexible lens mass profile modeling and use of stellar kinematics data~\citep{Birrer:2020tax}. Like \planck, it is consistent with the TRGB-based measurement \citep{freedman/etal:2019}, as well as with estimates derived from Baryon Acoustic Oscillations from galaxy clustering in combination with other non-CMB probes including Big Bang Nucleosynthesis and lensing (e.g.,~\citealp{Aubourg:2014yra,Addison:2017fdm,Abbott:2017smn,Cuceu:2019for}).

\begin{figure}[htp!]
    \centering
    \includegraphics[width=\columnwidth]{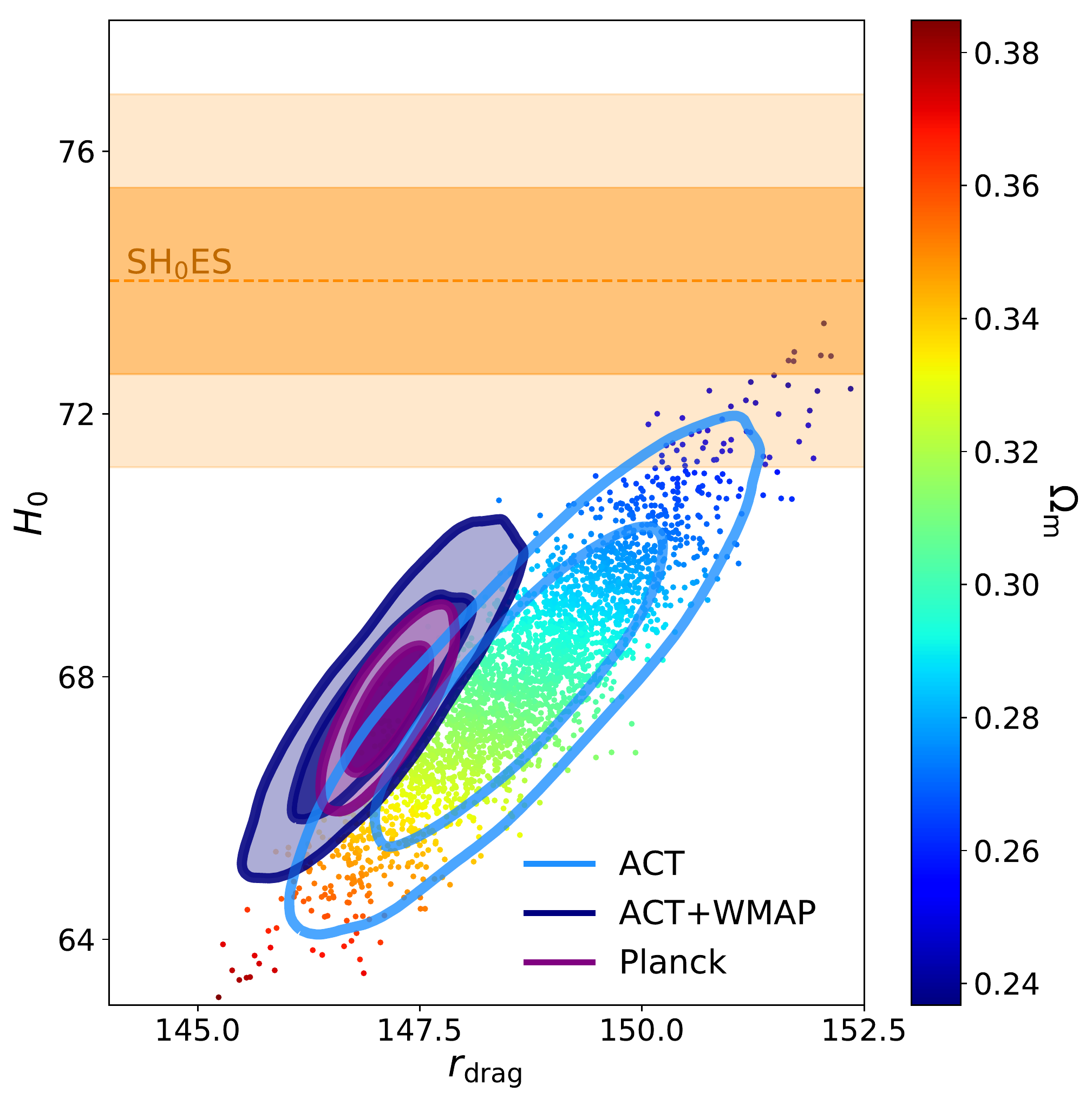}
    \caption{Correlation between the Hubble constant and the sound horizon at the baryon-drag epoch, $H_0-r_{\rm drag}$, showing the CMB-SH$_0$ES discrepancy (as in~\citealp{Knox:2019rjx}). Colored contours are the CMB 68 and 95\% constraints from ACT, ACT+WMAP and Planck; the horizontal band is the Cepheids-based measurement from~\cite{riess/etal:2019}. To reach the higher $H_0$ measured by SH$_0$ES, the CMB needs a lower value of the matter density, $\Omega_m$, which increases the sound horizon, as demonstrated for ACT-alone.}
    \label{fig:H0rs}
\end{figure}

ACT's measurement of $H_0$ comes from measuring the angular scale of the peaks in the temperature and polarization power spectra, coupled with the peak heights which constrain the cold dark matter density and baryon density. An explanation for how this constraint works is given in~\cite{Knox:2019rjx}. To summarize, the peak scale is set by the ratio of the sound horizon, $r_s^*$, to the angular diameter distance to the last-scattering surface, $D_A^*$, with $\theta^*=r_s^*/D_A^*$. The sound horizon is given by the integral of the sound speed, $c_s(t)$, over time before last scattering, $t^*$, with
\begin{equation}
    r_s^*= \int_0^{t^*} \frac{dt}{a(t)}c_s(t).
\end{equation}
It is set by the baryon and cold dark matter densities which affect the sound speed and the scale factor $a(t)$ before recombination, and are measured via the relative peak heights in the power spectra. Figure~\ref{fig:H0rs} shows how the sound horizon is anti-correlated with the total matter density.\footnote{Following~\citealp{Knox:2019rjx} we show $r_{\rm drag}$ rather than $r_s^{*}$, i.e., the sound horizon at the time of the baryon-drag epoch rather than the sound horizon at the time of CMB last-scattering. The two horizons differ by about 3~Mpc and exhibit the same trend.}

By measuring the peak scale, $\theta^*$, and given the sound horizon, one then finds the angular diameter distance to last scattering.
In a flat universe, this is given as a function of redshift, $z$, by
\begin{equation}
D^*_A(z)= \frac{c}{H_0(1+z)} \int_{0}^{z^*}\frac{dz'}{E(z')} \,,
\end{equation}
where $E(z)=(\Omega_r (1+z)^4+\Omega_m (1+z)^3+\Omega_\Lambda)^{0.5}$ is a function of radiation, matter and dark energy densities, and $c$ is the speed of light. For given densities, this then allows an estimate of the Hubble constant: a universe with a smaller distance to the surface of last scatter will have a larger Hubble constant today. Since the densities are not perfectly measured we see this interplay in the parameter correlations, highlighted in Fig.~\ref{fig:H0rs}: a lower matter density would increase the distance to recombination, so can be compensated with a higher Hubble constant to decrease the distance. An example is given in the Appendix~\ref{sec:H074} to show how a \LCDM\ universe with $H_0=74$~km/s/Mpc gives a poor fit to the data: the matter density is reduced to better fit the peak angles, but the peak heights cannot be sufficiently adjusted by varying the other parameters.\\

\section{Extensions to $\Lambda$CDM}
\label{sec:ext}

\begin{table*}[ht!]
\centering
\caption{Beyond $\Lambda$CDM parameters with 68\% confidence level or 95\% upper limits from ACT, ACT+WMAP, and ACT+Planck. \label{table:extensions}}
\centering
\begin{tabular}{lcccc}
\hline\hline
\rule{0pt}{3ex}Parameter & ACT & {\bf ACT+WMAP} & ACT+Planck & Planck\footnote{Planck alone results (TTTEEE  with the same $\tau$ prior) are reported for reference.} \\
\hline
\rule{0pt}{3ex}$\Omega_k$ & $-0.003_{-0.014}^{+0.022}$& ${\bf -0.001_{-0.010}^{+0.014}}$& $-0.018_{-0.010}^{+0.013}$ & $-0.037_{-0.014}^{+0.020}$ \\
$\Sigma m_\nu$[eV] & $<3.1$ & ${\bf <1.2}$ & $<0.54$ & $<0.37$\\
$N_{\rm eff}$ & $2.42\pm0.41$ & ${\bf 2.46 \pm 0.26}$ & $2.74 \pm 0.17$ & $2.97 \pm 0.19$ \\
$d n_s/d lnk$ & $0.069 \pm 0.029$ & ${\bf 0.0128 \pm 0.0081}$ & $0.0023\pm 0.0063$ & $-0.0067 \pm 0.0067$ \\
$Y_{\rm HE}$ & $0.211\pm 0.031$& ${\bf 0.220 \pm 0.018}$ & $0.232 \pm 0.011$ & $0.243\pm 0.013$\\
\hline\hline
\end{tabular}
\end{table*}

\begin{figure*}[htp!]
    \centering
    \includegraphics[width=\textwidth]{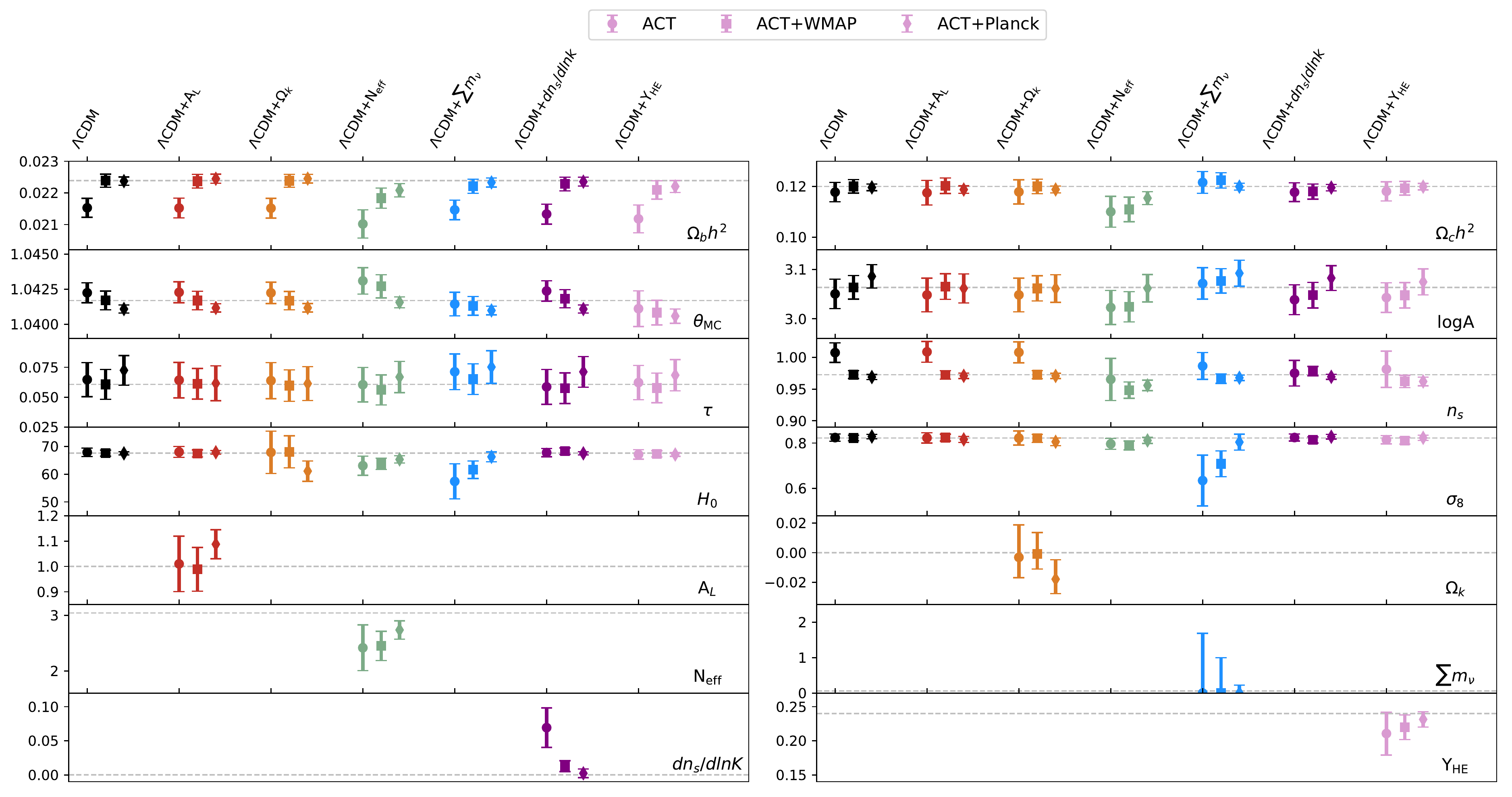}
    \caption{Constraints on the six $\Lambda$CDM and extended parameters, for single parameter beyond-$\Lambda$CDM models. The dashed horizontal lines are centered on our nominal ACT+WMAP $\Lambda$CDM results and on the standard model value for the extended parameters. Different colors span different models; circles, squares and diamonds compare ACT with ACT+WMAP and ACT+Planck.\\}
    \label{fig:exts}
\end{figure*}

The new ACT data provide us with the opportunity to explore a number of physically-motivated single parameter extensions beyond the basic $\Lambda$CDM model. In doing so we look in particular for models that might reduce the mild 2.3$\sigma$ offset between the ACT and \wmap\ data, and the slightly stronger 2.7$\sigma$ difference between the ACT and \planck\ best-fitting models. 

 A summary of the constraints on all of the extended model parameters is reported in Table~\ref{table:extensions} and shown in Fig.~\ref{fig:exts}. We do not find any significant deviations from $\Lambda$CDM, and highlight our main findings below. Figure~\ref{fig:exts} also shows how the basic six parameters move during exploration of extended models due to parameter degeneracies. We note that the Hubble constant is never driven to very high values. 
 
 Like for $\Lambda$CDM, our nominal ACT constraints are estimated using the CMB-only foreground-marginalized likelihood, described in C20. Recall that when extracting this CMB-only power we make use of the earlier ACT MBAC data which includes 220\,GHz observations and thus helps anchor the foreground model and reduces the associated uncertainties at high multipoles. Slightly different results would be obtained using the multi-frequency likelihood which does not include that MBAC data (an example of this is shown for the effective number of relativistic species, $N_{\rm eff}$, in Fig.~\ref{fig:neff_nrun}).

\subsection{Lensing-dependent parameters}

\begin{figure*}[htp!]
    \centering
    \includegraphics[width=\columnwidth]{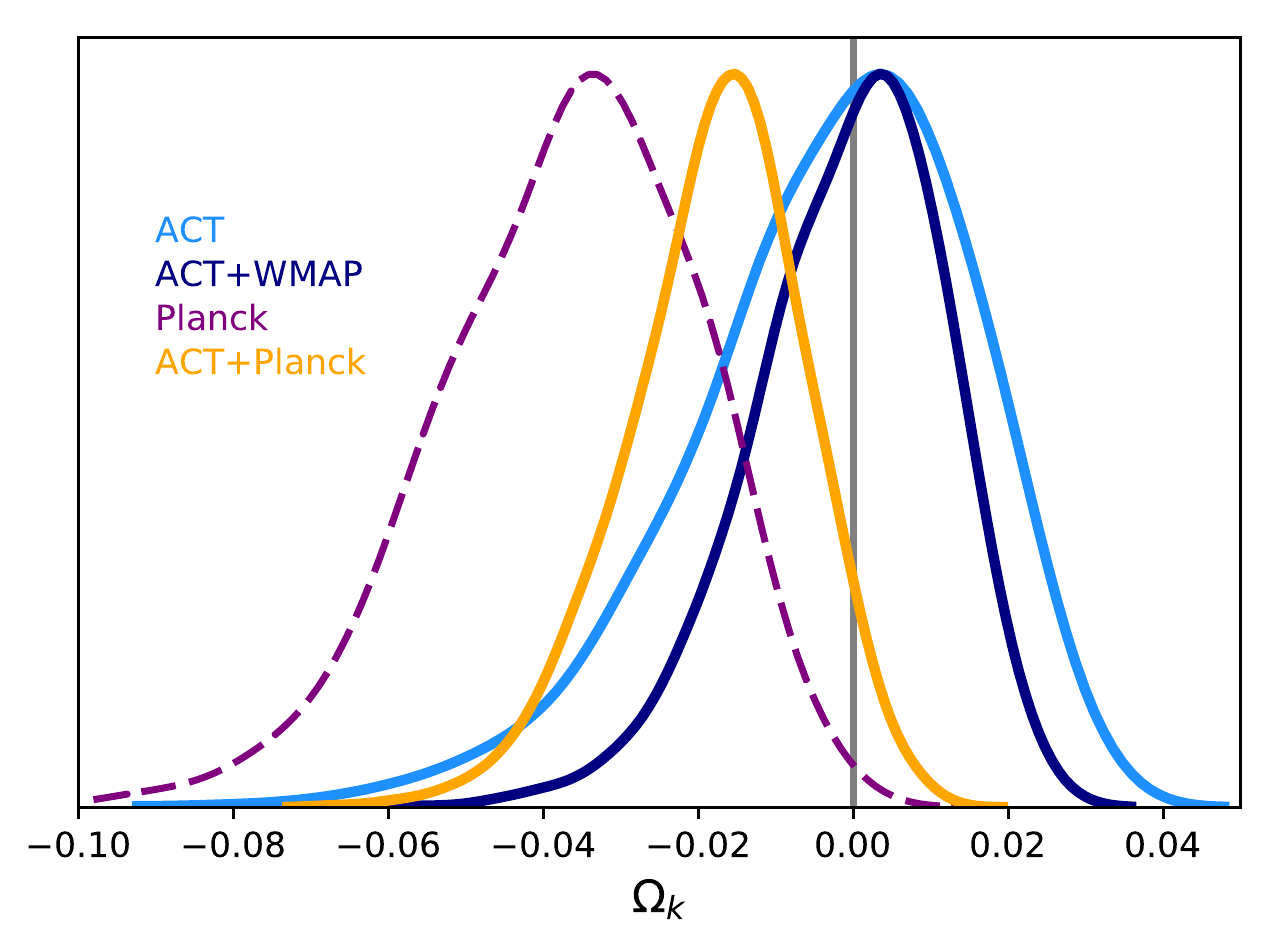}
    \includegraphics[width=\columnwidth]{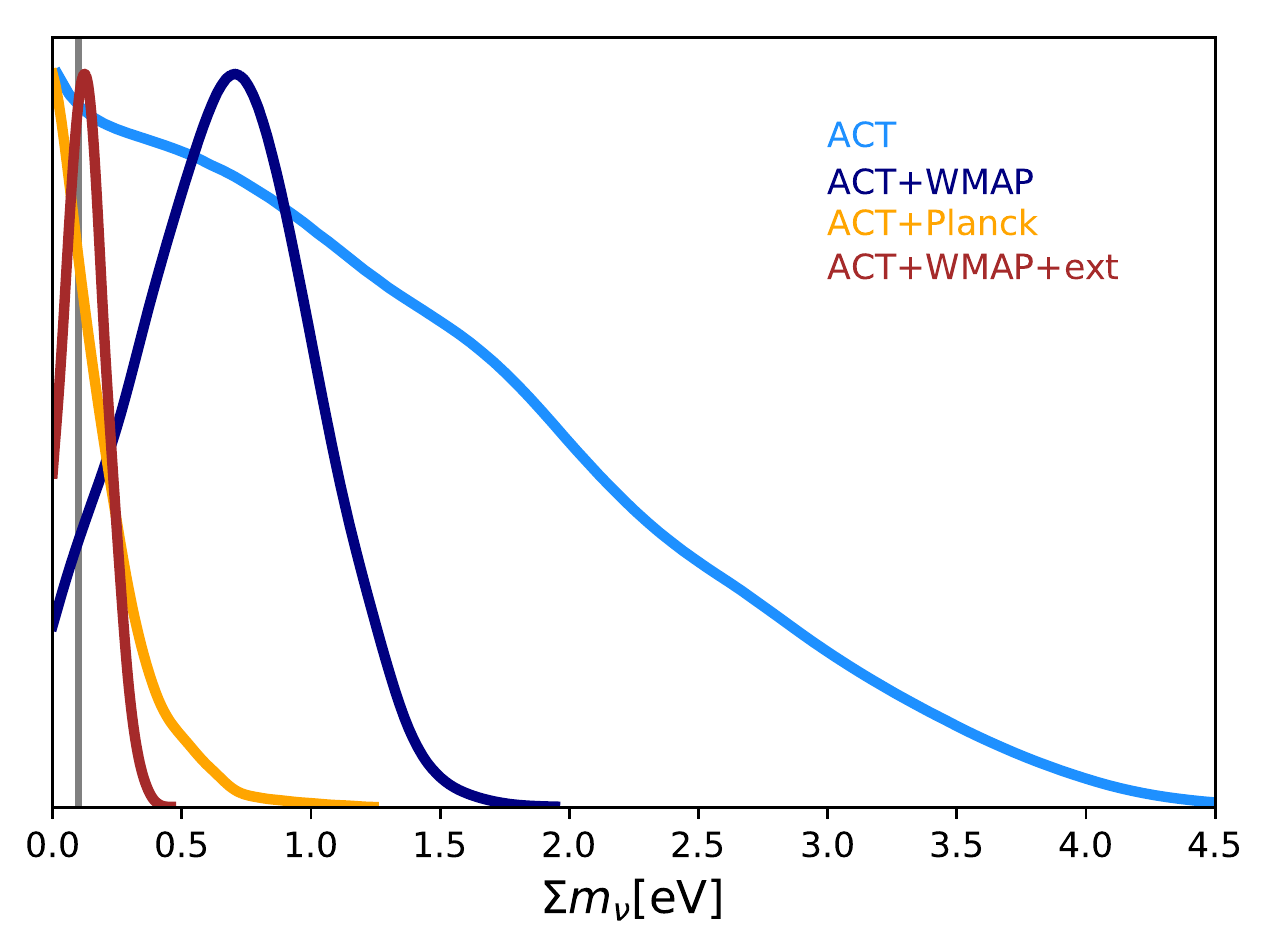}
    \caption{Constraints on the CMB-derived curvature parameter (left) and the sum of neutrino masses (right). These limits are driven by the lensing information in the power spectrum, as well as the primary CMB signal for neutrino mass. With the new ACT data we find no evidence of deviation from flatness, supporting the interpretation that the \planck\ primary CMB constraint (shown in purple for reference) is a statistical fluctuation. The neutrino mass is poorly constrained from ACT alone without the large-scale information. ACT+WMAP combined with external datasets (Planck lensing and BAO) gives $\sum m_\nu<0.27$\,eV at 95\% confidence compared to $<0.19$\,eV at 95\% confidence from \planck\ with the same data combination and same $\tau$ prior. The standard model value for a flat Universe and the expected neutrino mass in an inverted hierarchy are shown with vertical grey lines.\\}
    \label{fig:omegak_mnu}
\end{figure*}

The usual assumption in $\Lambda$CDM is that the geometry of the universe is flat, with no curvature. With the ACT data we find this to be a good assumption, with no evidence for spatial curvature from CMB alone:
\begin{equation}
\Omega_k=-0.001^{+0.014}_{-0.010} \quad {\rm ACT+WMAP}
\end{equation}
as shown in Fig.~\ref{fig:omegak_mnu}. This constraint on the curvature comes from the lensing information in the ACT power spectra: without lensing the curvature and the distance to the CMB, given by a combination of $\Omega_c h^2$, $\Omega_b h^2$, and $H_0$, are degenerate. The fact that ACT sees no evidence for an excess, or lack of, lensing compared to the \LCDM\ model, is reflected in the lack of preference for non-zero curvature. The \planck\ power spectrum data alone prefer a non-flat model at the 2--3$\sigma$ significance~\citep{planck2016-l06} (see also~\citealp{h2019curvature,DiValentino:2019}); this is connected to the deviation from unity of the expected lensing signal ($A_L$) and its significance is reduced with different sky fractions and \planck\ likelihood choices~\citep{Efstathiou_2020}. Our new measurement from ACT, coupled with the reconstructed lensing signal from \planck, lends additional support to the explanation that the preferred non-zero curvature in the \planck\ power spectrum is a statistical fluctuation. A tighter constraint on the curvature comes from combining the CMB data with measurements of the baryon acoustic oscillations scale from galaxy surveys (as done in e.g.,~\citealp{planck2016-l06}); since this measurement is dominated by the BAO data we do not repeat the constraint here.

ACT is also sensitive to the sum of the neutrino masses primarily through the degree of lensing in the power spectrum. The higher the neutrino mass, the more the amplitude of structure will be suppressed, and the smaller the lensing signal will be. The limit on the neutrino mass sum from ACT and \wmap\ is $\sum m_\nu<1.2$\,eV at 95\% confidence.
This upper limit is higher than \planck\ from the power spectrum alone, and is also connected to ACT's inferred lensing signal being slightly lower than \planck's. When combining with other external datasets (for example lensing potential measurements from \planck\,~\citep{planck2016-l08} and baryon acoustic oscillations from BOSS DR12 consensus,  6dFGS and SDSS MGS datasets~\citep{alam/etal:2017,Beutler_2011,Ross_2015}) to break degeneracies, in particular with the cold dark matter density, we find consistent constraints to \planck\, (see Fig.~\ref{fig:omegak_mnu}) with ACT+WMAP combined with external datasets giving $\sum m_{\nu}<0.27$\,eV at 95\% confidence.

\subsection{Primordial parameters}
\begin{figure*}[tp!]
    \centering
    \includegraphics[width=\columnwidth]{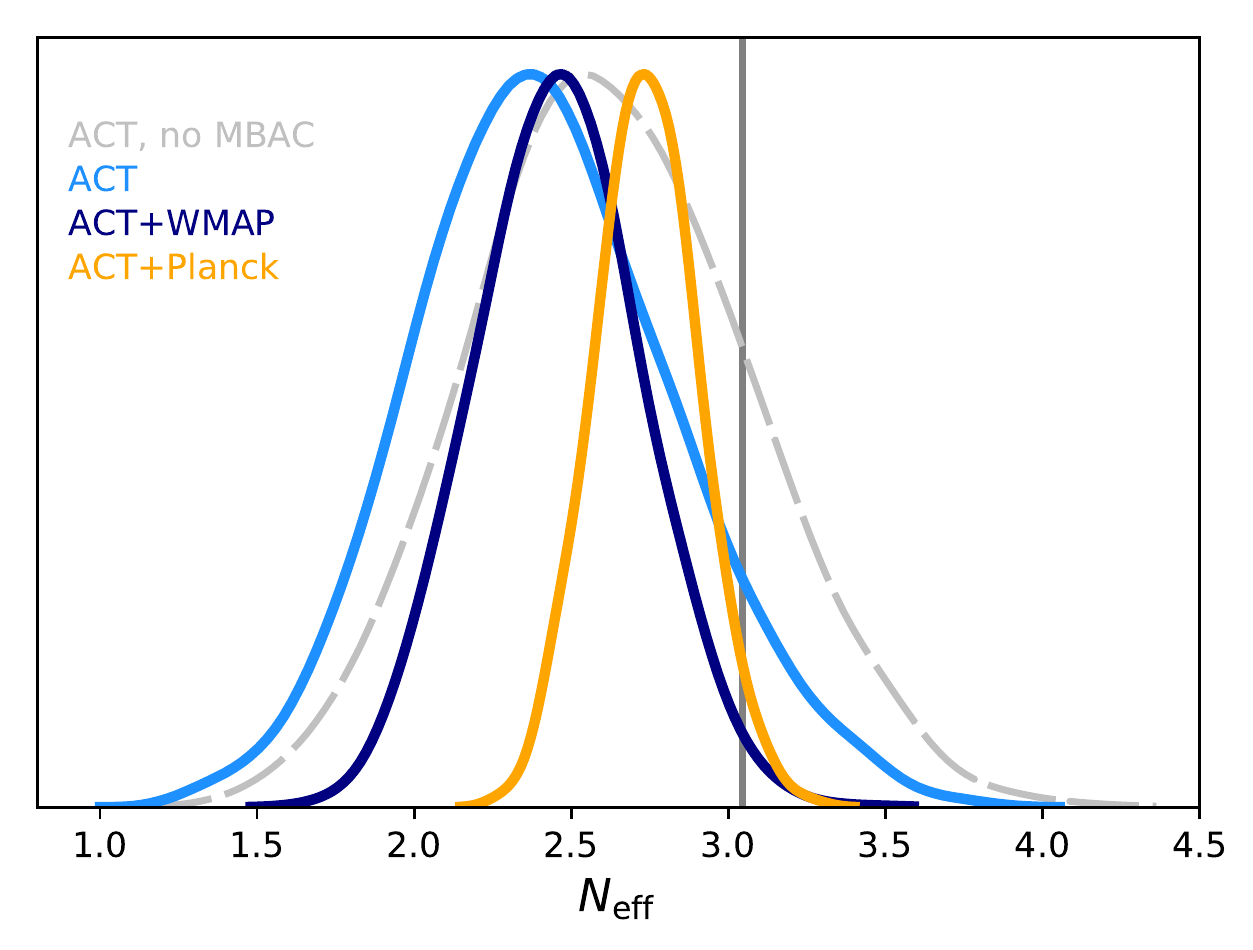}
    \includegraphics[width=\columnwidth]{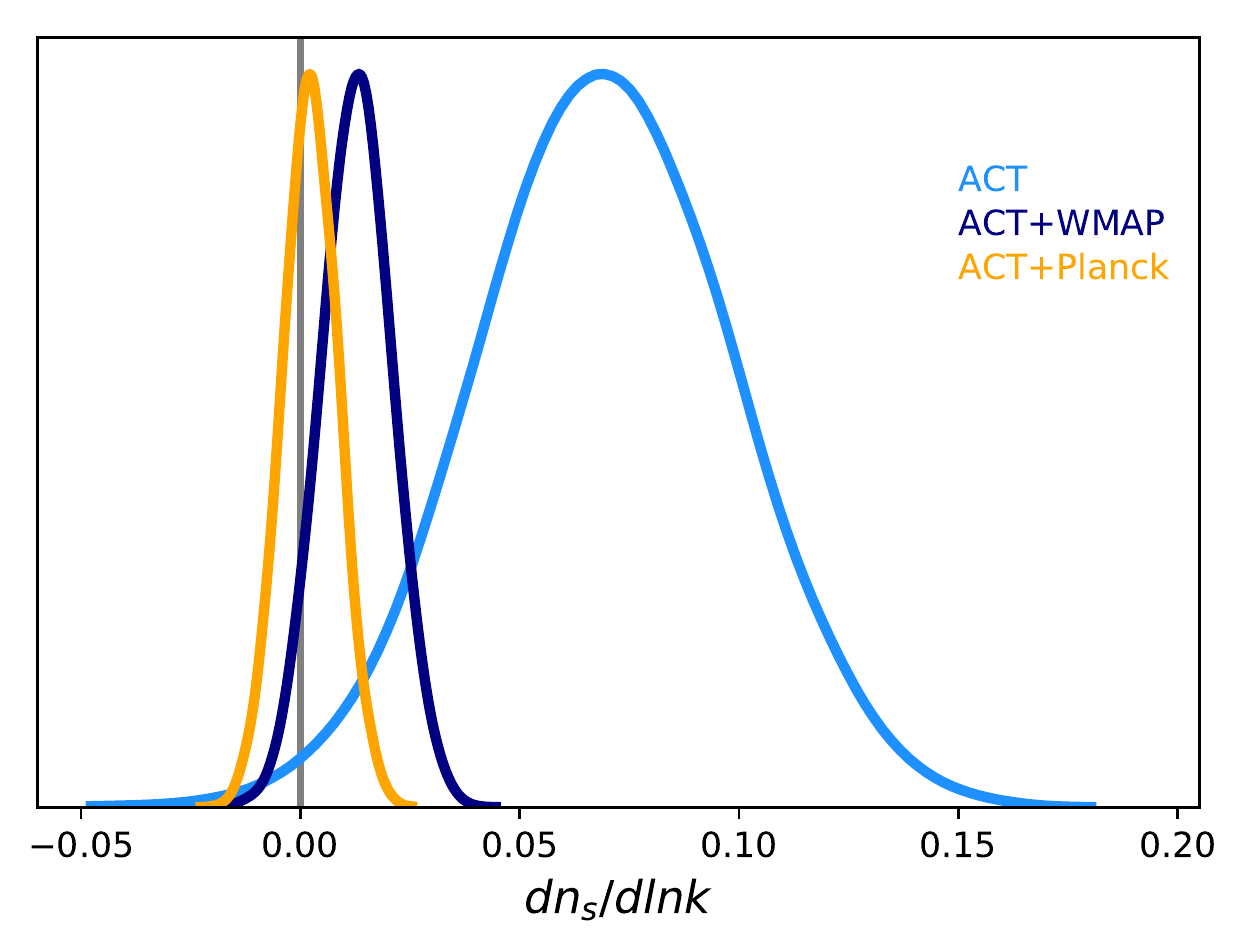}
    \caption{Constraints on the effective number of neutrino species (left) and the running of the spectral index (right). Increasing $N_{\rm eff}$ enhances the damping and shifts the peak positions; the ACT data prefer $N_{\rm eff}$ lower than the standard $3.046$, with less damping, but is still consistent with the standard model. The ACT data also prefer a positive running of the index at $2.4\sigma$, but including large scales from \wmap\ removes a preference for a scale-dependent index. The standard model values are shown with vertical grey lines.\\}
    \label{fig:neff_nrun}
\end{figure*}

Here we investigate the running of the spectral index, the effective number of relativistic species, and the primordial Helium abundance. The high-$n_s$ region allowed by the ACT data tilts the spectrum so that the ACT data by themselves show a mild preference for having less damping in the small-scale power spectrum than in the $\Lambda$CDM model. Since the dominant effect of all three of these extension parameters is to affect the degree of damping, this is reflected in the ACT data alone preferring a number of species $N_{\rm eff}$ that is less than 3.046, a running of the spectral index $d n_s/d lnk$ that is greater than zero at the $\sim$2.4$\sigma$ level, or a primordial Helium abundance, $Y_{\rm HE}$, that is less than the Big Bang Nucleosynthesis prediction of $0.247$~\citep{Pisanti:2007hk,Pitrou:2018cgg}.  

The  constraints are illustrated for $N_{\rm eff}$ in Fig.~\ref{fig:neff_nrun}. $N_{\rm eff}$ has a strong degeneracy with $n_s$ which will be limited if the large-scale data are included. We also find some impact from the degrees of freedom in our foreground modeling and the best-fit model moves depending on whether the MBAC data is used to constrain the foreground parameters. Although not directly correlated to foreground parameters, we find that the $N_{\rm eff}$ constraint is affected by other degeneracies in the multi-dimensional parameter space: by tightly anchoring the foreground parameters, MBAC is contributing to better measuring the amplitude of the spectrum and the calibration/polarization efficiency parameters which are correlated with $N_{\rm eff}$.

For all three primordial models, combining with \wmap\ or \planck\, results in a distribution that moves to a parameter space more consistent with $\Lambda$CDM predictions, at the 1.5--2.2$\sigma$ level for ACT+WMAP. We do not find evidence that any of these extensions significantly improve the goodness of fit of the combined ACT+WMAP dataset.
The data strongly disfavor a model with excess neutrinos, with $N_{\rm eff}=4$ ruled out at 6$\sigma$, and $N_{\rm eff}=3.5$ disfavored at 4$\sigma$. $N_{\rm eff}$ is correlated with the Hubble constant, since an enhanced $N_{\rm eff}$ will increase the expansion rate, reducing the sound horizon. To hold fixed the CMB peak positions one can decrease the angular diameter distance to the CMB by increasing the Hubble constant. Allowing $N_{\rm eff}$ to vary has the effect of lowering both $N_{\rm eff}$ and the inferred Hubble constant along the $N_{\rm eff}-H_0$ degeneracy line compared to the \LCDM\ values, in the opposite direction from the possible tension.

\subsection{Alternative models}
Are any of the tensions in the cosmological data due to a missing component of the model or a new process? A series of forthcoming papers will investigate some more extended models. 

A process that decreases the sound horizon can resolve the tension with the Cepheid-derived Hubble constant. Two examples are introducing a self-interaction between neutrinos that delays free-streaming until shortly before recombination \citep{cyr-racine/sigurdson:2014,Kreisch:2019yzn}, or introducing a period of early dark energy soon before recombination \citep{poulin/etal:2019,Hill:2020osr}. These models have distinct signatures in the small-scale CMB.

We know little about the physics of the dark sector. CMB observations can constrain the cold dark matter interactions with baryons~\citep{Gluscevic:2017ywp}. The signature is at small scales, beyond the resolution probed by \planck. In testing for deviations from fluctuations predicted by inflationary models, we can search not only for a running of the index, but also a model-independent deviation from a power law (e.g.,~\citealp{Hlozek_2012}).

\section{Summary and Outlook}
\label{sec:concl}
ACT has measured the temperature and polarization of almost half the sky at arcminute resolution. The data taken during 2013--2016 have been presented and analysed here as ACT DR4. ACT has over four times the data comprising DR4 from three more seasons of observations (2017--2019) which are in the process of analysis. During the 2020 season we are further expanding the frequency range taking more data with a lower-frequency channel. ACT is now producing maps over large areas of the sky, and we have developed optimal techniques for combining them with the \planck\ maps in order to produce definitive sky maps with an extended angular range compared to \planck\ alone. 

By fitting the $\Lambda$CDM parameters to the ACT DR4 deepest 15\% of the sky we have demonstrated that the standard cosmological model remains consistent with observations. Our measurement of the Hubble constant from ACT combined with \wmap\ agrees with that of \planck. We find no departures from Euclidean geometry and recover the expected amount of gravitational lensing in the power spectrum. The best-fit parameters from ACT-alone differ from those of \planck-alone by 2--3$\sigma$. This could be due to a statistical fluctuation, unknown systematic effects in our data or something new (noting that ACT and \planck\ measurements cover significantly different ranges of angular scales), and will be further explored with data already in-hand from ACT observations between 2017 and 2019. 
As we analyze more data, our cosmological constraints from polarization will improve to the point where they become competitive with, and more powerful than, temperature, giving additional consistency checks on the standard cosmological model.
Looking forward, ACT will continue to observe half the sky through 2021.

\section{\label{sec:ack}Acknowledgments}
This work was supported by the U.S. National Science Foundation through awards AST-0408698, AST-0965625, and AST-1440226 for the ACT project, as well as awards PHY-0355328, PHY-0855887 and PHY-1214379. Funding was also provided by Princeton University, the University of Pennsylvania, and a Canada Foundation for Innovation (CFI) award to UBC. ACT operates in the Parque Astron\'omico Atacama in northern Chile under the auspices of the Comisi\'on Nacional de Investigaci\'on (CONICYT). 
Computations were performed on the Niagara supercomputer at the SciNet HPC Consortium and on the Simons-Popeye cluster of the Flatiron Institute. SciNet is funded by the CFI under the auspices of Compute Canada, the Government of Ontario, the Ontario Research Fund---Research Excellence, and the University of Toronto.
Cosmological analyses were performed on the Hawk high-performance computing cluster at the Advanced Research Computing at Cardiff (ARCCA). We would like to thank the Scientific Computing Core (SCC) team at the Flatiron Institute, especially Nick Carriero, for their support. Flatiron Institute is supported by the Simons Foundation. Additional computations were performed on Hippo at the University of KwaZulu-Natal, on Tiger as part of Princeton Research Computing resources at Princeton University, on Feynman at Princeton University, and on Cori at NERSC. The development of multichroic detectors and lenses was supported by NASA grants NNX13AE56G and NNX14AB58G. Detector research at NIST was supported by the NIST Innovations in Measurement Science program. We thank Bert Harrop for his extensive efforts on the assembly of the detector arrays. The shops at Penn and Princeton have time and again built beautiful instrumentation on which ACT depends. We also thank Toby Marriage for numerous contributions.

SKC acknowledges support from the Cornell Presidential Postdoctoral Fellowship. RD thanks CONICYT for grant BASAL CATA AFB-170002. ZL, ES and JD are supported through NSF grant AST-1814971. KM and MHi acknowledge support from the National Research Foundation of South Africa. MDN acknowledges support from NSF award AST-1454881.
DH, AM, and NS acknowledge support from NSF grant numbers AST-1513618 and AST-1907657. EC acknowledges support from the STFC Ernest Rutherford Fellowship ST/M004856/2 and STFC Consolidated Grant ST/S00033X/1, and from the Horizon 2020 ERC Starting Grant (Grant agreement No 849169). NB acknowledges support from NSF grant AST-1910021. ML was supported by a Dicke Fellowship. LP gratefully acknowledges support from the Mishrahi and Wilkinson funds. RH acknowledges support as an Azrieli Global Scholar in CIfAR’s Gravity \& the Extreme Universe Program and as an Alfred. P. Sloan Research Fellow. RH is also supported by Canada’s NSERC Discovery Grants program and the Dunlap Institute, which was established with an endowment by the David Dunlap family and the University of Toronto.
We thank our many colleagues from ALMA, APEX, CLASS, and Polarbear/Simons Array who have helped us at critical junctures. Colleagues at AstroNorte and RadioSky provide logistical support and keep operations in Chile running smoothly.

Lastly, we gratefully acknowledge the many publicly available software packages that were essential for parts of this analysis. They include \texttt{CosmoMC} \citep{Lewis:2013hha,Lewis:2002ah}, \texttt{CAMB}~\citep{CAMB}, 
\texttt{healpy}~\citep{Healpix1}, \texttt{HEALPix}~\citep{Healpix2}, the \texttt{SLATEC}\footnote{http://www.netlib.org/slatec/guide} Fortran subroutine DRC3JJ.F9,  the  \texttt{SOFA} library \citep{SOFA:2019-07-22}, \texttt{libsharp}~\citep{reinecke/2013}, and 
\texttt{pixell}\footnote{https://github.com/simonsobs/pixell}. This research made use of \texttt{Astropy}\footnote{http://www.astropy.org}, a community-developed core Python package for Astronomy \citep{astropy:2013, astropy:2018}. We also acknowledge use of the \texttt{matplotlib}~\citep{Hunter:2007} package and the Python Image Library for producing plots in this paper.

\bibliographystyle{yahapj}
\bibliography{01_main,Planck_bib,choi_etal_bib}

\begin{thebibliography}{}
\providecommand\natexlab[1]{#1}
\providecommand\JournalTitle[1]{#1}

\bibitem[{Abbott {et~al.}(2018{\natexlab{a}})}]{Abbott:2017smn}
Abbott, T. M.~C., {et~al.} 2018{\natexlab{a}},
  \href{http://dx.doi.org/10.1093/mnras/sty1939}{\JournalTitle{Mon. Not. Roy.
  Astron. Soc.}, 480, 3879}

\bibitem[{Abbott {et~al.}(2018{\natexlab{b}})}]{Abbott:2017wau}
---. 2018{\natexlab{b}},
  \href{http://dx.doi.org/10.1103/PhysRevD.98.043526}{\JournalTitle{Phys.
  Rev.}, D98, 043526}

\bibitem[{{Abbott} {et~al.}(2018){Abbott}, {Abdalla}, {Allam}, {Amara},
  {Annis}, {Asorey}, {Avila}, {Ballester}, {Banerji}, {Barkhouse}, {Baruah},
  {Baumer}, {Bechtol}, {Becker}, {Benoit-L{\'e}vy}, {Bernstein}, {Bertin},
  {Blazek}, {Bocquet}, {Brooks}, {Brout}, {Buckley-Geer}, {Burke}, {Busti},
  {Campisano}, {Cardiel-Sas}, {Carnero Rosell}, {Carrasco Kind}, {Carretero},
  {Castander}, {Cawthon}, {Chang}, {Chen}, {Conselice}, {Costa}, {Crocce},
  {Cunha}, {D'Andrea}, {da Costa}, {Das}, {Daues}, {Davis}, {Davis}, {De
  Vicente}, {DePoy}, {DeRose}, {Desai}, {Diehl}, {Dietrich}, {Dodelson},
  {Doel}, {Drlica-Wagner}, {Eifler}, {Elliott}, {Evrard}, {Farahi}, {Fausti
  Neto}, {Fernandez}, {Finley}, {Flaugher}, {Foley}, {Fosalba}, {Friedel},
  {Frieman}, {Garc{\'\i}a-Bellido}, {Gaztanaga}, {Gerdes}, {Giannantonio},
  {Gill}, {Glazebrook}, {Goldstein}, {Gower}, {Gruen}, {Gruendl}, {Gschwend},
  {Gupta}, {Gutierrez}, {Hamilton}, {Hartley}, {Hinton}, {Hislop}, {Hollowood},
  {Honscheid}, {Hoyle}, {Huterer}, {Jain}, {James}, {Jeltema}, {Johnson},
  {Johnson}, {Kacprzak}, {Kent}, {Khullar}, {Klein}, {Kovacs}, {Koziol},
  {Krause}, {Kremin}, {Kron}, {Kuehn}, {Kuhlmann}, {Kuropatkin}, {Lahav},
  {Lasker}, {Li}, {Li}, {Liddle}, {Lima}, {Lin}, {L{\'o}pez-Reyes}, {MacCrann},
  {Maia}, {Maloney}, {Manera}, {March}, {Marriner}, {Marshall}, {Martini},
  {McClintock}, {McKay}, {McMahon}, {Melchior}, {Menanteau}, {Miller},
  {Miquel}, {Mohr}, {Morganson}, {Mould}, {Neilsen}, {Nichol}, {Nogueira},
  {Nord}, {Nugent}, {Nunes}, {Ogand o}, {Old}, {Pace}, {Palmese},
  {Paz-Chinch{\'o}n}, {Peiris}, {Percival}, {Petravick}, {Plazas}, {Poh},
  {Pond}, {Porredon}, {Pujol}, {Refregier}, {Reil}, {Ricker}, {Rollins},
  {Romer}, {Roodman}, {Rooney}, {Ross}, {Rykoff}, {Sako}, {Sanchez}, {Sanchez},
  {Santiago}, {Saro}, {Scarpine}, {Scolnic}, {Serrano}, {Sevilla-Noarbe},
  {Sheldon}, {Shipp}, {Silveira}, {Smith}, {Smith}, {Smith}, {Soares-Santos},
  {Sobreira}, {Song}, {Stebbins}, {Suchyta}, {Sullivan}, {Swanson}, {Tarle},
  {Thaler}, {Thomas}, {Thomas}, {Troxel}, {Tucker}, {Vikram}, {Vivas},
  {Walker}, {Wechsler}, {Weller}, {Wester}, {Wolf}, {Wu}, {Yanny}, {Zenteno},
  {Zhang}, {Zuntz}, {DES Collaboration}, {Juneau}, {Fitzpatrick}, {Nikutta},
  {Nidever}, {Olsen}, {Scott}, \& {NOAO Data Lab}}]{des_2018}
{Abbott}, T.~M.~C., {Abdalla}, F.~B., {Allam}, S., {et~al.} 2018,
  \href{http://dx.doi.org/10.3847/1538-4365/aae9f0}{\JournalTitle{\apjs}, 239,
  18}

\bibitem[{Addison {et~al.}(2016)Addison, Huang, Watts, Bennett, Halpern,
  Hinshaw, \& Weiland}]{Addison:2015wyg}
Addison, G., Huang, Y., Watts, D., {et~al.} 2016,
  \href{http://dx.doi.org/10.3847/0004-637X/818/2/132}{\JournalTitle{Astrophys.
  J.}, 818, 132}

\bibitem[{Addison {et~al.}(2018)Addison, Watts, Bennett, Halpern, Hinshaw, \&
  Weiland}]{Addison:2017fdm}
Addison, G., Watts, D., Bennett, C., {et~al.} 2018,
  \href{http://dx.doi.org/10.3847/1538-4357/aaa1ed}{\JournalTitle{Astrophys.
  J.}, 853, 119}

\bibitem[{{Aihara} {et~al.}(2019){Aihara}, {AlSayyad}, {Ando}, {Armstrong},
  {Bosch}, {Egami}, {Furusawa}, {Furusawa}, {Goulding}, {Harikane}, {Hikage},
  {Ho}, {Hsieh}, {Huang}, {Ikeda}, {Imanishi}, {Ito}, {Iwata}, {Jaelani},
  {Kakuma}, {Kawana}, {Kikuta}, {Kobayashi}, {Koike}, {Komiyama}, {Li},
  {Liang}, {Lin}, {Luo}, {Lupton}, {Lust}, {MacArthur}, {Matsuoka}, {Mineo},
  {Miyatake}, {Miyazaki}, {More}, {Murata}, {Namiki}, {Nishizawa}, {Oguri},
  {Okabe}, {Okamoto}, {Okura}, {Ono}, {Onodera}, {Onoue}, {Osato}, {Ouchi},
  {Shibuya}, {Strauss}, {Sugiyama}, {Suto}, {Takada}, {Takagi}, {Takata},
  {Takita}, {Tanaka}, {Terai}, {Toba}, {Uchiyama}, {Utsumi}, {Wang}, {Wang}, \&
  {Yamada}}]{hsc_2019}
{Aihara}, H., {AlSayyad}, Y., {Ando}, M., {et~al.} 2019,
  \href{http://dx.doi.org/10.1093/pasj/psz103}{\JournalTitle{\pasj}, 71, 114}

\bibitem[{{Alam} {et~al.}(2017){Alam}, {Ata}, {Bailey}, {Beutler}, {Bizyaev},
  {Blazek}, {Bolton}, {Brownstein}, {Burden}, {Chuang}, {Comparat}, {Cuesta},
  {Dawson}, {Eisenstein}, {Escoffier}, {Gil-Mar{\'\i}n}, {Grieb}, {Hand}, {Ho},
  {Kinemuchi}, {Kirkby}, {Kitaura}, {Malanushenko}, {Malanushenko}, {Maraston},
  {McBride}, {Nichol}, {Olmstead}, {Oravetz}, {Padmanabhan},
  {Palanque-Delabrouille}, {Pan}, {Pellejero-Ibanez}, {Percival}, {Petitjean},
  {Prada}, {Price-Whelan}, {Reid}, {Rodr{\'\i}guez-Torres}, {Roe}, {Ross},
  {Ross}, {Rossi}, {Rubi{\~n}o-Mart{\'\i}n}, {Saito}, {Salazar-Albornoz},
  {Samushia}, {S{\'a}nchez}, {Satpathy}, {Schlegel}, {Schneider},
  {Sc{\'o}ccola}, {Seo}, {Sheldon}, {Simmons}, {Slosar}, {Strauss}, {Swanson},
  {Thomas}, {Tinker}, {Tojeiro}, {Maga{\~n}a}, {Vazquez}, {Verde}, {Wake},
  {Wang}, {Weinberg}, {White}, {Wood-Vasey}, {Y{\`e}che}, {Zehavi}, {Zhai}, \&
  {Zhao}}]{alam/etal:2017}
{Alam}, S., {Ata}, M., {Bailey}, S., {et~al.} 2017,
  \href{http://dx.doi.org/10.1093/mnras/stx721}{\JournalTitle{\mnras}, 470,
  2617}

\bibitem[{{Albareti} {et~al.}(2017){Albareti}, {Allende Prieto}, {Almeida},
  {Anders}, {Anderson}, {Andrews}, {Arag{\'o}n-Salamanca},
  {Argudo-Fern{\'a}ndez}, {Armengaud}, {Aubourg}, {Avila-Reese}, {Badenes},
  {Bailey}, {Barbuy}, {Barger}, {Barrera-Ballesteros}, {Bartosz}, {Basu},
  {Bates}, {Battaglia}, {Baumgarten}, {Baur}, {Bautista}, {Beers}, {Belfiore},
  {Bershady}, {Bertran de Lis}, {Bird}, {Bizyaev}, {Blanc}, {Blanton},
  {Blomqvist}, {Bolton}, {Borissova}, {Bovy}, {Brand t}, {Brinkmann},
  {Brownstein}, {Bundy}, {Burtin}, {Busca}, {Orlando Camacho Chavez}, {Cano
  D{\'\i}az}, {Cappellari}, {Carrera}, {Chen}, {Cherinka}, {Cheung},
  {Chiappini}, {Chojnowski}, {Chuang}, {Chung}, {Cirolini}, {Clerc}, {Cohen},
  {Comerford}, {Comparat}, {Correa do Nascimento}, {Cousinou}, {Covey},
  {Crane}, {Croft}, {Cunha}, {Darling}, {Davidson}, {Dawson}, {Da Costa}, {Da
  Silva Ilha}, {Deconto Machado}, {Delubac}, {De Lee}, {De la Macorra}, {De la
  Torre}, {Diamond-Stanic}, {Donor}, {Downes}, {Drory}, {Du}, {Du Mas des
  Bourboux}, {Dwelly}, {Ebelke}, {Eigenbrot}, {Eisenstein}, {Elsworth},
  {Emsellem}, {Eracleous}, {Escoffier}, {Evans}, {Falc{\'o}n-Barroso}, {Fan},
  {Favole}, {Fernandez-Alvar}, {Fernand ez-Trincado}, {Feuillet}, {Fleming},
  {Font-Ribera}, {Freischlad}, {Frinchaboy}, {Fu}, {Gao}, {Garcia},
  {Garcia-Dias}, {Garcia-Hern{\'a}ndez}, {Garcia P{\'e}rez}, {Gaulme}, {Ge},
  {Geisler}, {Gillespie}, {Gil Marin}, {Girardi}, {Goddard}, {Gomez Maqueo
  Chew}, {Gonzalez-Perez}, {Grabowski}, {Green}, {Grier}, {Grier}, {Guo},
  {Guy}, {Hagen}, {Hall}, {Harding}, {Harley}, {Hasselquist}, {Hawley},
  {Hayes}, {Hearty}, {Hekker}, {Hernandez Toledo}, {Ho}, {Hogg},
  {Holley-Bockelmann}, {Holtzman}, {Holzer}, {Hu}, {Huber}, {Hutchinson},
  {Hwang}, {Ibarra-Medel}, {Ivans}, {Ivory}, {Jaehnig}, {Jensen}, {Johnson},
  {Jones}, {Jullo}, {Kallinger}, {Kinemuchi}, {Kirkby}, {Klaene}, {Kneib},
  {Kollmeier}, {Lacerna}, {Lane}, {Lang}, {Laurent}, {Law}, {Leauthaud}, {Le
  Goff}, {Li}, {Li}, {Li}, {Li}, {Liang}, {Liang}, {Lima}, {Lin}, {Lin}, {Lin},
  {Liu}, {Long}, {Lucatello}, {MacDonald}, {MacLeod}, {Mackereth}, {Mahadevan},
  {Maia}, {Maiolino}, {Majewski}, {Malanushenko}, {Malanushenko}, {Mallmann},
  {Manchado}, {Maraston}, {Marques-Chaves}, {Martinez Valpuesta}, {Masters},
  {Mathur}, {McGreer}, {Merloni}, {Merrifield}, {M{\'e}sz{\'a}ros}, {Meza},
  {Miglio}, {Minchev}, {Molaverdikhani}, {Montero-Dorta}, {Mosser}, {Muna},
  {Myers}, {Nair}, {Nandra}, {Ness}, {Newman}, {Nichol}, {Nidever},
  {Nitschelm}, {O'Connell}, {Oravetz}, {Oravetz}, {Pace}, {Padilla},
  {Palanque-Delabrouille}, {Pan}, {Parejko}, {Paris}, {Park}, {Peacock},
  {Peirani}, {Pellejero-Ibanez}, {Penny}, {Percival}, {Percival},
  {Perez-Fournon}, {Petitjean}, {Pieri}, {Pinsonneault}, {Pisani}, {Prada},
  {Prakash}, {Price-Jones}, {Raddick}, {Rahman}, {Raichoor}, {Barboza Rembold},
  {Reyna}, {Rich}, {Richstein}, {Ridl}, {Riffel}, {Riffel}, {Rix}, {Robin},
  {Rockosi}, {Rodr{\'\i}guez-Torres}, {Rodrigues}, {Roe}, {Roman Lopes},
  {Rom{\'a}n-Z{\'u}{\~n}iga}, {Ross}, {Rossi}, {Ruan}, {Ruggeri}, {Runnoe},
  {Salazar-Albornoz}, {Salvato}, {Sanchez}, {Sanchez}, {Sanchez-Gallego},
  {Santiago}, {Schiavon}, {Schimoia}, {Schlafly}, {Schlegel}, {Schneider},
  {Sch{\"o}nrich}, {Schultheis}, {Schwope}, {Seo}, {Serenelli}, {Sesar},
  {Shao}, {Shetrone}, {Shull}, {Silva Aguirre}, {Skrutskie}, {Slosar}, {Smith},
  {Smith}, {Sobeck}, {Somers}, {Souto}, {Stark}, {Stassun}, {Steinmetz},
  {Stello}, {Storchi Bergmann}, {Strauss}, {Streblyanska}, {Stringfellow},
  {Suarez}, {Sun}, {Taghizadeh-Popp}, {Tang}, {Tao}, {Tayar}, {Tembe},
  {Thomas}, {Tinker}, {Tojeiro}, {Tremonti}, {Troup}, {Trump}, {Unda-Sanzana},
  {Valenzuela}, {Van den Bosch}, {Vargas-Maga{\~n}a}, {Vazquez}, {Villanova},
  {Vivek}, {Vogt}, {Wake}, {Walterbos}, {Wang}, {Wang}, {Weaver}, {Weijmans},
  {Weinberg}, {Westfall}, {Whelan}, {Wilcots}, {Wild}, {Williams}, {Wilson},
  {Wood-Vasey}, {Wylezalek}, {Xiao}, {Yan}, {Yang}, {Ybarra}, {Yeche}, {Yuan},
  {Zakamska}, {Zamora}, {Zasowski}, {Zhang}, {Zhao}, {Zhao}, {Zheng}, {Zheng},
  {Zhou}, {Zhu}, {Zinn}, \& {Zou}}]{boss_dr13:2017}
{Albareti}, F.~D., {Allende Prieto}, C., {Almeida}, A., {et~al.} 2017,
  \href{http://dx.doi.org/10.3847/1538-4365/aa8992}{\JournalTitle{\apjs}, 233,
  25}

\bibitem[{{Astropy Collaboration} {et~al.}(2013){Astropy Collaboration},
  {Robitaille}, {Tollerud}, {Greenfield}, {Droettboom}, {Bray}, {Aldcroft},
  {Davis}, {Ginsburg}, {Price-Whelan}, {Kerzendorf}, {Conley}, {Crighton},
  {Barbary}, {Muna}, {Ferguson}, {Grollier}, {Parikh}, {Nair}, {Unther},
  {Deil}, {Woillez}, {Conseil}, {Kramer}, {Turner}, {Singer}, {Fox}, {Weaver},
  {Zabalza}, {Edwards}, {Azalee Bostroem}, {Burke}, {Casey}, {Crawford},
  {Dencheva}, {Ely}, {Jenness}, {Labrie}, {Lim}, {Pierfederici}, {Pontzen},
  {Ptak}, {Refsdal}, {Servillat}, \& {Streicher}}]{astropy:2013}
{Astropy Collaboration}, {Robitaille}, T.~P., {Tollerud}, E.~J., {et~al.} 2013,
  \href{http://dx.doi.org/10.1051/0004-6361/201322068}{\JournalTitle{\aap},
  558, A33}

\bibitem[{Aubourg {et~al.}(2015)}]{Aubourg:2014yra}
Aubourg, E., {et~al.} 2015,
  \href{http://dx.doi.org/10.1103/PhysRevD.92.123516}{\JournalTitle{Phys. Rev.
  D}, 92, 123516}

\bibitem[{{Bennett} {et~al.}(2013){Bennett}, {Larson}, {Weiland}, {Jarosik},
  {Hinshaw}, {Odegard}, {Smith}, {Hill}, {Gold}, {Halpern}, {Komatsu}, {Nolta},
  {Page}, {Spergel}, {Wollack}, {Dunkley}, {Kogut}, {Limon}, {Meyer}, {Tucker},
  \& {Wright}}]{bennett/etal:2013}
{Bennett}, C.~L., {Larson}, D., {Weiland}, J.~L., {et~al.} 2013,
  \href{http://dx.doi.org/10.1088/0067-0049/208/2/20}{\JournalTitle{\apjs},
  208, 20}

\bibitem[{Beutler {et~al.}(2011)Beutler, Blake, Colless, Jones, Staveley-Smith,
  Campbell, Parker, Saunders, \& Watson}]{Beutler_2011}
Beutler, F., Blake, C., Colless, M., {et~al.} 2011,
  \href{http://dx.doi.org/10.1111/j.1365-2966.2011.19250.x}{\JournalTitle{Monthly
  Notices of the Royal Astronomical Society}, 416, 3017–3032}

\bibitem[{Birrer {et~al.}(2020)}]{Birrer:2020tax}
Birrer, S., {et~al.} 2020, \href{http://arxiv.org/abs/2007.02941}{{\sffamily
  arXiv:2007.02941 [astro-ph.CO]}}

\bibitem[{Calabrese {et~al.}(2008)Calabrese, Slosar, Melchiorri, Smoot, \&
  Zahn}]{Calabrese:2008}
Calabrese, E., Slosar, A., Melchiorri, A., Smoot, G.~F., \& Zahn, O. 2008,
  \href{http://dx.doi.org/10.1103/PhysRevD.77.123531}{\JournalTitle{Phys. Rev.
  D}, 77, 123531}

\bibitem[{Calabrese {et~al.}(2017)}]{Calabrese:2017ypx}
Calabrese, E., {et~al.} 2017,
  \href{http://dx.doi.org/10.1103/PhysRevD.95.063525}{\JournalTitle{Phys. Rev.
  D}, 95, 063525}

\bibitem[{Choi {et~al.}(2020)}]{choi_atacama_2020}
Choi, S.~K., {et~al.} 2020, \href{http://arxiv.org/abs/2007.07289}{{\sffamily
  arXiv:2007.07289 [astro-ph.CO]}}

\bibitem[{{Cooke} {et~al.}(2016){Cooke}, {Pettini}, {Nollett}, \&
  {Jorgenson}}]{2016ApJ...830..148C}
{Cooke}, R.~J., {Pettini}, M., {Nollett}, K.~M., \& {Jorgenson}, R. 2016,
  \href{http://dx.doi.org/10.3847/0004-637X/830/2/148}{\JournalTitle{\apj},
  830, 148}

\bibitem[{Cuceu {et~al.}(2019)Cuceu, Farr, Lemos, \&
  Font-Ribera}]{Cuceu:2019for}
Cuceu, A., Farr, J., Lemos, P., \& Font-Ribera, A. 2019,
  \href{http://dx.doi.org/10.1088/1475-7516/2019/10/044}{\JournalTitle{JCAP},
  10, 044}

\bibitem[{{Cyr-Racine} \& {Sigurdson}(2014)}]{cyr-racine/sigurdson:2014}
{Cyr-Racine}, F.-Y., \& {Sigurdson}, K. 2014,
  \href{http://dx.doi.org/10.1103/PhysRevD.90.123533}{\JournalTitle{\prd}, 90,
  123533}

\bibitem[{{Darwish} {et~al.}(2020){Darwish}, {Madhavacheril}, {Sherwin},
  {Aiola}, {Battaglia}, {Beall}, {Becker}, {Bond}, {Calabrese}, {Choi},
  {Devlin}, {Dunkley}, {D{\"u}nner}, {Ferraro}, {Fox}, {Gallardo}, {Guan},
  {Halpern}, {Han}, {Hasselfield}, {Hill}, {Hilton}, {Hilton}, {Hincks}, {Ho},
  {Hubmayr}, {Hughes}, {Koopman}, {Kosowsky}, {Van Lanen}, {Louis}, {Lungu},
  {MacInnis}, {Maurin}, {McMahon}, {Moodley}, {Naess}, {Namikawa}, {Newburgh},
  {Nibarger}, {Niemack}, {Page}, {Partridge}, {Qu}, {Robertson}, {Schmitt},
  {Sehgal}, {Sif{\'o}n}, {Spergel}, {Staggs}, {Storer}, {van Engelen}, \&
  {Wollack}}]{darwish_atacama_2020}
{Darwish}, O., {Madhavacheril}, M.~S., {Sherwin}, B., {et~al.} 2020,
  \JournalTitle{arXiv e-prints},
  \href{http://arxiv.org/abs/2004.01139}{{\sffamily arXiv:2004.01139}}

\bibitem[{Das {et~al.}(2014)}]{Das:2013zf}
Das, S., {et~al.} 2014,
  \href{http://dx.doi.org/10.1088/1475-7516/2014/04/014}{\JournalTitle{JCAP},
  04, 014}

\bibitem[{{Datta} {et~al.}(2019){Datta}, {Aiola}, {Choi}, {Devlin}, {Dunkley},
  {D{\"u}nner}, {Gallardo}, {Gralla}, {Halpern}, {Hasselfield}, {Hilton},
  {Hincks}, {Ho}, {Hubmayr}, {Huffenberger}, {Hughes}, {Kosowsky},
  {L{\'o}pez-Caraballo}, {Louis}, {Lungu}, {Marriage}, {Maurin}, {McMahon},
  {Moodley}, {Naess}, {Nati}, {Niemack}, {Page}, {Partridge}, {Prince},
  {Staggs}, {Switzer}, {Wollack}, \& {Farren}}]{actpol_datta_2018}
{Datta}, R., {Aiola}, S., {Choi}, S.~K., {et~al.} 2019,
  \href{http://dx.doi.org/10.1093/mnras/sty2934}{\JournalTitle{\mnras}, 486,
  5239}

\bibitem[{{De Bernardis} {et~al.}(2016){De Bernardis}, {Stevens},
  {Hasselfield}, {Alonso}, {Bond}, {Calabrese}, {Choi}, {Crowley}, {Devlin},
  {Dunkley}, {Gallardo}, {Henderson}, {Hilton}, {Hlozek}, {Ho}, {Huffenberger},
  {Koopman}, {Kosowsky}, {Louis}, {Madhavacheril}, {McMahon}, {N{\ae}ss},
  {Nati}, {Newburgh}, {Niemack}, {Page}, {Salatino}, {Schillaci}, {Schmitt},
  {Sehgal}, {Sievers}, {Simon}, {Spergel}, {Staggs}, {van Engelen},
  {Vavagiakis}, \& {Wollack}}]{debernardis/2016}
{De Bernardis}, F., {Stevens}, J.~R., {Hasselfield}, M., {et~al.} 2016,
  \href{http://dx.doi.org/10.1117/12.2232824}{in \procspie, Vol. 9910,
  Observatory Operations: Strategies, Processes, and Systems VI}, 991014

\bibitem[{{de Jong, Jelte T. A.} {et~al.}(2017){de Jong, Jelte T. A.}, {Kleijn,
  Gijs A. Verdoes}, {Erben, Thomas}, {Hildebrandt, Hendrik}, {Kuijken, Konrad},
  {Sikkema, Gert}, {Brescia, Massimo}, {Bilicki, Maciej}, {Napolitano, Nicola
  R.}, {Amaro, Valeria}, {Begeman, Kor G.}, {Boxhoorn, Danny R.},
  {Buddelmeijer, Hugo}, {Cavuoti, Stefano}, {Getman, Fedor}, {Grado, Aniello},
  {Helmich, Ewout}, {Huang, Zhuoyi}, {Irisarri, Nancy}, {La Barbera,
  Francesco}, {Longo, Giuseppe}, {McFarland, John P.}, {Nakajima, Reiko},
  {Paolillo, Maurizio}, {Puddu, Emanuella}, {Radovich, Mario}, {Rifatto,
  Agatino}, {Tortora, Crescenzo}, {Valentijn, Edwin A.}, {Vellucci, Civita},
  {Vriend, Willem-Jan}, {Amon, Alexandra}, {Blake, Chris}, {Choi, Ami}, {Conti,
  Ian Fenech}, {Gwyn, Stephen D. J.}, {Herbonnet, Ricardo}, {Heymans,
  Catherine}, {Hoekstra, Henk}, {Klaes, Dominik}, {Merten, Julian}, {Miller,
  Lance}, {Schneider, Peter}, \& {Viola, Massimo}}]{kids_2017}
{de Jong, Jelte T. A.}, {Kleijn, Gijs A. Verdoes}, {Erben, Thomas}, {et~al.}
  2017,
  \href{http://dx.doi.org/10.1051/0004-6361/201730747}{\JournalTitle{A\&A},
  604, A134}

\bibitem[{{Dey} {et~al.}(2019){Dey}, {Schlegel}, {Lang}, {Blum}, {Burleigh},
  {Fan}, {Findlay}, {Finkbeiner}, {Herrera}, {Juneau}, {Landriau}, {Levi},
  {McGreer}, {Meisner}, {Myers}, {Moustakas}, {Nugent}, {Patej}, {Schlafly},
  {Walker}, {Valdes}, {Weaver}, {Y{\`e}che}, {Zou}, {Zhou}, {Abareshi},
  {Abbott}, {Abolfathi}, {Aguilera}, {Alam}, {Allen}, {Alvarez}, {Annis},
  {Ansarinejad}, {Aubert}, {Beechert}, {Bell}, {BenZvi}, {Beutler}, {Bielby},
  {Bolton}, {Brice{\~n}o}, {Buckley-Geer}, {Butler}, {Calamida}, {Carlberg},
  {Carter}, {Casas}, {Castander}, {Choi}, {Comparat}, {Cukanovaite}, {Delubac},
  {DeVries}, {Dey}, {Dhungana}, {Dickinson}, {Ding}, {Donaldson}, {Duan},
  {Duckworth}, {Eftekharzadeh}, {Eisenstein}, {Etourneau}, {Fagrelius},
  {Farihi}, {Fitzpatrick}, {Font-Ribera}, {Fulmer}, {G{\"a}nsicke},
  {Gaztanaga}, {George}, {Gerdes}, {Gontcho}, {Gorgoni}, {Green}, {Guy},
  {Harmer}, {Hernand ez}, {Honscheid}, {Huang}, {James}, {Jannuzi}, {Jiang},
  {Joyce}, {Karcher}, {Karkar}, {Kehoe}, {Kneib}, {Kueter-Young}, {Lan},
  {Lauer}, {Le Guillou}, {Le Van Suu}, {Lee}, {Lesser}, {Perreault Levasseur},
  {Li}, {Mann}, {Marshall}, {Mart{\'\i}nez-V{\'a}zquez}, {Martini}, {du Mas des
  Bourboux}, {McManus}, {Meier}, {M{\'e}nard}, {Metcalfe},
  {Mu{\~n}oz-Guti{\'e}rrez}, {Najita}, {Napier}, {Narayan}, {Newman}, {Nie},
  {Nord}, {Norman}, {Olsen}, {Paat}, {Palanque-Delabrouille}, {Peng},
  {Poppett}, {Poremba}, {Prakash}, {Rabinowitz}, {Raichoor}, {Rezaie},
  {Robertson}, {Roe}, {Ross}, {Ross}, {Rudnick}, {Safonova}, {Saha},
  {S{\'a}nchez}, {Savary}, {Schweiker}, {Scott}, {Seo}, {Shan}, {Silva},
  {Slepian}, {Soto}, {Sprayberry}, {Staten}, {Stillman}, {Stupak}, {Summers},
  {Sien Tie}, {Tirado}, {Vargas-Maga{\~n}a}, {Vivas}, {Wechsler}, {Williams},
  {Yang}, {Yang}, {Yapici}, {Zaritsky}, {Zenteno}, {Zhang}, {Zhang}, {Zhou}, \&
  {Zhou}}]{desi_2018}
{Dey}, A., {Schlegel}, D.~J., {Lang}, D., {et~al.} 2019,
  \href{http://dx.doi.org/10.3847/1538-3881/ab089d}{\JournalTitle{\aj}, 157,
  168}

\bibitem[{{Di Valentino} {et~al.}(2019){Di Valentino}, {Melchiorri}, \&
  {Silk}}]{DiValentino:2019}
{Di Valentino}, E., {Melchiorri}, A., \& {Silk}, J. 2019,
  \href{http://dx.doi.org/10.1038/s41550-019-0906-9}{\JournalTitle{Nature
  Astronomy}, 484}

\bibitem[{{D{\"u}nner} {et~al.}(2013){D{\"u}nner}, {Hasselfield}, {Marriage},
  {Sievers}, {Acquaviva}, {Addison}, {Ade}, {Aguirre}, {Amiri}, {Appel},
  {Barrientos}, {Battistelli}, {Bond}, {Brown}, {Burger}, {Calabrese},
  {Chervenak}, {Das}, {Devlin}, {Dicker}, {Bertrand Doriese}, {Dunkley},
  {Essinger-Hileman}, {Fisher}, {Gralla}, {Fowler}, {Hajian}, {Halpern},
  {Hern{\'a}ndez-Monteagudo}, {Hilton}, {Hilton}, {Hincks}, {Hlozek},
  {Huffenberger}, {Hughes}, {Hughes}, {Infante}, {Irwin}, {Baptiste Juin},
  {Kaul}, {Klein}, {Kosowsky}, {Lau}, {Limon}, {Lin}, {Louis}, {Lupton},
  {Marsden}, {Martocci}, {Mauskopf}, {Menanteau}, {Moodley}, {Moseley},
  {Netterfield}, {Niemack}, {Nolta}, {Page}, {Parker}, {Partridge}, {Quintana},
  {Reid}, {Sehgal}, {Sherwin}, {Spergel}, {Staggs}, {Swetz}, {Switzer},
  {Thornton}, {Trac}, {Tucker}, {Warne}, {Wilson}, {Wollack}, \&
  {Zhao}}]{dunner_atacama_2013}
{D{\"u}nner}, R., {Hasselfield}, M., {Marriage}, T.~A., {et~al.} 2013,
  \href{http://dx.doi.org/10.1088/0004-637X/762/1/10}{\JournalTitle{\apj}, 762,
  10}

\bibitem[{Efstathiou \& Gratton(2019)}]{Efstathiou:2019mdh}
Efstathiou, G., \& Gratton, S. 2019,
  \href{http://arxiv.org/abs/1910.00483}{{\sffamily arXiv:1910.00483
  [astro-ph.CO]}}

\bibitem[{Efstathiou \& Gratton(2020)}]{Efstathiou_2020}
---. 2020,
  \href{http://dx.doi.org/10.1093/mnrasl/slaa093}{\JournalTitle{Monthly Notices
  of the Royal Astronomical Society: Letters}, 496, L91–L95}

\bibitem[{{Freedman} {et~al.}(2019){Freedman}, {Madore}, {Hatt}, {Hoyt},
  {Jang}, {Beaton}, {Burns}, {Lee}, {Monson}, {Neeley}, {Phillips}, {Rich}, \&
  {Seibert}}]{freedman/etal:2019}
{Freedman}, W.~L., {Madore}, B.~F., {Hatt}, D., {et~al.} 2019,
  \href{http://dx.doi.org/10.3847/1538-4357/ab2f73}{\JournalTitle{\apj}, 882,
  34}

\bibitem[{{Galli} {et~al.}(2014){Galli}, {Benabed}, {Bouchet}, {Cardoso},
  {Elsner}, {Hivon}, {Mangilli}, {Prunet}, \& {Wandelt}}]{galli/etal:2014}
{Galli}, S., {Benabed}, K., {Bouchet}, F., {et~al.} 2014,
  \href{http://dx.doi.org/10.1103/PhysRevD.90.063504}{\JournalTitle{\prd}, 90,
  063504}

\bibitem[{Gluscevic \& Boddy(2018)}]{Gluscevic:2017ywp}
Gluscevic, V., \& Boddy, K.~K. 2018,
  \href{http://dx.doi.org/10.1103/PhysRevLett.121.081301}{\JournalTitle{Phys.
  Rev. Lett.}, 121, 081301}

\bibitem[{{G{\'o}rski} {et~al.}(2005{\natexlab{a}}){G{\'o}rski}, {Hivon},
  {Banday}, {Wand elt}, {Hansen}, {Reinecke}, \& {Bartelmann}}]{healpix_2005}
{G{\'o}rski}, K.~M., {Hivon}, E., {Banday}, A.~J., {et~al.} 2005{\natexlab{a}},
  \href{http://dx.doi.org/10.1086/427976}{\JournalTitle{\apj}, 622, 759}

\bibitem[{{G{\'o}rski} {et~al.}(2005{\natexlab{b}}){G{\'o}rski}, {Hivon},
  {Banday}, {Wandelt}, {Hansen}, {Reinecke}, \& {Bartelmann}}]{Healpix2}
---. 2005{\natexlab{b}},
  \href{http://dx.doi.org/10.1086/427976}{\JournalTitle{\apj}, 622, 759}

\bibitem[{{Grace} {et~al.}(2014){Grace}, {Beall}, {Bond}, {Cho}, {Datta},
  {Devlin}, {D{\"u}nner}, {Fox}, {Gallardo}, {Hasselfield}, {Henderson},
  {Hilton}, {Hincks}, {Hlozek}, {Hubmayr}, {Irwin}, {Klein}, {Koopman}, {Li},
  {Lungu}, {Newburgh}, {Nibarger}, {Niemack}, {Maurin}, {McMahon}, {Naess},
  {Page}, {Pappas}, {Schmitt}, {Sievers}, {Staggs}, {Thornton}, {Van Lanen}, \&
  {Wollack}}]{grace_atacama_2014}
{Grace}, E., {Beall}, J., {Bond}, J.~R., {et~al.} 2014, Society of
  Photo-Optical Instrumentation Engineers (SPIE) Conference Series, Vol. 9153,
  {ACTPol: on-sky performance and characterization}, 915310

\bibitem[{{Hamaker} \& {Bregman}(1996)}]{hamaker/bregman:1996}
{Hamaker}, J.~P., \& {Bregman}, J.~D. 1996, \JournalTitle{\aaps}, 117, 161

\bibitem[{Handley(2019)}]{h2019curvature}
Handley, W. 2019, Curvature tension: evidence for a closed universe,
  \href{http://arxiv.org/abs/1908.09139}{{\sffamily arXiv:1908.09139
  [astro-ph.CO]}}

\bibitem[{Hasselfield {et~al.}(2013)Hasselfield, Moodley, Bond, Das, Devlin,
  Dunkley, Dunner, Fowler, Gallardo, Gralla, Hajian, Halpern, Hincks, Marriage,
  Marsden, Niemack, Nolta, Page, Partridge, Schmitt, Sehgal, Sievers, Staggs,
  Swetz, Switzer, \& Wollack}]{Hasselfield_atacama_2013}
Hasselfield, M., Moodley, K., Bond, J.~R., {et~al.} 2013,
  \href{http://dx.doi.org/10.1088/0067-0049/209/1/17}{\JournalTitle{The
  Astrophysical Journal Supplement Series}, 209, 17}

\bibitem[{{Henning} {et~al.}(2018){Henning}, {Sayre}, {Reichardt}, {Ade},
  {Anderson}, {Austermann}, {Beall}, {Bender}, {Benson}, {Bleem}, {Carlstrom},
  {Chang}, {Chiang}, {Cho}, {Citron}, {Corbett Moran}, {Crawford}, {Crites},
  {de Haan}, {Dobbs}, {Everett}, {Gallicchio}, {George}, {Gilbert},
  {Halverson}, {Harrington}, {Hilton}, {Holder}, {Holzapfel}, {Hoover}, {Hou},
  {Hrubes}, {Huang}, {Hubmayr}, {Irwin}, {Keisler}, {Knox}, {Lee}, {Leitch},
  {Li}, {Lowitz}, {Manzotti}, {McMahon}, {Meyer}, {Mocanu}, {Montgomery},
  {Nadolski}, {Natoli}, {Nibarger}, {Novosad}, {Padin}, {Pryke}, {Ruhl},
  {Saliwanchik}, {Schaffer}, {Sievers}, {Smecher}, {Stark}, {Story}, {Tucker},
  {Vanderlinde}, {Veach}, {Vieira}, {Wang}, {Whitehorn}, {Wu}, \&
  {Yefremenko}}]{henning/etal:2018}
{Henning}, J.~W., {Sayre}, J.~T., {Reichardt}, C.~L., {et~al.} 2018,
  \href{http://dx.doi.org/10.3847/1538-4357/aa9ff4}{\JournalTitle{\apj}, 852,
  97}

\bibitem[{Hikage {et~al.}(2019)}]{Hikage:2018qbn}
Hikage, C., {et~al.} 2019,
  \href{http://dx.doi.org/10.1093/pasj/psz010}{\JournalTitle{Publ. Astron. Soc.
  Jap.}, 71, Publications of the Astronomical Society of Japan, Volume 71,
  Issue 2, April 2019, 43, https://doi.org/10.1093/pasj/psz010}

\bibitem[{Hill {et~al.}(2020)Hill, McDonough, Toomey, \&
  Alexander}]{Hill:2020osr}
Hill, J.~C., McDonough, E., Toomey, M.~W., \& Alexander, S. 2020,
  \href{http://arxiv.org/abs/2003.07355}{{\sffamily arXiv:2003.07355
  [astro-ph.CO]}}

\bibitem[{Hilton {et~al.}(2020)}]{hilton_atacama_2020}
Hilton, M., {et~al.} 2020, \href{http://arxiv.org/abs/2009.11043}{{\sffamily
  arXiv:2009.11043 [astro-ph.CO]}}

\bibitem[{Hinshaw {et~al.}(2013)Hinshaw, Larson, Komatsu, Spergel, Bennett,
  Dunkley, Nolta, Halpern, Hill, Odegard, \& et~al.}]{Hinshaw_2013}
Hinshaw, G., Larson, D., Komatsu, E., {et~al.} 2013,
  \href{http://dx.doi.org/10.1088/0067-0049/208/2/19}{\JournalTitle{The
  Astrophysical Journal Supplement Series}, 208, 19}

\bibitem[{Hlozek {et~al.}(2012)Hlozek, Dunkley, Addison, Appel, Bond, Carvalho,
  Das, Devlin, Dünner, Essinger-Hileman, \& et~al.}]{Hlozek_2012}
Hlozek, R., Dunkley, J., Addison, G., {et~al.} 2012,
  \href{http://dx.doi.org/10.1088/0004-637x/749/1/90}{\JournalTitle{The
  Astrophysical Journal}, 749, 90}

\bibitem[{Hunter(2007)}]{Hunter:2007}
Hunter, J.~D. 2007,
  \href{http://dx.doi.org/10.1109/MCSE.2007.55}{\JournalTitle{Computing in
  Science \& Engineering}, 9, 90}

\bibitem[{{IAU SOFA Board}(2019)}]{SOFA:2019-07-22}
{IAU SOFA Board}. 2019, IAU SOFA Software Collection

\bibitem[{{Ivezi{\'c}} {et~al.}(2019){Ivezi{\'c}}, {Kahn}, {Tyson}, {Abel},
  {Acosta}, {Allsman}, {Alonso}, {AlSayyad}, {Anderson}, {Andrew}, \&
  et~al.}]{lsst_2019}
{Ivezi{\'c}}, {\v Z}., {Kahn}, S.~M., {Tyson}, J.~A., {et~al.} 2019,
  \href{http://dx.doi.org/10.3847/1538-4357/ab042c}{\JournalTitle{\apj}, 873,
  111}

\bibitem[{Knox \& Millea(2020)}]{Knox:2019rjx}
Knox, L., \& Millea, M. 2020,
  \href{http://dx.doi.org/10.1103/PhysRevD.101.043533}{\JournalTitle{Phys. Rev.
  D}, 101, 043533}

\bibitem[{Koopman {et~al.}(2016)Koopman, Austermann, Cho, Coughlin, Duff,
  Gallardo, Hasselfield, Henderson, Ho, Hubmayr, Irwin, Li, McMahon, Nati,
  Niemack, Newburgh, Page, Salatino, Schillaci, Schmitt, Simon, Vavagiakis,
  Ward, \& Wollack}]{koopman_spie_2016}
Koopman, B., Austermann, J., Cho, H.-M., {et~al.} 2016,
  \href{http://dx.doi.org/10.1117/12.2231912}{in Proc. {SPIE}, ed. W.~S.
  Holland \& J.~Zmuidzinas}, 99142T

\bibitem[{Kreisch {et~al.}(2019)Kreisch, Cyr-Racine, \&
  Doré}]{Kreisch:2019yzn}
Kreisch, C.~D., Cyr-Racine, F.-Y., \& Doré, O. 2019,
  \href{http://arxiv.org/abs/1902.00534}{{\sffamily arXiv:1902.00534
  [astro-ph.CO]}}

\bibitem[{{Lesgourgues}(2011)}]{CLASS}
{Lesgourgues}, J. 2011, \JournalTitle{arXiv e-prints}, arXiv:1104.2932

\bibitem[{Lewis(2013)}]{Lewis:2013hha}
Lewis, A. 2013,
  \href{http://dx.doi.org/10.1103/PhysRevD.87.103529}{\JournalTitle{\prd}, 87,
  103529}

\bibitem[{Lewis \& Bridle(2002)}]{Lewis:2002ah}
Lewis, A., \& Bridle, S. 2002,
  \href{http://dx.doi.org/10.1103/PhysRevD.66.103511}{\JournalTitle{\prd}, 66,
  103511}

\bibitem[{{Lewis} {et~al.}(2000){Lewis}, {Challinor}, \& {Lasenby}}]{CAMB}
{Lewis}, A., {Challinor}, A., \& {Lasenby}, A. 2000,
  \href{http://dx.doi.org/10.1086/309179}{\JournalTitle{\apj}, 538, 473}

\bibitem[{Louis {et~al.}(2017)Louis, Grace, Hasselfield, Lungu, Maurin,
  Addison, Ade, Aiola, Allison, Amiri, Angile, Battaglia, Beall, de~Bernardis,
  Bond, Britton, Calabrese, Cho, Choi, Coughlin, Crichton, Crowley, Datta,
  Devlin, Dicker, Dunkley, Dünner, Ferraro, Fox, Gallardo, Gralla, Halpern,
  Henderson, Hill, Hilton, Hilton, Hincks, Hlozek, Ho, Huang, Hubmayr,
  Huffenberger, Hughes, Infante, Irwin, Kasanda, Klein, Koopman, Kosowsky, Li,
  Madhavacheril, Marriage, McMahon, Menanteau, Moodley, Munson, Naess, Nati,
  Newburgh, Nibarger, Niemack, Nolta, Nuñez, Page, Pappas, Partridge, Rojas,
  Schaan, Schmitt, Sehgal, Sherwin, Sievers, Simon, Spergel, Staggs, Switzer,
  Thornton, Trac, Treu, Tucker, Engelen, Ward, \& Wollack}]{louis_atacama_2017}
Louis, T., Grace, E., Hasselfield, M., {et~al.} 2017,
  \href{http://dx.doi.org/10.1088/1475-7516/2017/06/031}{\JournalTitle{Journal
  of Cosmology and Astroparticle Physics}, 2017, 031}

\bibitem[{{Ludwig}(1973)}]{ludwig_1973}
{Ludwig}, A.~C. 1973,
  \href{http://dx.doi.org/10.1109/TAP.1973.1140406}{\JournalTitle{IEEE
  Transactions on Antennas and Propagation}, 21, 116}

\bibitem[{Madhavacheril {et~al.}(2019)}]{Madhavacheril:2019nfz}
Madhavacheril, M.~S., {et~al.} 2019,
  \href{http://arxiv.org/abs/1911.05717}{{\sffamily arXiv:1911.05717
  [astro-ph.CO]}}

\bibitem[{{Makarov} {et~al.}(2014){Makarov}, {Prugniel}, {Terekhova},
  {Courtois}, \& {Vauglin}}]{hyperleda_2014}
{Makarov}, D., {Prugniel}, P., {Terekhova}, N., {Courtois}, H., \& {Vauglin},
  I. 2014,
  \href{http://dx.doi.org/10.1051/0004-6361/201423496}{\JournalTitle{\aap},
  570, A13}

\bibitem[{Naess {et~al.}(2014)Naess, Hasselfield, McMahon, Niemack, Addison,
  Ade, Allison, Amiri, Battaglia, Beall, de~Bernardis, Bond, Britton,
  Calabrese, Cho, Coughlin, Crichton, Das, Datta, Devlin, Dicker, Dunkley,
  Dünner, Fowler, Fox, Gallardo, Grace, Gralla, Hajian, Halpern, Henderson,
  Hill, Hilton, Hilton, Hincks, Hlozek, Ho, Hubmayr, Huffenberger, Hughes,
  Infante, Irwin, Jackson, Kasanda, Klein, Koopman, Kosowsky, Li, Louis, Lungu,
  Madhavacheril, Marriage, Maurin, Menanteau, Moodley, Munson, Newburgh,
  Nibarger, Nolta, Page, Pappas, Partridge, Rojas, Schmitt, Sehgal, Sherwin,
  Sievers, Simon, Spergel, Staggs, Switzer, Thornton, Trac, Tucker, Uehara,
  Engelen, Ward, \& Wollack}]{naess_atacama_2014}
Naess, S., Hasselfield, M., McMahon, J., {et~al.} 2014,
  \href{http://dx.doi.org/10.1088/1475-7516/2014/10/007}{\JournalTitle{Journal
  of Cosmology and Astroparticle Physics}, 2014, 007}

\bibitem[{Naess {et~al.}(2020)}]{naess_atacama_2020}
Naess, S., {et~al.} 2020, \href{http://arxiv.org/abs/2007.07290}{{\sffamily
  arXiv:2007.07290 [astro-ph.IM]}}

\bibitem[{{N{\ae}ss}(2019)}]{naess_mm_2019}
{N{\ae}ss}, S.~K. 2019,
  \href{http://dx.doi.org/10.1088/1475-7516/2019/12/060}{\JournalTitle{\jcap},
  2019, 060}

\bibitem[{Namikawa {et~al.}(2020)}]{Namikawa:2020ffr}
Namikawa, T., {et~al.} 2020,
  \href{http://dx.doi.org/10.1103/PhysRevD.101.083527}{\JournalTitle{Phys. Rev.
  D}, 101, 083527}

\bibitem[{Natale {et~al.}(2020)Natale, Pagano, Lattanzi, Migliaccio, Colombo,
  Gruppuso, Natoli, \& Polenta}]{Natale:2020owc}
Natale, U., Pagano, L., Lattanzi, M., {et~al.} 2020,
  \href{http://arxiv.org/abs/2005.05600}{{\sffamily arXiv:2005.05600
  [astro-ph.CO]}}

\bibitem[{{Pagano} {et~al.}(2019){Pagano}, {Delouis}, {Mottet}, {Puget}, \&
  {Vibert}}]{2019arXiv190809856P}
{Pagano}, L., {Delouis}, J.~M., {Mottet}, S., {Puget}, J.~L., \& {Vibert}, L.
  2019, \JournalTitle{arXiv e-prints},
  \href{http://arxiv.org/abs/1908.09856}{{\sffamily arXiv:1908.09856}}

\bibitem[{Pisanti {et~al.}(2008)Pisanti, Cirillo, Esposito, Iocco, Mangano,
  Miele, \& Serpico}]{Pisanti:2007hk}
Pisanti, O., Cirillo, A., Esposito, S., {et~al.} 2008,
  \href{http://dx.doi.org/10.1016/j.cpc.2008.02.015}{\JournalTitle{Comput.
  Phys. Commun.}, 178, 956}

\bibitem[{Pitrou {et~al.}(2018)Pitrou, Coc, Uzan, \& Vangioni}]{Pitrou:2018cgg}
Pitrou, C., Coc, A., Uzan, J.-P., \& Vangioni, E. 2018,
  \href{http://dx.doi.org/10.1016/j.physrep.2018.04.005}{\JournalTitle{Phys.
  Rept.}, 754, 1}

\bibitem[{{Planck Collaboration}(2020)}]{Plancknpipe}
{Planck Collaboration}. 2020, \JournalTitle{arXiv e-prints},
  \href{http://arxiv.org/abs/2007.04997}{{\sffamily arXiv:2007.04997}}

\bibitem[{{Planck Collaboration I}(2018)}]{planck2016-l01}
{Planck Collaboration I}. 2018, \JournalTitle{\aap, submitted},
  \href{http://arxiv.org/abs/1807.06205}{{\sffamily arXiv:1807.06205}}

\bibitem[{{Planck Collaboration V}(2018)}]{planck2016-l05}
{Planck Collaboration V}. 2018, \JournalTitle{\aap, submitted},
  \href{http://arxiv.org/abs/1907.12875}{{\sffamily arXiv:1907.12875}}

\bibitem[{{Planck Collaboration VI}(2018)}]{planck2016-l06}
{Planck Collaboration VI}. 2018, \JournalTitle{\aap, submitted},
  \href{http://arxiv.org/abs/1807.06209}{{\sffamily arXiv:1807.06209}}

\bibitem[{{Planck Collaboration VIII}(2018)}]{planck2016-l08}
{Planck Collaboration VIII}. 2018, \JournalTitle{\aap, in press},
  \href{http://arxiv.org/abs/1807.06210}{{\sffamily arXiv:1807.06210}}

\bibitem[{{Planck Collaboration XI}(2016)}]{planck2014-a13}
{Planck Collaboration XI}. 2016,
  \href{http://dx.doi.org/10.1051/0004-6361/201526926}{\JournalTitle{\aap},
  594, A11}

\bibitem[{{Planck Collaboration XV}(2014)}]{planck2013-p08}
{Planck Collaboration XV}. 2014,
  \href{http://dx.doi.org/10.1051/0004-6361/201321573}{\JournalTitle{\aap},
  571, A15}

\bibitem[{{Poulin} {et~al.}(2019){Poulin}, {Smith}, {Karwal}, \&
  {Kamionkowski}}]{poulin/etal:2019}
{Poulin}, V., {Smith}, T.~L., {Karwal}, T., \& {Kamionkowski}, M. 2019,
  \href{http://dx.doi.org/10.1103/PhysRevLett.122.221301}{\JournalTitle{\prl},
  122, 221301}

\bibitem[{{Price-Whelan} {et~al.}(2018){Price-Whelan}, {Sip{\H{o}}cz},
  {G{\"u}nther}, {Lim}, {Crawford}, {Conseil}, {Shupe}, {Craig}, {Dencheva},
  {Ginsburg}, {VanderPlas}, {Bradley}, {P{\'e}rez-Su{\'a}rez}, {de Val-Borro},
  {Paper Contributors}, {Aldcroft}, {Cruz}, {Robitaille}, {Tollerud},
  {Coordination Committee}, {Ardelean}, {Babej}, {Bach}, {Bachetti}, {Bakanov},
  {Bamford}, {Barentsen}, {Barmby}, {Baumbach}, {Berry}, {Biscani}, {Boquien},
  {Bostroem}, {Bouma}, {Brammer}, {Bray}, {Breytenbach}, {Buddelmeijer},
  {Burke}, {Calderone}, {Cano Rodr{\'\i}guez}, {Cara}, {Cardoso}, {Cheedella},
  {Copin}, {Corrales}, {Crichton}, {D{\textquoteright}Avella}, {Deil},
  {Depagne}, {Dietrich}, {Donath}, {Droettboom}, {Earl}, {Erben}, {Fabbro},
  {Ferreira}, {Finethy}, {Fox}, {Garrison}, {Gibbons}, {Goldstein}, {Gommers},
  {Greco}, {Greenfield}, {Groener}, {Grollier}, {Hagen}, {Hirst}, {Homeier},
  {Horton}, {Hosseinzadeh}, {Hu}, {Hunkeler}, {Ivezi{\'c}}, {Jain}, {Jenness},
  {Kanarek}, {Kendrew}, {Kern}, {Kerzendorf}, {Khvalko}, {King}, {Kirkby},
  {Kulkarni}, {Kumar}, {Lee}, {Lenz}, {Littlefair}, {Ma}, {Macleod},
  {Mastropietro}, {McCully}, {Montagnac}, {Morris}, {Mueller}, {Mumford},
  {Muna}, {Murphy}, {Nelson}, {Nguyen}, {Ninan}, {N{\"o}the}, {Ogaz}, {Oh},
  {Parejko}, {Parley}, {Pascual}, {Patil}, {Patil}, {Plunkett}, {Prochaska},
  {Rastogi}, {Reddy Janga}, {Sabater}, {Sakurikar}, {Seifert}, {Sherbert},
  {Sherwood-Taylor}, {Shih}, {Sick}, {Silbiger}, {Singanamalla}, {Singer},
  {Sladen}, {Sooley}, {Sornarajah}, {Streicher}, {Teuben}, {Thomas},
  {Tremblay}, {Turner}, {Terr{\'o}n}, {van Kerkwijk}, {de la Vega}, {Watkins},
  {Weaver}, {Whitmore}, {Woillez}, {Zabalza}, \& {Contributors}}]{astropy:2018}
{Price-Whelan}, A.~M., {Sip{\H{o}}cz}, B.~M., {G{\"u}nther}, H.~M., {et~al.}
  2018, \href{http://dx.doi.org/10.3847/1538-3881/aabc4f}{\JournalTitle{\aj},
  156, 123}

\bibitem[{{Reinecke} \& {Seljebotn}(2013)}]{reinecke/2013}
{Reinecke}, M., \& {Seljebotn}, D.~S. 2013,
  \href{http://dx.doi.org/10.1051/0004-6361/201321494}{\JournalTitle{\aap},
  554, A112}

\bibitem[{{Riess} {et~al.}(2019){Riess}, {Casertano}, {Yuan}, {Macri}, \&
  {Scolnic}}]{riess/etal:2019}
{Riess}, A.~G., {Casertano}, S., {Yuan}, W., {Macri}, L.~M., \& {Scolnic}, D.
  2019, \href{http://dx.doi.org/10.3847/1538-4357/ab1422}{\JournalTitle{\apj},
  876, 85}

\bibitem[{Ross {et~al.}(2015)Ross, Samushia, Howlett, Percival, Burden, \&
  Manera}]{Ross_2015}
Ross, A.~J., Samushia, L., Howlett, C., {et~al.} 2015,
  \href{http://dx.doi.org/10.1093/mnras/stv154}{\JournalTitle{Monthly Notices
  of the Royal Astronomical Society}, 449, 835–847}

\bibitem[{{Ruze}(1966)}]{ruze_antenna_1966}
{Ruze}, J. 1966, \JournalTitle{IEEE Proceedings}, 54, 633

\bibitem[{{Schmid} {et~al.}(2006){Schmid}, {Joos}, \& {Tschan}}]{schmid_2006}
{Schmid}, H.~M., {Joos}, F., \& {Tschan}, D. 2006,
  \href{http://dx.doi.org/10.1051/0004-6361:20053273}{\JournalTitle{\aap}, 452,
  657}

\bibitem[{{Sievers} {et~al.}(2013){Sievers}, {Hlozek}, {Nolta}, {Acquaviva},
  {Addison}, {Ade}, {Aguirre}, {Amiri}, {Appel}, {Barrientos}, {Battistelli},
  {Battaglia}, {Bond}, {Brown}, {Burger}, {Calabrese}, {Chervenak}, {Crichton},
  {Das}, {Devlin}, {Dicker}, {Bertrand Doriese}, {Dunkley}, {D{\"u}nner},
  {Essinger-Hileman}, {Faber}, {Fisher}, {Fowler}, {Gallardo}, {Gordon},
  {Gralla}, {Hajian}, {Halpern}, {Hasselfield}, {Hern{\'a}ndez-Monteagudo},
  {Hill}, {Hilton}, {Hilton}, {Hincks}, {Holtz}, {Huffenberger}, {Hughes},
  {Hughes}, {Infante}, {Irwin}, {Jacobson}, {Johnstone}, {Baptiste Juin},
  {Kaul}, {Klein}, {Kosowsky}, {Lau}, {Limon}, {Lin}, {Louis}, {Lupton},
  {Marriage}, {Marsden}, {Martocci}, {Mauskopf}, {McLaren}, {Menanteau},
  {Moodley}, {Moseley}, {Netterfield}, {Niemack}, {Page}, {Page}, {Parker},
  {Partridge}, {Plimpton}, {Quintana}, {Reese}, {Reid}, {Rojas}, {Sehgal},
  {Sherwin}, {Schmitt}, {Spergel}, {Staggs}, {Stryzak}, {Swetz}, {Switzer},
  {Thornton}, {Trac}, {Tucker}, {Uehara}, {Visnjic}, {Warne}, {Wilson},
  {Wollack}, {Zhao}, \& {Zunckel}}]{sievers/etal:2013}
{Sievers}, J.~L., {Hlozek}, R.~A., {Nolta}, M.~R., {et~al.} 2013,
  \href{http://dx.doi.org/10.1088/1475-7516/2013/10/060}{\JournalTitle{\jcap},
  2013, 060}

\bibitem[{{Spergel} {et~al.}(2003){Spergel}, {Verde}, {Peiris}, {Komatsu},
  {Nolta}, {Bennett}, {Halpern}, {Hinshaw}, {Jarosik}, {Kogut}, {Limon},
  {Meyer}, {Page}, {Tucker}, {Weiland}, {Wollack}, \&
  {Wright}}]{wmap_spergel_2003}
{Spergel}, D.~N., {Verde}, L., {Peiris}, H.~V., {et~al.} 2003,
  \href{http://dx.doi.org/10.1086/377226}{\JournalTitle{\apjs}, 148, 175}

\bibitem[{Thornton {et~al.}(2016)Thornton, Ade, Aiola, Angilè, Amiri, Beall,
  Becker, Cho, Choi, Corlies, Coughlin, Datta, Devlin, Dicker, Dünner, Fowler,
  Fox, Gallardo, Gao, Grace, Halpern, Hasselfield, Henderson, Hilton, Hincks,
  Ho, Hubmayr, Irwin, Klein, Koopman, Li, Louis, Lungu, Maurin, McMahon,
  Munson, Naess, Nati, Newburgh, Nibarger, Niemack, Niraula, Nolta, Page,
  Pappas, Schillaci, Schmitt, Sehgal, Sievers, Simon, Staggs, Tucker, Uehara,
  Lanen, Ward, \& Wollack}]{thornton_atacama_2016}
Thornton, R.~J., Ade, P. A.~R., Aiola, S., {et~al.} 2016,
  \href{http://dx.doi.org/10.3847/1538-4365/227/2/21}{\JournalTitle{The
  Astrophysical Journal Supplement Series}, 227, 21}

\bibitem[{{Thornton} {et~al.}(2016){Thornton}, {Ade}, {Aiola}, {Angil{\`e}},
  {Amiri}, {Beall}, {Becker}, {Cho}, {Choi}, {Corlies}, {Coughlin}, {Datta},
  {Devlin}, {Dicker}, {D{\"u}nner}, {Fowler}, {Fox}, {Gallardo}, {Gao},
  {Grace}, {Halpern}, {Hasselfield}, {Henderson}, {Hilton}, {Hincks}, {Ho},
  {Hubmayr}, {Irwin}, {Klein}, {Koopman}, {Li}, {Louis}, {Lungu}, {Maurin},
  {McMahon}, {Munson}, {Naess}, {Nati}, {Newburgh}, {Nibarger}, {Niemack},
  {Niraula}, {Nolta}, {Page}, {Pappas}, {Schillaci}, {Schmitt}, {Sehgal},
  {Sievers}, {Simon}, {Staggs}, {Tucker}, {Uehara}, {van Lanen}, {Ward}, \&
  {Wollack}}]{thornton/2016}
{Thornton}, R.~J., {Ade}, P.~A.~R., {Aiola}, S., {et~al.} 2016,
  \href{http://dx.doi.org/10.3847/1538-4365/227/2/21}{\JournalTitle{\apjs},
  227, 21}

\bibitem[{Torrado \& Lewis(2020)}]{Cobaya}
Torrado, J., \& Lewis, A. 2020,
  \href{http://arxiv.org/abs/2005.05290}{{\sffamily arXiv:2005.05290
  [astro-ph.IM]}}

\bibitem[{{Wong} {et~al.}(2019){Wong}, {Suyu}, {Chen}, {Rusu}, {Millon},
  {Sluse}, {Bonvin}, {Fassnacht}, {Taubenberger}, {Auger}, {Birrer}, {Chan},
  {Courbin}, {Hilbert}, {Tihhonova}, {Treu}, {Agnello}, {Ding}, {Jee},
  {Komatsu}, {Shajib}, {Sonnenfeld}, {Bland ford}, {Koopmans}, {Marshall}, \&
  {Meylan}}]{wong/etal:2019}
{Wong}, K.~C., {Suyu}, S.~H., {Chen}, G. C.~F., {et~al.} 2019,
  \JournalTitle{arXiv e-prints},
  \href{http://arxiv.org/abs/1907.04869}{{\sffamily arXiv:1907.04869}}

\bibitem[{Zonca {et~al.}(2019)Zonca, Singer, Lenz, Reinecke, Rosset, Hivon, \&
  Gorski}]{Healpix1}
Zonca, A., Singer, L., Lenz, D., {et~al.} 2019,
  \href{http://dx.doi.org/10.21105/joss.01298}{\JournalTitle{Journal of Open
  Source Software}, 4, 1298}

\bibitem[{{Zuntz} {et~al.}(2015){Zuntz}, {Paterno}, {Jennings}, {Rudd},
  {Manzotti}, {Dodelson}, {Bridle}, {Sehrish}, \& {Kowalkowski}}]{cosmosis2015}
{Zuntz}, J., {Paterno}, M., {Jennings}, E., {et~al.} 2015,
  \href{http://dx.doi.org/10.1016/j.ascom.2015.05.005}{\JournalTitle{Astronomy
  and Computing}, 12, 45}

\end{thebibliography}
\newpage
\appendix
\section{A. The constraining power of ACT DR4}
\label{sec:ACTmodes}
In this Appendix we report signal-to-noise (S/N) estimates for the ACT, \wmap\ and \planck\ spectra, supplementing Fig.~20 in C20 that compares the ACT DR4 power spectrum error bars with those from \planck\ and \wmap. To assess the overall significance of the data we compute the $\chi^2$ of the TT, TE and EE spectra compared to null (i.e., how significant the measurement is when compared to no signal) and define S/N as its square root, given in Table \ref{tab:sn}. We estimate this for both \planck\ and ACT using their CMB-only likelihoods which account for the uncertainty due to foregrounds, calibration, polarization efficiency and beams. The equivalent for \wmap\ is estimated from the full \wmap\ likelihood.

\begin{table}[ht!]
\caption{Signal-to-noise estimates for ACT, WMAP and Planck CMB spectra.}
\vspace{-0.1in}
\begin{center}
\begin{tabular}{r|ccc|cccc|cccc}
\hline
\hline
\rule{0pt}{3ex}S/N & \multicolumn{3}{c|}{WMAP} & \multicolumn{4}{c|}{Planck} & \multicolumn{4}{c}{ACT} \\
& TT & TE & EE & TT & TE & EE & TT+TE+EE & TT & TE & EE & TT+TE+EE \\
\hline
\rule{0pt}{3ex}& 273 & 23 & 12 & 606 & 182 & 236 & 646 & 175 & 76 & 112 & 185 \\
\hline
\hline
\end{tabular}
\end{center}
\label{tab:sn}
\end{table}

We find that: (i) the S/N for all measurements is greatest in TT; (ii) \planck\ has overall greater constraining power than ACT in TE and EE (even if limited to the same $\ell$ range, i.e. $\ell>300$); (iii) ACT has roughly comparable constraining power in EE and TE and comparable to \wmap\ in TT; (iv) ACT's current overall constraining power is about 1/3 that of \planck's.

\begin{figure*}[htbp!]
    \centering
    \includegraphics[width=0.45\columnwidth]{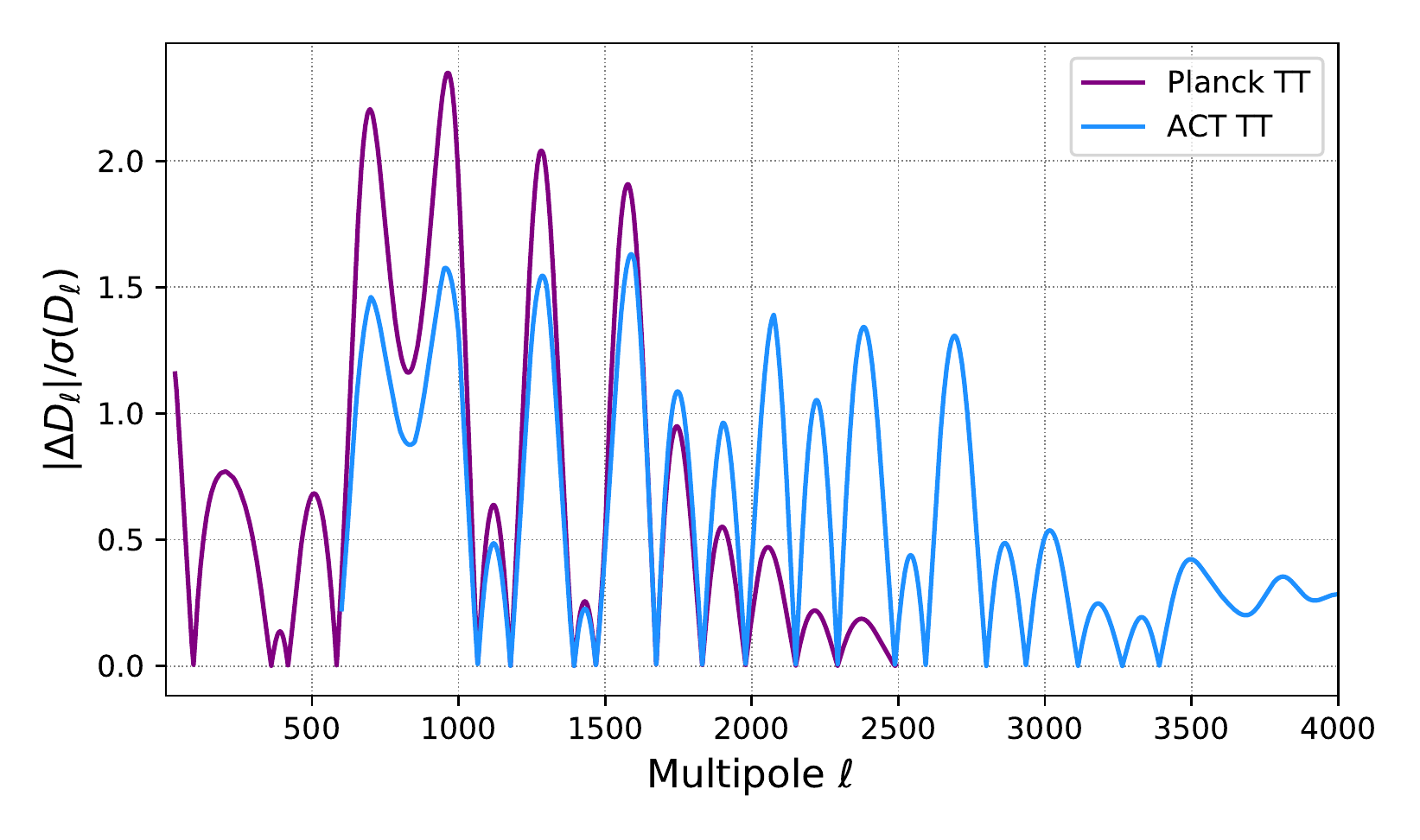}
    \includegraphics[width=0.45\columnwidth]{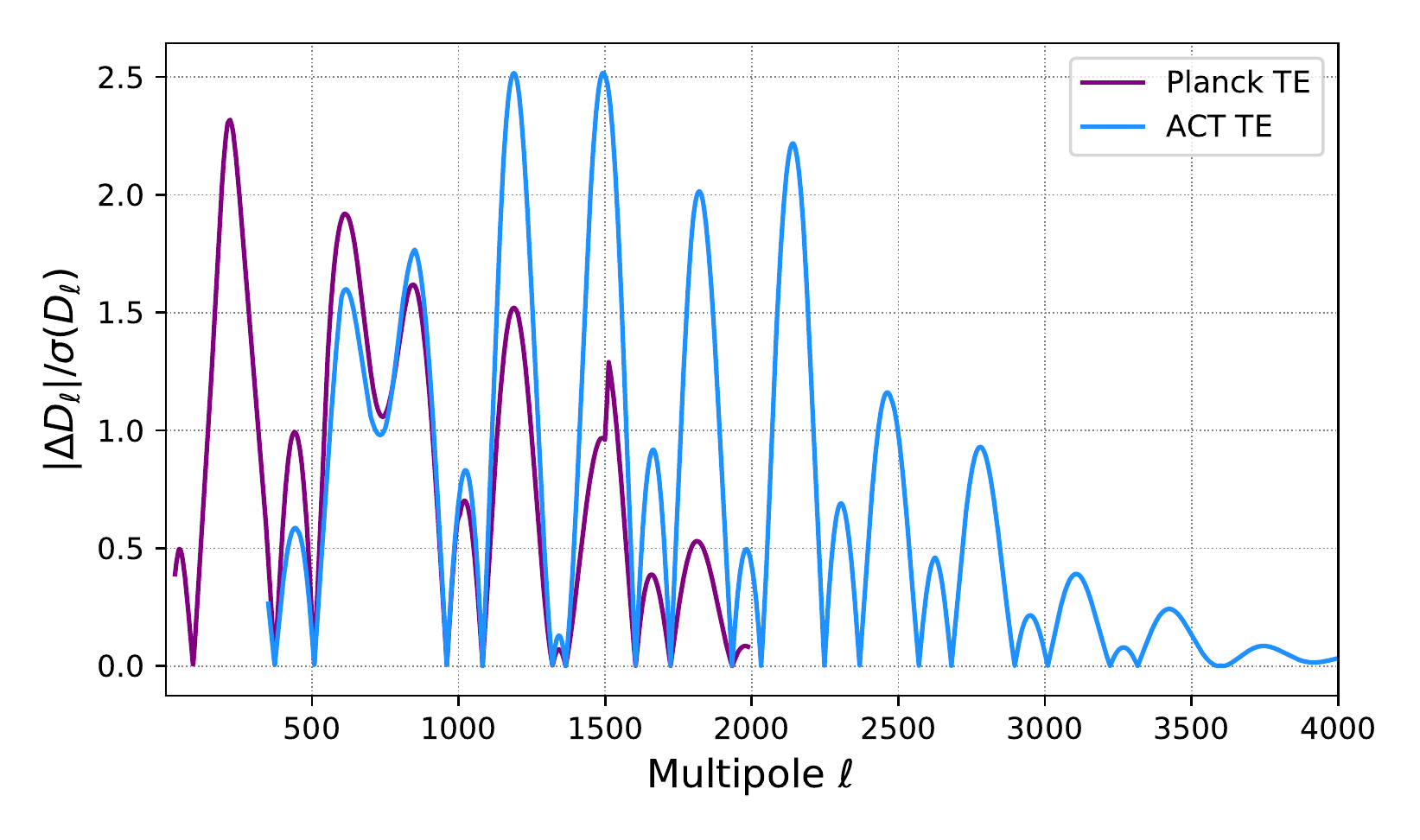}
    \includegraphics[width=0.45\columnwidth]{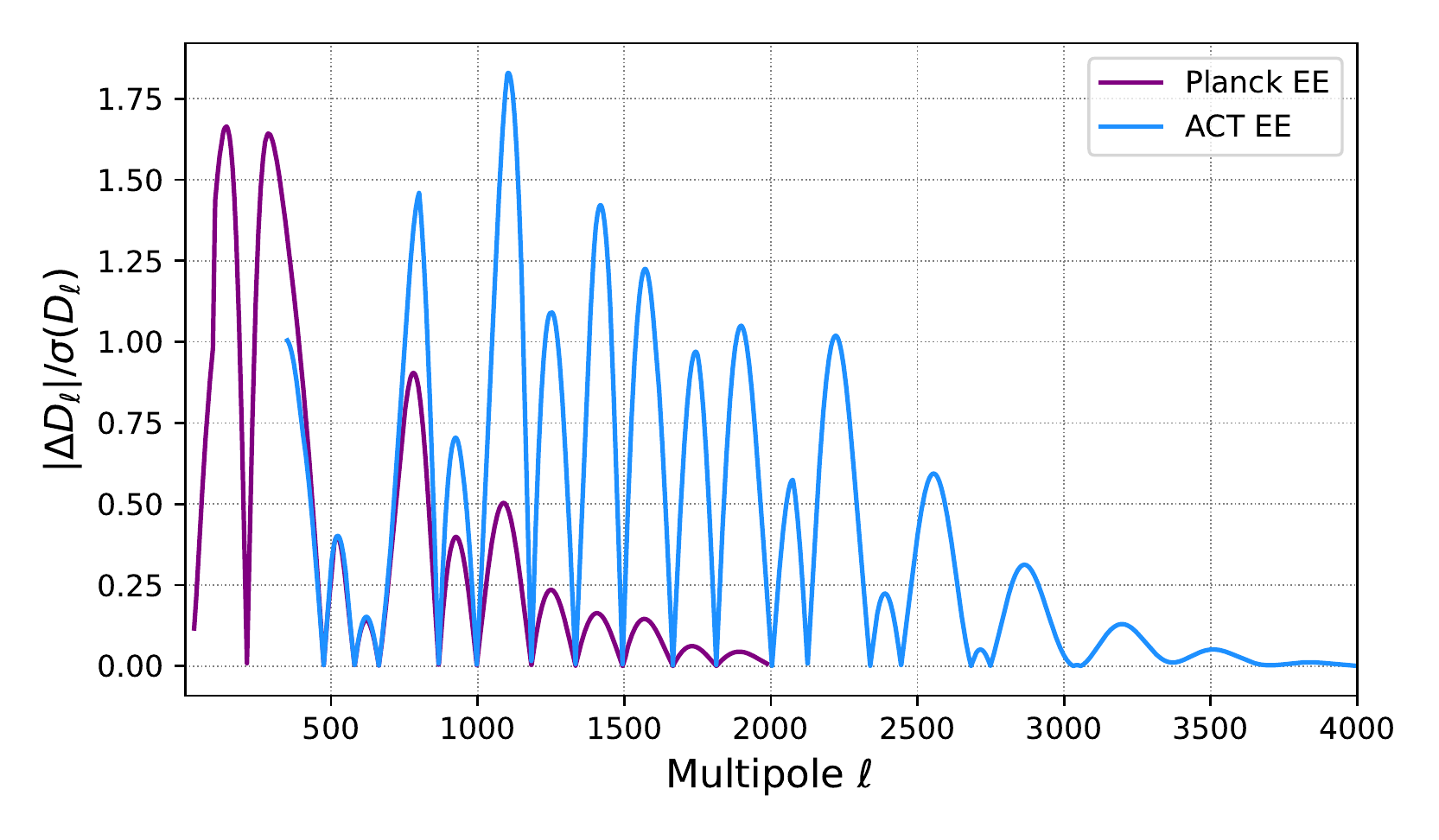}
    \caption{ACT and Planck signal-to-noise contributing to the estimate of the Hubble constant as function of multipole. The relative constraining power of each experiment is shown with the difference between two theoretical models with widely-spaced Hubble constants (63 and 76 km/s/Mpc, both consistent with the \wmap\ data to within $2\sigma$) normalized by the power spectrum error bar. This highlights the new information in the ACT DR4 data and their complementary with Planck: the largest contribution from ACT comes from the TE spectrum in the range $1000<\ell<2500$ and the small scales in TT at $\ell>2000$ and EE at $\ell>1000$ provide additional information.}
    \label{fig:contraint_model}
\end{figure*}

What is not captured in the above is the constraining power of each spectrum as a function of multipole. We explore this by taking two \LCDM\ cosmological models with the most different inferred Hubble constants that fit the \wmap\ data to within 2$\sigma$ of the best-fitting model. We then take the difference between these two models, and in Fig.~\ref{fig:contraint_model} show the difference normalized by the error on the spectra measured by \planck\ and ACT. This then gives a relative measure of the S/N for estimating the Hubble constant, the parameter that we are particularly interested in, as a function of multipole.

In comparing \planck\ to ACT, we find that the \planck\ data have the most constraining power at angular scales $\ell<1800$, with the ACT data constraining at smaller scales. In TE, the transition is at $\ell \sim 1000$, and in EE at $\ell \sim 600$. Figure~\ref{fig:contraint_model} also shows the relative importance of TT, TE and EE. Despite the TT S/N itself being higher than TE and EE, it is the TE that has higher S/N for ACT in terms of determining the Hubble constant, and other cosmological parameters. This is consistent with the \LCDM\ parameter constraints reported in C20, and with previous studies in e.g., \citet{galli/etal:2014}.

\section{B. How the ACT data disfavor a Hubble constant $H_0=74$~km/s/Mpc within \LCDM.}\label{sec:H074}
\begin{figure}[htp!]
    \centering
    \includegraphics[width=0.5\columnwidth]{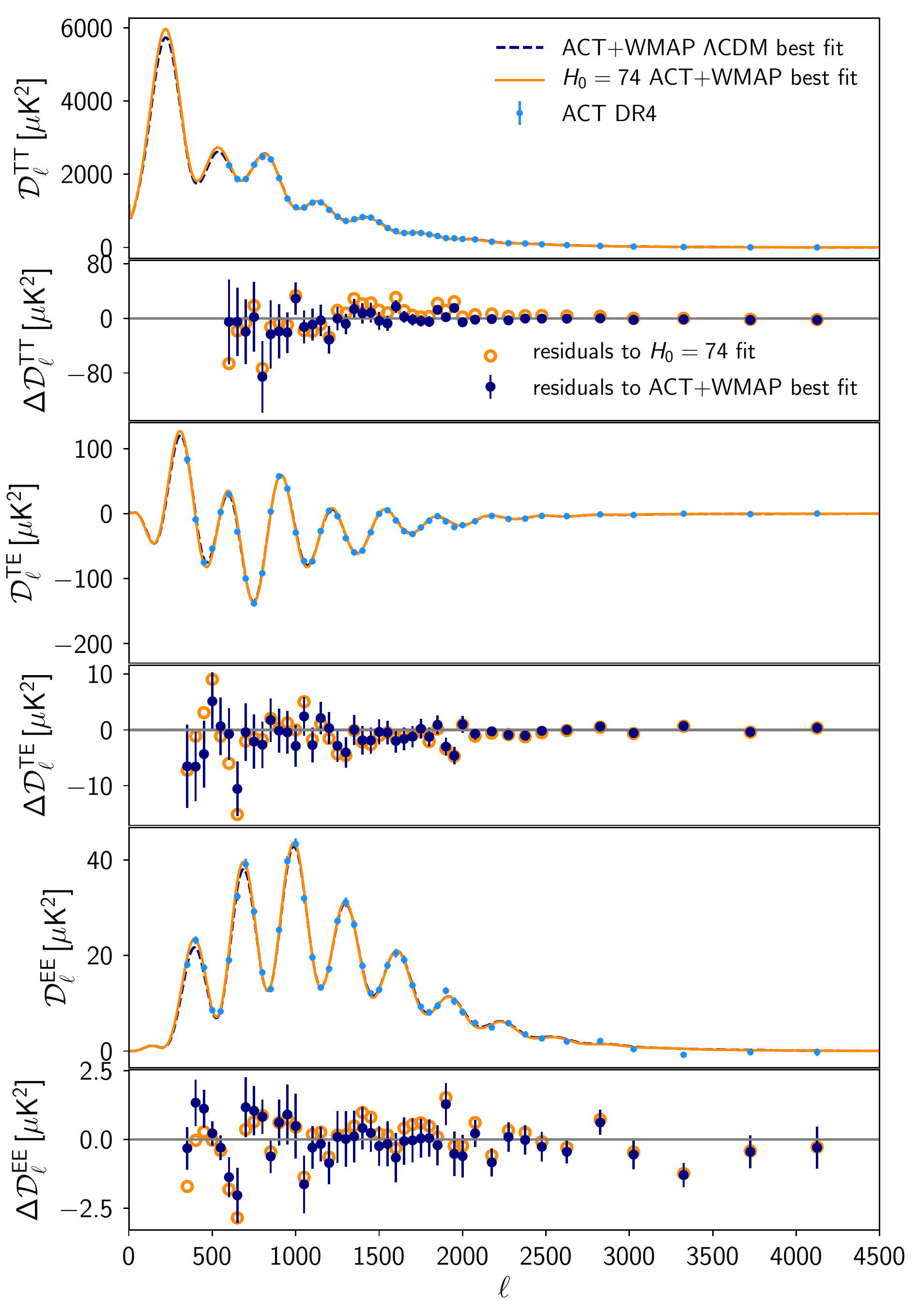}
    \caption{Comparison between the ACT+WMAP best-fit $\Lambda$CDM model and the ACT+WMAP best-fit model with $H_0$ fixed to 74~km/s/Mpc. The residuals show significant degradation of the fit to the ACT data.}
    \label{fig:H074}
\end{figure}

The latest Cepheid-based estimate of the Hubble constant gives $H_0=74.03 \pm 1.42$~km/s/Mpc ~\citep{riess/etal:2019}. Here we show how such a high value of $H_0$ would compare with our data and our nominal \LCDM\ fit. Figure~\ref{fig:H074} compares the ACT+WMAP best-fitting \LCDM\ model and its residuals compared to the ACT data, to the equivalent for a \LCDM\ fit where $H_0$ has been fixed to $H_0=74$~km/s/Mpc. 

As described in \S~\ref{subsec:hubble}, such a model will have a lower matter density to fit the well-measured peak scale. This high-$H_0$ model also has a higher tilt and baryon density, attempting to compensate for effects of the changed matter density, resulting in a theory model that has more power in the first peaks and less power at small-scales than the best-fitting model. Overall this high-$H_0$ fit degrades the ACT fit by $\Delta \chi^2=55$ (48 in TT, 16 in TE, 7 in EE), with excess power in the residuals at TT and EE $\ell>1000$.

\end{document}